\pdfoutput=1 
\documentclass[11pt, twoside]{mathdoc}
\usepackage{distoperators}

\usepackage[title]{appendix}
\usepackage{layouts}
\usepackage{IEEEtrantools}
\usepackage[shortlabels, inline]{enumitem}
\usepackage{bm,mathrsfs,bbm}
\usepackage[font=small]{caption}
\usepackage{array,booktabs,multirow}
\usepackage{tikz}
\usetikzlibrary{bayesnet}
\usetikzlibrary{arrows}
\usepackage{placeins}
\usepackage{multirow}

\usepackage{graphicx}
\usepackage{lscape}
\usepackage{arydshln}
\usepackage{caption}
\usepackage{subcaption}
\usepackage{flafter}

\usepackage{algorithm}
\usepackage{algpseudocode}

\definecolor{arr2flatcolour}{HTML}{4477AA}
\definecolor{arr2minncolour}{HTML}{EE6677}
\definecolor{gaussiancolour}{HTML}{228833}
\definecolor{minnesotacolour}{HTML}{CCBB44}
\definecolor{rhscolour}{HTML}{AA3377}

\theoremstyle{plain}

\newtheorem{theorem}{Theorem}[section]

\theoremstyle{definition}
\newtheorem{definition}[theorem]{Definition}

\theoremstyle{remark}

\DeclareMathOperator{\rmse}{RMSE}

\DeclareMathOperator{\VAR}{VAR}
\DeclareMathOperator{\QVP}{QVP}
\DeclareMathOperator{\NCQVP}{NC-QVP}
\DeclareMathOperator{\CQR}{CQR}
\DeclareMathOperator{\BQR}{BQR}
\DeclareMathOperator{\HSBQR}{HSBQR}
\DeclareMathOperator{\QVAR}{QVAR}
\DeclareMathOperator{\QIRF}{QIRF}
\DeclareMathOperator{\IP}{IP}
\DeclareMathOperator{\CISS}{CISS}
\DeclareMathOperator{\MCMC}{MCMC}
\DeclareMathOperator{\QS}{QS}
\DeclareMathOperator{\BRW}{BRW}

\newcommand{\GIG}[1]{\text{\normalfont GiG}\left(#1\right)}
\newcommand{\IG}[1]{\text{\normalfont IG}\left(#1\right)}
\newcommand{\vect}[1]{\text{\normalfont vec}\left(#1\right)}

\newcommand{\norm}[1]{\left\lVert#1\right\rVert}

\DeclareUnicodeCharacter{2212}{-}

\title{Joint Quantile Shrinkage: A State-Space Approach toward Non-Crossing Bayesian Quantile Models}
\author[1]{David Kohns}
\author[2]{Tibor Szendrei}
\affil[1]{Department of Computer Science, Aalto University}
\affil[2]{National Institute of Economic and Social Research, London}

\makeatletter
\def\runauthor{Kohns and Szendrei}
\let\inserttitle\@title
\def\runtitle{The QVP prior}
\makeatother

\begin{document}
\maketitle
\thispagestyle{empty}
\begin{abstract}
    Crossing of fitted conditional quantiles is a prevalent problem for quantile regression models. We propose a new  Bayesian modelling framework that penalises multiple quantile regression functions toward the desired non-crossing space. We achieve this by estimating multiple quantiles jointly with a prior on variation across quantiles, a fused shrinkage prior with quantile adaptivity. The posterior is derived from a decision-theoretic general Bayes perspective, whose form yields a natural state-space interpretation aligned with Time-Varying Parameter ($\mathrm{TVP}$) models. Taken together our approach leads to a Quantile-Varying Parameter ($\QVP$) model, for which we develop efficient sampling algorithms. 
We demonstrate that our proposed modelling framework provides superior parameter recovery and predictive performance compared to competing Bayesian and frequentist quantile regression estimators in simulated experiments and a real-data application to multivariate quantile estimation in macroeconomics.

\end{abstract}
\section{Introduction}
Quantile regression estimates the conditional quantile function of a response variable given a set of covariates. It is a powerful tool for inference on the relationship between response and covariates, especially in the presence of non-linearity in the covariates' impacts across the distribution of the response. While independent estimation of quantiles has been the norm in the applied literature \citep{koenker2005}, the presence of non-monotonically increasing fitted quantile functions, referred to as quantile crossing, remains largely unaddressed. 
The probability of observing the crossing problem increases with the number of conditional quantiles estimated and the dimensionality of the covariate set \citet{wang2024composite}. Many constrained optimisation solutions have been suggested in the literature, yet full probabilistic inference remains a challenge, particularly when retaining the assumption of non-parametrically modelling the error distribution. In this paper, we propose a general framework to joint quantile regression where we use the connection between a constrained multiple quantile objective function and an implied negative log posterior to motivate a novel prior structure that penalises crossing of fitted quantiles. We name the suggested prior approach the quantile-varying-parameter ($\QVP$) prior due to the connection to time-varying-parameters ($\mathrm{TVP}$) models popular in the state space literature.  Compared to previous approaches, crossing is penalised via the structure of the prior instead of the structure of the likelihood. In particular, in this paper we:
\begin{enumerate}
    \item Establish that the $\QVP$ prior can be regarded as a Bayesian adaptation of the fused quantile lasso model of \citet{jiang2013interquantile} and the composite quantile regression estimator of \citet{zou2008composite} and establish the prior's shrinkage properties
    \item Provide a state-space representation of the $\QVP$ prior, which allows for efficient posterior sampling algorithms 
    \item Compare the the $\QVP$ prior to popular alternatives for Bayesian and frequentist quantile regression approaches in terms of parameter recovery and prediction accuracy in simulated and real-world data experiments
\end{enumerate}
The simulated and real data experiments confirm superior inference and prediction performance with our joint prior approach compared to commonly used Bayesian and frequentist models. For the real-world data application, we extend the methods to the Quantile vector-autoregressive (QVAR) model with Euro Area data presented in \citet{chavleishvili2024forecasting}. 
%
%
\subsection{Structure of the Paper}
In section \ref{subsec:background} we begin by discussing previous approaches to quantile regression as background to this work.
In Section~\ref{sec:motivation} we show that one can view a probabilistic generalisation of the non-crossing constraint as a prior that penalises differences across coefficients of quantiles. \ref{sec:likelihood}, shows that the pseudo-likelihood can be  derived from the objective function of interest. 
We show that this likelihood has a convenient representation as a mixture of normals.
Section~\ref{sec:mcmc} derives an efficient posterior sampling algorithm.
Section~\ref{sec:alt-parameterisation} presents an alternative representation of the $\QVP$ model that offers improved sampling efficiency and shrinkage properties when the data imply low amount of quantile variation.
In Section~\ref{sec:savs}, we discuss post-processing methods for achieving exact sparsity of the $\QVP$ parameter posteriors for improved inference when sparsity in the quantile coefficient vector is suspected.
In Section~\ref{sec:theoretical-properties}, we investigate the theoretical shrinkage properties of the $\QVP$ prior.
We investigate finite sample performance in Section~\ref{sec:simulation}.
We apply the methods presented to multivariate target data, where we   estimate quantile vector autoregressive models ($\QVAR$) with real world data in Section~\ref{sec:application}.
We conclude in Section~\ref{sec:conclusion}.
\subsection{Background and Previous Work} \label{subsec:background}
\citet{yu2001bayesian} established that the commonly used tick-loss quantile regression objective function implies an asymmetric-Laplace distribution ($\mathcal{ALD}$) \citep{kotz2001asymmetric} as a likelihood function. Treated as a working-likelihood,\footnote{Inference on a set of quantile regression coefficients $\beta_q$ for a quantile index $q$, where the percentile $\tau_q \in (0,1)$, is asymptotically equivalent to frequentist treatment of the quantile objective, when treating the $\mathcal{ALD}$ as a working likelihood under conditions discussed in \citet{sriram2013posterior}.} Bayesian inference on quantile regression models has become ubiquitous. The more recent focus being on priors for high dimensional problems \citep{kohns2024horseshoe,alhamzawi2015model,li2010bayesian}. Similar to the literature on normal observation models, those priors are designed to heavily shrink coefficient of noise variables to zero \citep{polson2010shrink}. Yet  these approaches assume that quantile functions are independent, and therefore do not  address the problem of crossing of conditional quantiles.

Many probabilistic methods have been put forward to address the issue of crossing fitted quantile functions. These generally fall into the class of semi-parametric \citep{reich2011bayesian,reich2012spatiotemporal,reich2013bayesian,kottas2001bayesian,yang2017joint}, fully non-parametric \citep{scaccia2003bayesian,taddy2010bayesian}, empirical likelihood \citep{lancaster2010bayesian,yang2012bayesian,yang2015quantile} as well as two-step methods \citep{reich2013bayesian,rodrigues2017regression}. 

A common perspective taken in the semi-parametric quantile literature is to centre the quantile process for a set of finite number of  percentiles on a fully parametric model which is linked piece-wise via some valid quantile function, such as the normal quantile function \citep{reich2013bayesian}. This approach has some notable drawbacks. One such drawback is that the likelihood may not be available in closed form \citep{reich2011bayesian} necessitating approximation methods. An additional drawback is that conditional posteriors are not available in closed form, thus prohibiting efficient updating via Gibbs MCMC methods \citep{reich2012spatiotemporal,reich2013bayesian}. Computational complexity is also the bottleneck for the empirical likelihood and full non-parametric methods, even in moderate dimensions \citep{rodrigues2017regression}.
\citet{rodrigues2017regression} propose a computationally more convenient approach, in which the first stage of estimating individual quantiles are estimated with an $\mathcal{ALD}$ likelihood. These are combined with a Gaussian process into a valid joint density in a secon step. Their approach maintains valid frequentist coverage. Yet quality of inference heavily depends on the first stage, and joint estimation in the first step has been shown to significantly improve inference even for individual quantiles \citep{bondell2010noncrossing}.

Closer to our approach are the methods presented in \citet{wu2021bayesian} as well as \citet{wang2024composite} where quantiles are estimated jointly, and information is shared via a prior on the differences. While \citet{wu2021bayesian} take a moment-based approach following \citet{chernozhukov2003mcmc}, \citet{wang2024composite} model the smoothness of neighbouring quantile curves via basis function expansions. In contrast, our proposed $\QVP$ approach connects the logic of difference penalisation of linear quantile models to the non-crossing constrained objective function presented in \citet{bondell2010noncrossing}. Additionally, the $\QVP$ framework allows for efficient posterior computation via Gibbs sampling due to the availability of standard conditional posteriors. While we maintain the assumption that the quantile function is linear in parameters, the method can be extended to use smoothing splines as in \citet{bondell2010noncrossing}. 

Estimation frameworks that only allow location shifts of the quantile function are referred to as composite quantile regression ($\CQR$) models \citep{zou2008composite}. Here only the parameter on the intercept identifies differences across quantile functions which leads to non-crossing fitted quantiles. In contrast to the above methods, the $\QVP$ allows for a unifying framework in which the process that models quantile variation is centred on the coefficient vector implied by the $\CQR$ model.    

The frequentist literature has proposed many solutions to the quantile crossing problem, where one of the simplest solutions is to sort the fitted quantiles post-estimation \citep{chernozhukov2010quantile}. 
In this paper, interest resides in both prediction and inference on  quantile regression coefficients. We therefore do not consider sorting any further.
Most closely related to the QVP framework in the frequentist literature is \citet{szendrei2023fused} who show that, one can re-formulate the exact non-crossing objective function of \citet{bondell2010noncrossing} as a fused lasso model with a particular formulation of the penalisation constant on the fused term.


\section{A Penalised Likelihood Approach to Joint Quantiles}\label{sec:obj_to_joint_estim}
\subsection{Notation}
The symbol $\sim$ is used both as a sampling statement as well as signifying a variable's probability density function or a likelihood function, synonymously written as $p\left(\right)$. This makes the notation close to probabilistic programming languages such as Stan \citep{carpenter_stan_2017}. Denote by $\mathcal{S}$ a non-empty sample space on which the $\sigma$-algebra $\mathcal{M}$ is defined. Then, by $P\left(X|Y\right)$ we denote the probability of $X=x$ given $Y=y$, where $(X,Y) \subseteq D$. We suppress the differentiation between random and ordinary bound variables for readability. We refer to all unspecified parameters in definitions of conditional probability density functions by $\vartheta$. 
\subsection{Motivation}\label{sec:motivation}
Let $X = (X_1,\dotsc,X_K)^T$ be a set of covariates which are related to a response vector $y$. Let $D \subseteq \mathbbm{R}^K$ be a closed convex polytope represented as the convex hull of $\mathcal{T}$~points in $K$~dimensions. We are interested in a regression at $\mathcal{Q}$~quantile levels $0<\tau_1<\dotsc<\tau_{\mathcal{Q}}<1$, where $\mathcal{Q}$ is a finite integer. Denote by $Q_{\tau_q}(X,\beta_q)$ the $q$\textsuperscript{th} quantile of $y$ as a function of $X$, given some set of quantile-specific regression coefficients $\beta_q$, such that $P\left(y\leq Q_{\tau_q}(X,\beta_q) \vert X, \beta_q \right) = \tau_q$. The classic solution to quantile function estimation approach is to presume quantile functions to be independent, regardless of $\mathcal{Q}$. However, this neglects that the quantile coefficients, $\beta_q$, are often correlated across  quantiles. Such information sharing can drastically improve inference, even if the modeller is only interested in inference on a single quantile \citep{bondell2010noncrossing,jiang2013interquantile}. The key interest in this paper, is to implement information sharing across  quantiles via priors on the differences of the quantile coefficients. Importantly, the quantile coefficient process is centred on a quantile invariant coefficient vector - akin to the composite quantile model \citep{zou2008composite}. By doing so, as differences are adaptively shrunk to zero, the model reduces to the composite quantile regression model, which estimates parallel quantile functions. In this way the model penalises quantile crossing.

To motivate the functional form of the prior, consider the following objective function: 
\begin{IEEEeqnarray}{rl}\label{eq:general-objective-function}
    \sum_{q=1}^{\mathcal{Q}}\sum_{t=1}^{\mathcal{T}} \rho_{\tau_q} & \left( y_t - \alpha_q - x_t^T\beta_0 - x_t^T\beta_q\right) + \mathrm{pen}\left( \left\{\beta_q\right\}_{q=1}^{\mathcal{Q}}\right)  
    \\
    \mathrm{s.t.}~ & x_t^T\beta_q + \alpha_q \geq x_t^T\beta_{q-1} + \alpha_{q-1} \; \forall q = 2,\dotsc,\mathcal{Q}, t = 1,\dotsc,\; \mathcal{T}, \label{eq:NC-constraint-facevalue}
\end{IEEEeqnarray}
where $\rho_{\tau_q}(u) = u(\tau_{q}-I(u<0))$ is the tick-loss function and the constraint in Equation~\ref{eq:NC-constraint-facevalue} ensures monotonicity of the conditional quantile functions. Here, $\alpha_q\in \mathbbm{R}$ is an intercept specific to each quantile. $\beta_0 \in \mathbbm{R}^{K}$ are the quantile invariant coefficients, and $\beta_q \in \mathbbm{R}^{K}$ capture the variation in the effects of covariates across quantiles. Denote by $x_t$ the $t$-th row of X. If $\alpha$ is a monotone function of $\tau$, then the penalty term included, shrinks the model to uniformly parallel, or pair-wise parallel lines. \citet{szendrei2023fused} show that when recasting the data domain to $x_t \in \left[-1,1\right]^{K}$,\footnote{Recasting the data domain to $x_t \in \left[-1,1\right]^{K}$, rather than $x_t \in \left[0,1\right]^{K}$ as in \citet{bondell2010noncrossing}, has the advantage that negative ($\beta_q - \beta_{q-1}<0$) and positive ($\beta_q - \beta_{q-1}>0$) differences are treated symmetrically.} Objective Function~\ref{eq:general-objective-function} with the constraints in Equation~\ref{eq:NC-constraint-facevalue} is equivalent to a fused lasso type quantile regression problem in which differences in $\beta$ across $q$ are penalised by the $L$-1 norm:
\begin{IEEEeqnarray}{rl}\label{eq:motivating-obj}
    \sum_{q=1}^{\mathcal{Q}}\sum_{t=1}^{\mathcal{T}} \rho_{\tau_q} & \left( y_t - \alpha_q - x_t^T\beta_0 - x_t^T\beta_q\right) + \sum_{q=2}^{\mathcal{Q}} \lambda_q \left( \norm{\beta_q - \beta_{q-1}}_1 -\frac{\alpha_q-\alpha_{q-1}}{\varsigma}\right), \label{eq:fused-objective-function}
\end{IEEEeqnarray}
where $\lambda_q$ is a quantile specific shrinkage parameter, $\varsigma$ is a hyperparameter regulating the degree of ``tightness'' of the non-crossing constrains in \citet{szendrei2023fused} and $\norm{x}_j$ refers to the $j\textsuperscript{th}$ norm of x. Setting $\varsigma=1$ recovers the non-crossing constraints of \citet{bondell2010noncrossing}. 

In the following we offer a full probabilistic solution to this problem along with several improvements, where we allow the quantile specific penalties to be freely estimated. The imposed penalisation pushes the $\beta_q$ parameters toward the desired constrained space.
\begin{figure}[h]
    \centering
    \begin{subfigure}[t]{0.48\textwidth}
        \centering
        \includegraphics[width=1\textwidth]{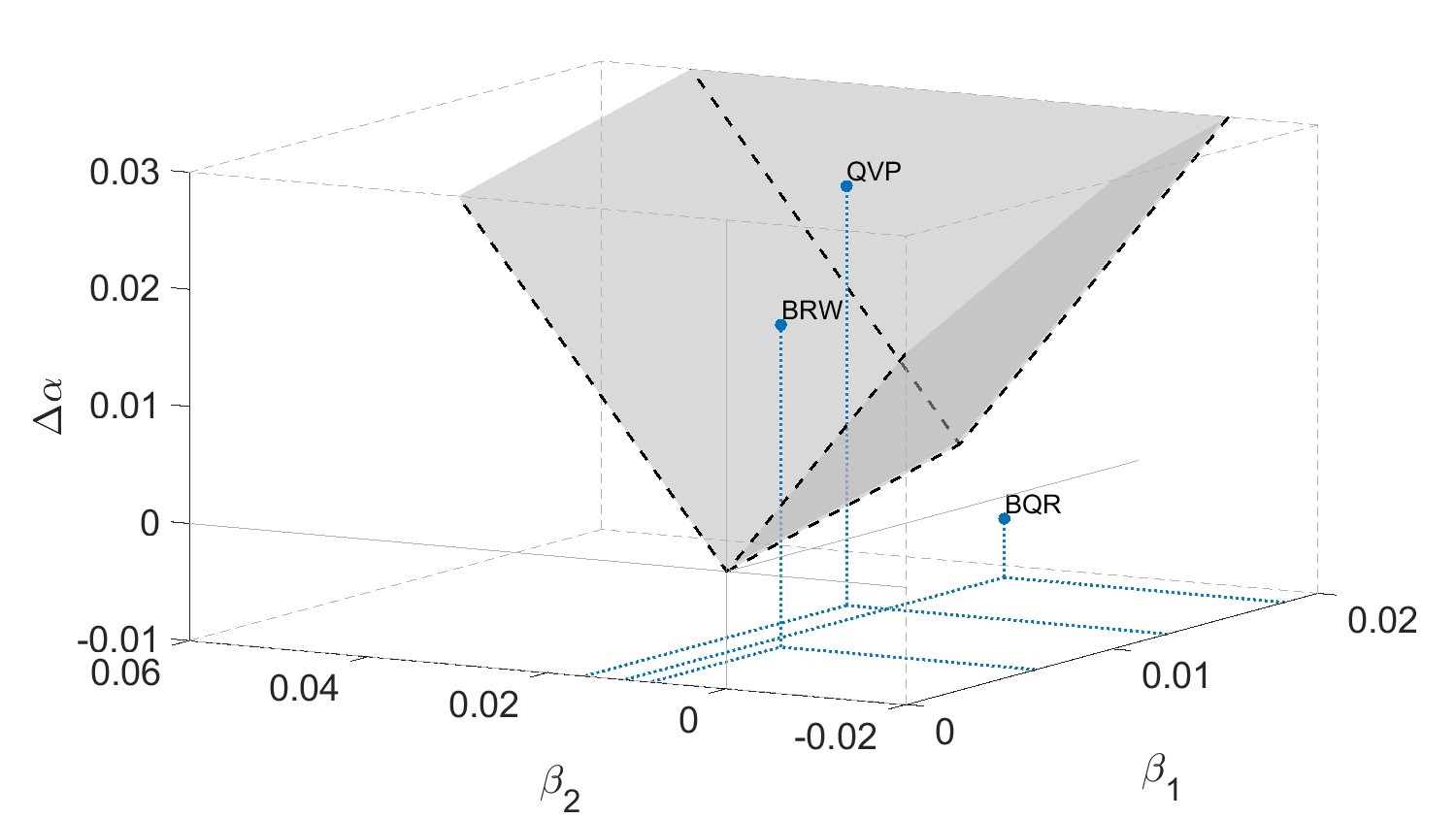}
    \end{subfigure}%
    \hfill
    \begin{subfigure}[t]{0.48\textwidth}
        \centering
        \includegraphics[width=1\textwidth]{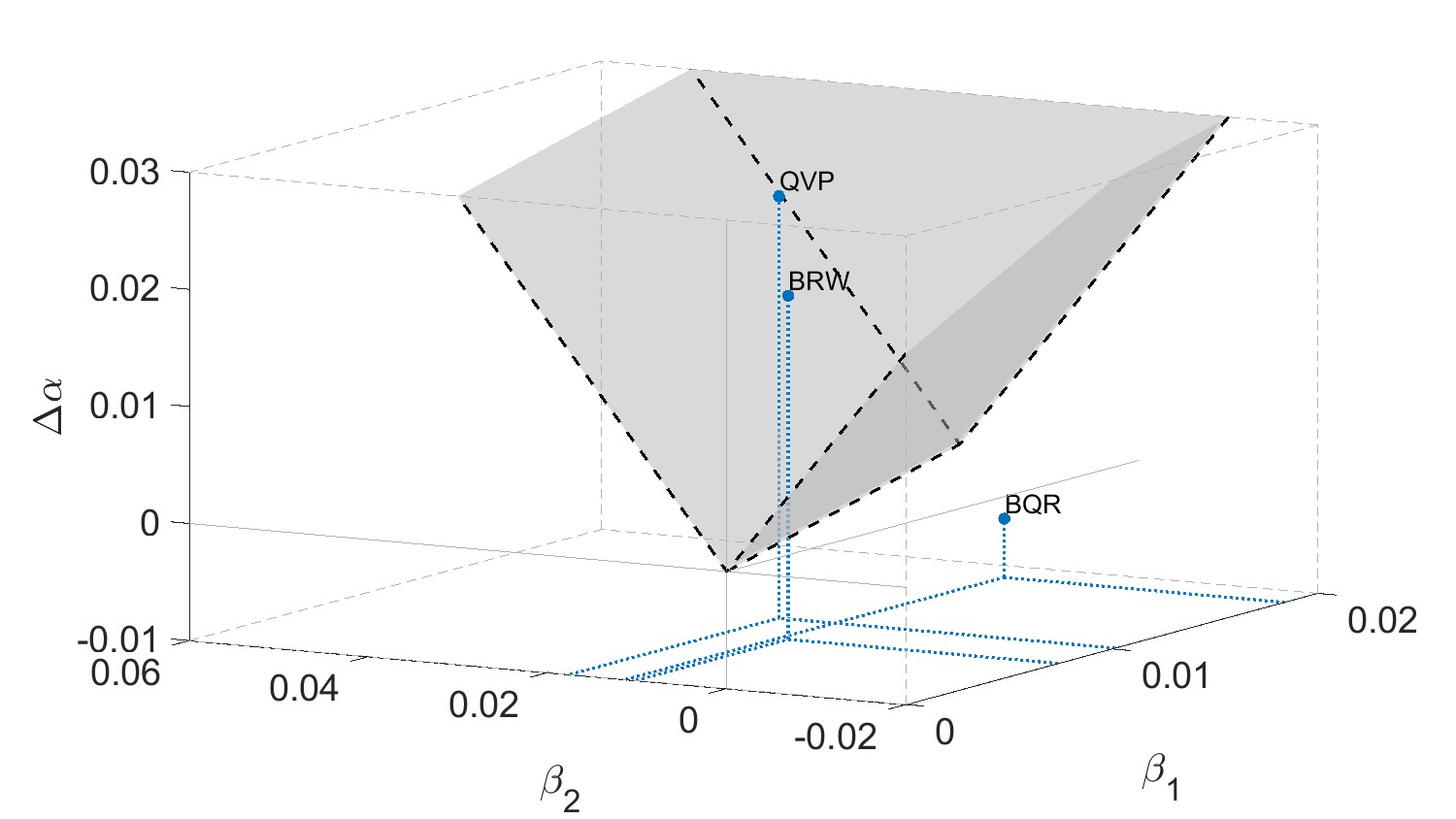}
    \end{subfigure}
    \caption{Means of the difference $\Delta\alpha = \alpha_{q}-\alpha_{q-1}$ and $(\beta_{q},\beta_{q-1})$ (blue points) for the $50^{\mathrm{th}}$ and $51^{\mathrm{st}}$ quantiles for the $\mathrm{QVP}$, $\mathrm{BQR}$ and $\mathrm{BRW}$ models. The grey area is the space in which the quantile curves are non-crossing. In the left figure, $\mathcal{Q}=2$, and in the right $\mathcal{Q}=20$ (19 equidistant quantiles and the $51^{\mathrm{st}}$ quantile).}
    \label{fig:nc-zone}
\end{figure}
To visually motivate that this is the case, we plot in Figure~\ref{fig:nc-zone}, the posterior means of the proposed $\mathrm{QVP}$ along-side  the $\mathrm{BQR}$ model with uninformative priors which estimates quantiles independently, as well as the frequentist model of \citet{bondell2010noncrossing}, $\mathrm{BRW}$. The grey area which indicates the space for which two adjacent quantile regression functions are non-crossing. We estimate the models for the median and $51^{\mathrm{st}}$ quantile. While the $\mathrm{BQR}$ leads to posterior point-estimates outside the space of non-crossing, the penalisation implied by the $\mathrm{QVP}$
, and naturally the $\mathrm{BRW}$ model, lead  to coefficient estimates inside the non-crossing area. It would be expected that as quantiles are chosen to be further apart ($\Delta\alpha$ increases, volume of grey area increases), one would obtain estimates of all models in the non-crossing area. However, when the modeller wants to  get an accurate a picture of coefficient-heterogeneity across quantiles, many quantiles are typically estimated which reduces a given pair of quantiles $\Delta\alpha$, thus decreasing the space in the parameter domain for which the quantile curves are non-crossing. We will show that the proposed $\mathrm{QVP}$ scales well with the amount of estimated quantiles.
\subsection{Likelihood}\label{sec:likelihood}

Let $\vartheta$ contain parameters: $\{\alpha_q,\beta_0,\beta_q\}$. Assume prior beliefs about $\vartheta$ are represented by $p(\vartheta)$, then a valid and coherent update of $p(\vartheta)$, following the general belief updating framework of \citet{bissiri2016general}, is to the posterior $p(\vartheta|X)$:
\begin{equation}
    p(\vartheta|X)\propto \exp\left(-\ell\left(\vartheta,x\right)\right)p(\vartheta),
\end{equation}
where $\ell\left(\vartheta,X\right)=\sum_{q=1}^{\mathcal{Q}}\sum_{t=1}^{\mathcal{T}} \rho_{\tau_q} \left( y_t - \alpha_q - x_t^T\beta_0 - x_t^T\beta_q\right)$.\footnote{The general updating of beliefs framework of \citet{bissiri2016general} includes an unknown scalar multiplying the loss function in order to calibrate the amount of relative information from the data vis-a-vis the prior. We follow the recommendations of Bayesian quantile literature \citep{yu2001bayesian,li2010bayesian} to assume that this scale is equal to 1. \citet{mclatchie2025predictive} show that large enough data, the scale only marginally influences inference.} Exponentiation of the loss function is equivalent up to a constant of proportionality to the commonly employed asymmetric-Laplace ($\mathcal{ALD}$) working likelihood used for probabilistic quantile regression modelling:
\begin{IEEEeqnarray}{rl}
    l\left(\vartheta,X\right) &\;= \exp\left(-\sum^{\mathcal{Q}}_{q=1}\sum^{\mathcal{T}}_{t=1}\rho_{\tau_q}(y_t- \alpha_q -x_t^T\beta_0 -x_t^T\beta_{q})\right) \\
    & \; \propto \prod_{q=1}^\mathcal{Q} \prod_{t=1}^T \mathcal{ALD}\left(y_t\vert\alpha_q,\beta_0,\beta_q,x_t\right) \\
    & \; = \left( \prod_{q=1}^{\mathcal{Q}}\prod_{t=1}^{\mathcal{T}} \tau_q(1-\tau_q) \right) \exp\left(-\sum_{q=1}^{
    \mathcal{Q}}\sum_{t=1}^{\mathcal{T}} \rho_{\tau_q}\left(y_t - \alpha_q -x_t^T\beta_0 -x_t^T\beta_q\right)\right).
\end{IEEEeqnarray}
Hence, minimising the expected multiple quantile loss is equivalent to maximising the combination of the individual $\mathcal{ALD}$ likelihoods of each quantile, where quantiles are assumed to be exchangeable, conditional on $\left(\alpha_q,\beta_0,\beta_q\right)$. For computational convenience, we make use of the fact that the $\mathcal{ALD}$ can be written as a mixture of normal distributions, with the scale parameter having an exponential distribution, following \citet{kozumi2011gibbs}:
\begin{equation}
    \begin{split} 
    l\left(\vartheta,X\right) & \propto \int_0^{\infty} \frac{1}{\sqrt{2\zeta_q^2\omega_q}} \exp\left(-\sum_{q=1}^{\mathcal{Q}}\sum_{t=1}^{\mathcal{T}}(y_t - \alpha_q- x_t^T\beta_0 -x_{t}^T\beta_q - \theta_q \omega_{q,t})^2/(2\omega_{q,t}\zeta_q^2)\right) \\
    &  \times \prod_{t=1}^{\mathcal{T}} \mathrm{e}^{\omega_{i,t} } d\omega_{i,t}, \quad \forall i \in \{1,\dots,\mathcal{Q}\},\; \forall t \in {1,\dotsc,\mathcal{T}} \\ 
    \end{split}
\end{equation}
where $\theta_q = (1-2\tau_q)/(\tau_q(1-\tau_q))$, $\zeta_q^2 = 2/(\tau_q(1-\tau_q))$, and $\omega_{q,t}\sim \mathrm{exp}\left(\sigma^{y}_q\right)$. 
To vectorise across $q$, define $\boldsymbol{y} = \mathbbm{1}_{\mathcal{Q}} \otimes y$, $\boldsymbol{X} = \mathbbm{I}_{\mathcal{Q}} \otimes  X$, where $\mathbbm{1}_{\mathcal{Q}}$ denotes a $\mathcal{Q}$-dimensional vector of ones, and let $\mathbbm{I}_Q$ be the identity matrix of dimension $Q\times Q$. Further, we denote the location adjustment due to data augementation by $\mu_{q,t} = \theta_q\omega_{q,t}$ . Denote $\omega_q = (\omega_{q,1},\dotsc,\omega_{q,\mathcal{T}})^T$, $\boldsymbol{\alpha} = (\alpha_1,\dotsc,\alpha_1,\dotsc,\alpha_\mathcal{Q},\dotsc,\alpha_\mathcal{Q}) \in \mathbbm{R}^{\mathcal{Q} \mathcal{T}}$, $\boldsymbol{\mu} = (\mu_{1,1},\dotsc,\mu_{1,\mathcal{T}},\dotsc,\mu_{\mathcal{Q},1},\dotsc,\mu_{\mathcal{Q},\mathcal{T}})^T$, then the joint likelihood implied is:
\begin{equation} \label{eq:integrated_likelihood}
 \boldsymbol{y}  \sim \int_{0}^{\infty} \mvn\left( \boldsymbol{\alpha} + \boldsymbol{X} \left(\mathbbm{1}_{\mathcal{Q}} \otimes \beta_0\right) +  \boldsymbol{X}\boldsymbol{\beta} + \boldsymbol{\mu}, \boldsymbol{\Omega}\right) \times e^{−\boldsymbol{\omega} } d\boldsymbol{\omega},
\end{equation}
where $\boldsymbol{\beta} = (\beta_1^T,\dotsc,\beta_{\mathcal{Q}}^T)^T$ captures the heterogeneity induced by the covariates, and $\boldsymbol{\Omega} =\text{diag}(\omega_1,\dotsc,\omega_{\mathcal{Q}})$.  $\mvn()$ stands for the multivariate normal distribution. 
\subsection{The QVP Prior}\label{sec:qvp-prior}
Consider again the penalised quantile objective of Equation~\ref{eq:fused-objective-function}. A probabilistic generalisation can be found by the following discrete state space representation: 
\begin{IEEEeqnarray}{rl}
     y_{t} & = \alpha_q + x_t^{T}\beta_q + \mu_{q,t} + \epsilon_{q,t}^y,\; \epsilon^y_{q,t}\sim\normal\left(0,\theta_q^2\sigma^y_q\omega_{q,t}\right), \; t = 1,\dotsc, \mathcal{T} \label{eq:observation-equation-centred} \\
     \beta_q & = \beta_{q-1} + \epsilon^{\beta}_q,\; \epsilon^{\beta}_q \sim \mvn\left(0,\Sigma_{q}\right),  \; q  = 1,\dotsc,\mathcal{Q} \label{eq:state-equation-centred} \\
     \beta_0 & \sim \mvn\left(0,\Sigma_0\right) \label{eq:starting-condition-cented}, \; \alpha \propto \mathbbm{1}_{Q\mathcal{T}},
\end{IEEEeqnarray}
where $\beta_0$ is the quantile invariant vector and $\Sigma_q = \text{diag}(\sigma_{q,1}^2,\dotsc,\sigma_{q,K}^2)$. $\Sigma_q$ controls the variability of the coefficients between quantiles and therefore how strongly correlated the coefficients are.\footnote{Allowing for non-zero off diagonals would allow for any quantile  coefficient to affect any of the other estimated quantiles directly. While relevant, we leave investigation of the properties of this modelling approach to future research.} Modelling $\epsilon_q^{\beta}$  with Laplacian densities would create the exact Bayesian equivalent to the penalisation implied by the lasso penalty in the objective function. Namely, when $\beta_q - \beta_{q-1} \sim \mathcal{MAL}\left(0, \Sigma_q  \right)$, or equivilantly, $\beta_q \sim \mathcal{MAL}\left(\beta_{q-1}, \Sigma_q  \right)$, where $\mathcal{MAL}\left( \right)$ stands for the multivariate Laplace distribution \citep{kotz2001asymmetric}. A similar logic is presented in the derivation of the univariate lasso prior of \citep{park2008bayesian}. We, however, follow the more recent literature on shrinkage priors that show superior shrinkage properties with normal kernels \citep{carvalho_handling_2009,piironen_hyperprior_2017}. To adhere to the convention of the state-space literature, we will refer to Equation~\ref{eq:observation-equation-centred} as the observation equation, and Equation~\ref{eq:state-equation-centred} as the state equation for quantile $\tau_q$ respectively.

The state vector $\beta_0$ plays a special role in this model setup. It is both the initialisation of the state process, and equivalent to the quantile invariant vector in Objective~\ref{eq:fused-objective-function}. This can be easily verified by backward substitution of the state equation into the observation equation. In fact, when $\mathbf{\beta}=\boldsymbol{0}$, then $\beta_0$ is equivalent to the composite quantile regression vector, considered in \citet{zou2008composite}. The significance of this for the shrinkage properties will be further investigated in Section~\ref{sec:theoretical-properties}.

The state-space representation results in a particular structure to the joint prior. Write the state process in Equation~$\ref{eq:state-equation-centred}$ stacked across $\mathcal{Q}$ in matrix form as:
\begin{equation}
    H\boldsymbol{\beta} = \tilde{\boldsymbol{\beta}} + \boldsymbol{\epsilon^{\beta}},
\end{equation}
where $\boldsymbol{\epsilon^{\beta}} \sim \mvn(0,\boldsymbol{\Sigma})$, $\boldsymbol{\Sigma} = \mathrm{diag}(\Sigma_1,\dotsc,\Sigma_{\mathcal{Q}})$, $\boldsymbol{\tilde{\beta}} = (\beta_0^T,0,\dotsc,0)^T$ and
\begin{equation}
    \boldsymbol{H} = 
    \begin{pmatrix}
        \mathbbm{I}_{K} & 0 & 0 & \dotsc & 0 \\ 
        -\mathbbm{I}_{K} & \mathbbm{I}_{K} & 0 & \dotsc & 0 \\ 
        0 & -\mathbbm{I}_{K} & \mathbbm{I}_{K} & \dotsc & 0 \\
        \vdots &  & \ddots & \ddots & \vdots \\
        0 & 0 & \dotsc & -\mathbbm{I}_{K} & \mathbbm{I}_{K} \\
    \end{pmatrix}.
\end{equation}
$\boldsymbol{H}$ is a $\mathcal{Q}K \times \mathcal{Q}K$ difference matrix, which is invertible since $|\boldsymbol{H}| = 1$. It is straightforward to show that $\boldsymbol{H}^{-1}\boldsymbol{\tilde{\beta}} = \mathbbm{1}_{\mathcal{Q}} \otimes \beta_0$. Then, the joint prior for $\boldsymbol{\beta}$, conditional on $\left(\beta_0,\boldsymbol{\Sigma}\right)$ is
\begin{equation}
    (\boldsymbol{\beta}\mid \beta_0,\boldsymbol{\Sigma}) \sim \mvn\left(\mathbbm{1}_{\mathcal{Q}} \otimes \beta_0,(\boldsymbol{H}^T\boldsymbol{\Sigma}^{-1}\boldsymbol{H})^{-1}\right),
\end{equation}
Due to the close connection to joints priors for time-varying parameter regression models in which $\beta$ is indexed by time, we call this prior the joint quantile-varying parameter ($\QVP$) prior.
And by standard manipulations, the posterior is given by
\begin{equation} \label{eq:Posterior_beta_centred}
    \left(\boldsymbol{\beta} \mid \boldsymbol{y},\dotsc\right) \propto \mvn\left(\overline{\boldsymbol{\beta}}, \boldsymbol{\text{K}_{\beta}}^{-1}\right),
\end{equation}
where
\begin{equation}
    \begin{split}
        \boldsymbol{\text{K}_{\beta}} & = \boldsymbol{H}^T\boldsymbol{\Sigma}^{-1}\boldsymbol{H} + \boldsymbol{X}^T\boldsymbol{\Omega}^{-1}\boldsymbol{X} \\
        \overline{\boldsymbol{\beta}} & = \boldsymbol{\text{K}_{\beta}}^{-1} \left(\boldsymbol{H}^T\boldsymbol{\Sigma}^{-1}\boldsymbol{H}\left(\mathbbm{1}_{\mathcal{Q}} \otimes \beta_0\right) + \boldsymbol{X}^T\boldsymbol{\Omega}^{-1}\boldsymbol{y^{*}}\right), 
    \end{split}
\end{equation}
and $\boldsymbol{y^{*}}  =  \boldsymbol{y} - \boldsymbol{\alpha} - \boldsymbol{\mu}$.
\subsection{Fused Horseshoe Prior}
The priors on $\boldsymbol{\Sigma}$ determine the amount of quantile variation. We model explicitly: 1) adaptivity of shrinkage on the difference in coefficients across quantiles, specific to each covariate, 2) global regularisation of each $\beta_q-\beta_{q-1}$~difference vector. Therefore, it is natural to follow the global-local prior literature where the prior hierarchy on $\sigma_{q,j}$ employs a mixture of fat-tailed distributions with singularity at 0 to allow for both large and small changes in coefficients. A plethora of priors can be considered \citep{polson_half-cauchy_2012}, however, we consider here the horseshoe prior \citep{carvalho_handling_2009}, adapted to fused shrinkage. In particular, define $\sigma^2_{q,j} = \nu_q^2\lambda_{q,j}^2$,\footnote{Setting double-Laplace priors for $\lambda_{q,j}$, the exact Bayesian interpretation of the absolute deviation penalisation of the motivating objective function in Equation~\ref{eq:motivating-obj} can be recovered.} then the fused horseshoe prior is
\begin{equation}\label{eq:prior-differences-centred}
    \nu_q\sim C_{+}\left(0,1/\sqrt{\mathcal{T}}\right),\quad \lambda_{q,j}\sim C_{+}(0,1),
\end{equation}
where $C_+()$ stands for the half Cauchy distribution. Following \citet{piironen_hyperprior_2017}, we scale the global parameter by the number of data points. While $\nu_q$ controls overall approximate sparsity of differences in quantiles, the local scales $\lambda_{q,j}$ control the local adaptivity in how much the differences are shrunk dependent on the quantile level as well as the covariate.
\subsection{Prior on $\beta_0$}
We again set a horseshoe prior. Let $\Sigma_0 = \nu^2_0\text{diag}(\lambda^2_{0,1},\dotsc,\lambda^2_{0,K})$, then
\begin{equation} \label{eq:prior_level}
\begin{split}
    \beta_0 & \sim \mvn\left(0,\Sigma_0\right) \\
    \pi\left(\nu_0\right) & \sim C_{+}\left(0,1/\sqrt{\mathcal{T}\mathcal{Q}}\right),\quad \pi\left(\lambda_{0,j}\right) \sim C_{+}\left(0,1\right). 
\end{split}
\end{equation}
This prior may also induce approximate sparsity in the quantile invariant vector.

We summarise the $\QVP$ prior model with the following definition:
\begin{definition}\label{def:qvp_centred}
The $\QVP$ prior defined over $\mathcal{Q}$ quantiles with quantile-adaptive horseshoe  regularisation yields the following model structure, where $q = 1,\dotsc,\mathcal{Q}$ and $j = 1,\dotsc,K$:
    \begin{IEEEeqnarray}{rl}
        \boldsymbol{y} \;& \sim\mvn\left(\boldsymbol{\alpha} + \boldsymbol{\mu} + \boldsymbol{X}\boldsymbol{\beta},\boldsymbol{\Omega}\right)\label{eq:qvp_likelihood} \\
        \boldsymbol{\beta} \;&\sim \mvn\left(\mathbbm{1}_{\mathcal{Q}}\otimes\beta_0, \left(\boldsymbol{H}^T\boldsymbol{\Sigma}^{-1}\boldsymbol{H}\right)^{-1}\right)\label{eq:qvp_beta_prior} \\
                \beta_0 \;&\sim \mvn\left(0, \Sigma_0\right)\label{eq:qvp_beta_0_prior} \\
        \boldsymbol{\Sigma}\;&=\mathrm{diag}\left(\sigma^2_{1,1},\dotsc,\sigma^2_{1,K},\dotsc,\sigma^2_{\mathcal{Q},1},\dotsc,\sigma^2_{\mathcal{Q},K}\right),\; \sigma^2_{q,j} = \nu_q^2\lambda^2_{q,j} \\
        \Sigma_{0}\;&=\mathrm{diag}\left(\sigma^2_{0,1},\dotsc,\sigma^2_{0,K}\right),\; \sigma^2_{0,j} = \nu_0^2\lambda^2_{0,j} \\
        \nu_q\;& \sim C_+\left(0,\frac{1}{\sqrt{\mathcal{T}}}\right),\;        \lambda_{q,j}\;\sim C_+\left(0,1\right),\; \nu_0 \sim C_{+}\left(0,\frac{1}{\sqrt{\mathcal{T}\mathcal{Q}}}\right) \\
        \mu_{q,t}\;&= \theta_q\omega_{q,t},\;\omega_{q,t}\sim \mathrm{exp}\left(\sigma^{y}_q\right) \\
        \boldsymbol{\alpha}\;&\propto\mathbbm{1}_{\mathcal{Q}\mathcal{T}} \\
        \sigma^y_q\;&\sim p\left(\right),
    \end{IEEEeqnarray}
\end{definition}
where $p()$ stands for some probability density. With uninformative priors $\boldsymbol{\alpha} \propto 1$,
 any location shift in the quantile function of $\boldsymbol{\beta}$ is determined by the data only. Notice, that this deviates from incrementing the posterior with the difference prior on $alpha_q-\alpha_{q-1}$ in \ref{eq:fused-objective-function}, which would influence only the conditional posterior of the quantile specific global-shrinkage. For this and the additional reason that shrinking toward the quantile invariant vector favours shrinking toward parallel quantile curves, we opt for uninformative priors. However for completeness, we present simulation evidence in Appendix~\ref{subsubsec:alpha-predictions} with the full difference prior on $\alpha_q-\alpha_{q-1}$. The results are virtually identical. Finally, we set the relatively uninformative prior of $\sigma^y_q \sim \IG{0.1,0.1}$, where $\IG{\underline{a},\underline{b}}$ stands for the inverse-Gamma distribution with rate $\underline{a}$ and scale $\underline{b}$. 
%
\section{MCMC Sampling Algorithm for QVP}\label{sec:mcmc}
\begin{figure}[h!]
    \centering
    \includegraphics[width=0.5\linewidth]{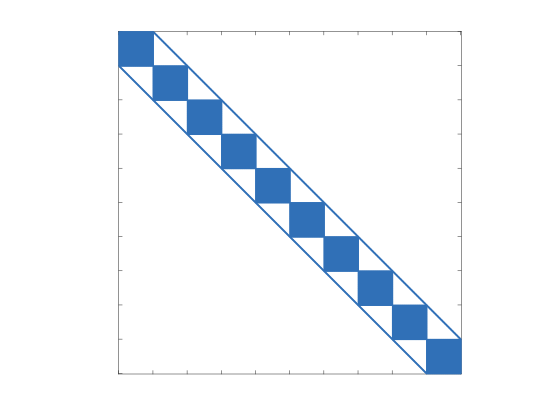}
    \caption{Structure of Posterior Precision Matrix, $\text{K}_{\beta}$}
    \label{fig:precision_Kbeta}
\end{figure}

The priors outlined above results in known conditional posterior distributions and therefore an efficient Gibbs sampling scheme which iterates through the following updates:
\begin{enumerate}
    \item $\boldsymbol{\beta}^{(s)} \sim p(\boldsymbol{\beta}\mid \boldsymbol{y},\boldsymbol{\Sigma}^{(s-1)},\beta_0^{(s-1)},\boldsymbol{\alpha}^{(s-1)},\boldsymbol{\mu}^{(s-1)},\boldsymbol{\Omega}^{(s-1)} )$,
    \item $\boldsymbol{\Sigma}^{(s)} \sim p(\boldsymbol{\Sigma}\mid \boldsymbol{y},\boldsymbol{\beta}^{(s)},\beta_0^{(s-1)},\boldsymbol{\alpha}^{(s-1)},\boldsymbol{\mu}^{(s-1)}, \boldsymbol{\Omega}^{(s-1)})$, 
    \item $\beta_0^{(s)} \sim p(\beta_0\mid \boldsymbol{y},\boldsymbol{\beta}^{(s)},\boldsymbol{\Sigma}^{(s)},\boldsymbol{\alpha}^{(s-1)},\boldsymbol{\mu}^{(s-1)},\boldsymbol{\Omega}^{(s-1)} )$, 
    \item $\boldsymbol{\alpha}^{(s)} \sim p(\boldsymbol{\alpha} \mid \boldsymbol{y},\boldsymbol{\beta}^{(s)},\boldsymbol{\Sigma}^{(s)},\beta_0^{(s)},\boldsymbol{\mu}^{(s-1)},\boldsymbol{\Omega}^{(s-1)} )$, 
    \item $\boldsymbol{\mu}^{(s)} \sim p(\boldsymbol{\mu} \mid \boldsymbol{y},\boldsymbol{\beta}^{(s)},\boldsymbol{\Sigma}^{(s)},\beta_0^{(s)},\boldsymbol{\alpha}^{(s),},\boldsymbol{\Omega}^{(s-1)})$, 
    \item $\boldsymbol{\Omega}^{(s)} \sim p(\Omega\mid \boldsymbol{y},\boldsymbol{\beta}^{(s)},\boldsymbol{\Sigma}^{(s)},\beta_0^{(s)},\boldsymbol{\alpha}^{(s)},\boldsymbol{\mu}^{(s)} )$, ,
\end{enumerate}
for $s = (1,\dotsc,S)$ until convergence. The conditional posteriors are given in Appendix~\ref{app:qvp-posteriors}.
The main computational bottleneck in sampling from Posterior~\ref{eq:Posterior_beta_centred} is the inversion of the $\mathcal{Q}K \times \mathcal{Q} K$-dimensional full covariance matrix $K_\beta^{-1}$ which can easily become high-dimensional. Computing the Cholesky factor for this covariance matrix will involve $\mathcal{O}\left((\mathcal{Q} K)^3\right)$ operations. 
Notice that the precision matrix, $\text{K}_{\beta}$, on the other hand has a band structure, which will typically look like in Figure~\ref{fig:precision_Kbeta}. This motivates a more efficient sampling algorithm that utilises the sparse nature of the matrix. In particular, computing the Cholesky factor of the precision matrix only involves $\mathcal{O}\left(\mathcal{Q} K\right)$ operations, which can be sped up in practice with sparse matrix routines available in most programming languages.
Hence, to obtain draws from the conditional posterior of $\boldsymbol{\beta}$, we make use of the following steps:
\begin{enumerate}
    \item Compute Cholesky factor $\text{C}^{(s)}$ of $\text{K}_{\beta}^{(s)}$
    \item Generate $\text{Z}^{(s)} \sim \mvn(0,\mathbbm{I}_{\mathcal{Q} \cdot K})$
    \item Set the $\text{s}^{\text{th}}$ draw, $\boldsymbol{\beta}^{(\text{s})}$, to $\overline{\boldsymbol{\beta}}^{(s)} + ({\text{C}^{(s)}}^T)^{-1}\text{Z}^{(s)}$,
\end{enumerate}
where $\boldsymbol{\overline{\beta}^{(s)}}$ and $({\text{C}^{(s)}}^T)^{-1}$ can be found efficiently by solving linear equations. The rest of the sampling steps are standard and further explained in Appendix~\ref{app:qvp-posteriors}.
%

%

\section{Alternative Parameterisations of the QVP}\label{sec:alt-parameterisation}
A challenge within the multiple quantile estimation literature is the identification of quantile variation of coefficients. Viewed from the state-space representation in Equations~\ref{eq:observation-equation-centred}-\ref{eq:state-equation-centred}, selection of quantile variation can be seen as a variance selection problem. This entails boundary estimation which is computationally difficult and known to lead to slow convergence of $\mathrm{MCMC}$ samplers \citep{fruhwirth2010stochastic}, even with horseshoe priors \citep{bitto2019achieving}. For this reason, we formulate the state-space in its non-centred form in the spirit of \citet{fruhwirth2010stochastic}. This shifts the variance selection problem to a standard conjugate variable selection problem. To see this, re-write the State-equation \ref{eq:state-equation-centred} as:
\begin{equation}
    \beta_{q,j} = \beta_{0,j} + \sigma_{q,j}\tilde{\beta}_{q,j},
\end{equation}
\begin{equation}
    \tilde{\beta}_{q,j} = \tilde{\beta}_{q-1,j} + \tilde{\epsilon}_{q,j}, \quad \tilde{\epsilon}_{q,j} \sim \normal(0,1),
\end{equation}
with initial condition $\tilde{\beta}_{0,j} = 0$. Using this transformation the Observation-equation \ref{eq:observation-equation-centred} can be equivalently written as:
\begin{equation} \label{eq:NC_model}
 y_{t} = \alpha_q + x^{T}\beta_0 + x^{T}_t\text{diag}(\sigma_{q,1},\dotsc,\sigma_{q,K})\tilde{\beta}_q + \epsilon^y_{q,t}.   
\end{equation}
Re-writing $x^{T}_t\text{diag}(\sigma_{q,1},\dotsc,\sigma_{q,K})\tilde{\beta}_q$ as $\tilde{x}_{q,t}^{T}\sigma_q$ where $\tilde{x}_{q,t} = \tilde{\beta}_q^{T}\text{diag}(x_t)$, the state standard deviations may be viewed simply as regression coefficients, motivating a shift of the domain of $\sigma_{q,j}$ from the positive only to the entire real line. 
Doing so avoids the boundary estimation issues and additionally results in conditionally conjugate posteriors, allowing for efficient Gibbs sampling.
For simplicity, we employ horseshoe priors with a normal kernel for $\beta_0$ and $\sigma_q$. 
The prior for $\beta_0$ remains the same as in Equation~\ref{eq:prior_level} and $(\sigma_1^T,\dotsc,\sigma_\mathcal{Q}^T)^T = \boldsymbol{\sigma}$ now takes the following form:
\begin{IEEEeqnarray}{rl}
    \boldsymbol{\sigma} & \sim \mvn\left(0,\boldsymbol{\tilde{\Sigma}}\right) \\
        \tilde{\nu}_q & \sim C_{+}\left(0,1/\sqrt{\mathcal{T}}\right), \quad q = 1,\dotsc,\mathcal{Q} \\
        \tilde{\lambda}_{q,j} & \sim C_{+}\left(0,1\right), \quad q = 1,\dotsc,\mathcal{Q},\; j = 1,\dotsc,K
\end{IEEEeqnarray}
where $\tilde{\Sigma}_q = \tilde{\nu}_q^2\mathrm{diag}({\tilde{\lambda}_{q,1}}^2,\dotsc,{\tilde{\lambda}_{q,K}}^2)$ and $\boldsymbol{\tilde{\Sigma}} = \mathrm{diag}(\tilde{\Sigma}_1,\dotsc,\tilde{\Sigma}_{\mathcal{Q}})$. It can be shown, that a normal prior on the scale, $\sigma_q$, implies a generalised inverse-Gaussian prior on the variance whose properties for state-space models are studied in \citet{cadonna2020triple}.\footnote{In fact, the horseshoe prior on the scale can be shown to be nested by the more general triple-gamma prior \citep{cadonna2020triple} framework} This results in a higher concentration rate of the marginal prior on $\sigma_{q,j}$ near the origin and lower rate of tail-decay, which is desirable for variable selection type inference tasks \citep{polson_half-cauchy_2012}. 
\subsection{MCMC Sampling algorithm for Non-Centred QVP}
In order to draw inference on the non-centred $\QVP$ model in Equation~\ref{eq:NC_model}, we can again make use of efficient updating via conditional posteriors: 
\begin{enumerate}
    \item ${\boldsymbol{\beta}_0}^{(s)} \sim p(\boldsymbol{\beta}_0\mid \boldsymbol{y},\boldsymbol{\tilde{\beta}}^{(s-1)},\boldsymbol{\sigma}^{(s-1)},\boldsymbol{\tilde{\Sigma}}^{(s-1)},\Sigma_0^{(s-1)},\boldsymbol{\alpha}^{(s-1)},\boldsymbol{\mu}^{(s-1)},\boldsymbol{\Omega}{(s-1)} )$, 
    \item $\Sigma^{(s)}_0 \sim p(\Sigma_0\mid \boldsymbol{y},\boldsymbol{\tilde{\beta}}^{(s-1)},\boldsymbol{\sigma}^{(s-1)},\boldsymbol{\tilde{\Sigma}}^{(s-1)},\boldsymbol{\beta}_0^{(s)},\boldsymbol{\alpha}^{(s-1)},\boldsymbol{\mu}^{(s-1)}, \boldsymbol{\Omega}{(s-1)})$, 
    \item ${\boldsymbol{\tilde{\beta}}}^{(s)} \sim p({\boldsymbol{\tilde{\beta}}}\mid \boldsymbol{y},\boldsymbol{\sigma}^{(s-1)},\boldsymbol{\tilde{\Sigma}}^{(s-1)},\boldsymbol{{\beta}}_0^{(s)},\Sigma_0^{(s)},\boldsymbol{\alpha}^{(s-1)},\boldsymbol{\mu}^{(s-1)},\boldsymbol{\Omega}{(s-1)} )$, 
    \item $\boldsymbol{\sigma}^{(s)} \sim p(\boldsymbol{\sigma} | \boldsymbol{y},\boldsymbol{\tilde{\beta}}^{(s)},\boldsymbol{\tilde{\Sigma}}^{(s-1)},\boldsymbol{{\beta}}_0^{(s)},\Sigma_0^{(s)},\boldsymbol{\alpha}^{(s-1)},\boldsymbol{\mu}^{(s-1)},\boldsymbol{\Omega}{(s-1)} )$, 
    \item $\boldsymbol{\tilde{\Sigma}}^{(s)} \sim p(\boldsymbol{\tilde{\Sigma}}\mid \boldsymbol{y},\boldsymbol{\tilde{\beta}}^{(s)},\boldsymbol{\sigma}^{(s)},\boldsymbol{{\beta}}_0^{(s)},\Sigma_0^{(s)},\boldsymbol{\alpha}^{(s-1)},\boldsymbol{\mu}^{(s-1)}, \boldsymbol{\Omega}{(s-1)})$, 
    \item $\boldsymbol{\alpha}^{(s)} \sim p(\boldsymbol{\alpha} \mid \boldsymbol{y},\boldsymbol{\tilde{\beta}}^{(s)},\boldsymbol{\sigma}^{(s)},\boldsymbol{\tilde{\Sigma}}^{(s)},\boldsymbol{{\beta}}_0^{(s)},\Sigma_0^{(s)},\boldsymbol{\mu}^{(s-1)},\boldsymbol{\Omega}{(s-1)} )$, 
    \item $\boldsymbol{\mu}^{(s)} \sim p(\boldsymbol{\mu} \mid \boldsymbol{y},\boldsymbol{\tilde{\beta}}^{(s)},\boldsymbol{\sigma}^{(s)},\boldsymbol{\tilde{\Sigma}}^{(s)},\boldsymbol{{\beta}}_0^{(s)},\Sigma_0^{(s)},\boldsymbol{\alpha}^{(s),},\boldsymbol{\Omega}{(s-1)})$, 
    \item $\boldsymbol{\Omega}{(s)} \sim p(\Omega\mid \boldsymbol{y},\boldsymbol{\tilde{\beta}}^{(s)},\boldsymbol{\sigma}^{(s)},\boldsymbol{\tilde{\Sigma}}^{(s)},\boldsymbol{{\beta}}_0^{(s)},\Sigma_0^{(s)},\boldsymbol{\alpha}^{(s)},\boldsymbol{\mu}^{(s)} )$,
\end{enumerate}
for $s=(1,\dotsc,S)$ until convergence. Compared to the sampling steps in the centred QVP model, sampling individually the quantile invariant vector $\beta_0$ and associated hyper-parameters, adds two further sampling blocks: one for the $\boldsymbol{\sigma}$ and $\boldsymbol{\tilde{\Sigma}}$, and modified sampling steps for $\beta_0$, which we discuss in turn.
%

%
To update $\beta_0$, the relavant likelihood and prior contributions are proportional to:
\begin{equation}
    \prod_{q=1}^Q \mvn\left(\alpha_q + X\beta_0 + \tilde{X}_q\sigma_q + \mu_q,\Omega_q\right) \times \mvn\left(0,\Sigma_0\right),
\end{equation}
where $\tilde{X}_q = X \odot \left(\mathbbm{1}_T\otimes \tilde{\beta}_q^T\right)$, and $\Omega_q = \text{diag}(\omega_q)$. Since conditional on $\beta_0$, all likelihood contributions across quantiles are exchangeable, the posterior of $\beta_0$ may be efficiently updated one quantile at a time.\footnote{This follows from basic probability theory in that $p(\theta|Y_1,Y_2)\propto p(Y_2|Y_1,\theta)\times p(\theta|Y_1)$} The conditional posterior for $\beta_0$ is thus normal:
\begin{equation}\label{eq:NC-beta0-posterior}
    \beta_0|\boldsymbol{Y},\vartheta \propto \mvn\left(\overline{\beta}_0,K^{-1}_{\beta_0}\right),
\end{equation}
where $K_{\beta_0} = (\sum_{q=1}^{\mathcal{Q}}X^T\boldsymbol{\Omega}^{-1}_qX) + \Sigma_0^{-1}$ and $\overline{\beta}_0  = K^{-1}_{\beta_0}(\sum_{q=1}^{\mathcal{Q}}X^T\Omega_q^{-1}(y - \alpha_q - \mu_q - \tilde{X}_q\sigma_q))$. See Appendix~\ref{app:qvp-posteriors} for further derivation of the posterior moments.

Due to the non-centred representation of the state-space, the prior for $\boldsymbol{\tilde{\beta}}$ simplifies to $\boldsymbol{\tilde{\beta}} \sim \mvn\left(0,\left(\boldsymbol{H}^T\boldsymbol{H}\right)^{-1}\right)$. Define $\boldsymbol{\check{X}} = \boldsymbol{X}\text{diag}(\boldsymbol{\sigma})$ and $\boldsymbol{\beta_0} = \mathbbm{1}_{\mathcal{Q}} \otimes \beta_0$, then the posterior for $\boldsymbol{\tilde{\beta}}$ is conditionally normal:
\begin{equation}
    \boldsymbol{\tilde{\beta}} | \boldsymbol{Y},\vartheta \propto \mvn\left(\overline{\boldsymbol{\tilde{\beta}}},\tilde{K}_{\tilde{\beta}}^{-1}\right),
\end{equation}
where $\tilde{K}_{\tilde{\beta}} = (\boldsymbol{\check{X}}^{T}\boldsymbol{\Omega}^{-1}\boldsymbol{\check{X}} + \boldsymbol{H}^T\boldsymbol{H})$ and $\boldsymbol{\tilde{\beta}} = \tilde{K}_{\tilde{\beta}}^{-1}(\boldsymbol{\check{X}}^{T}\boldsymbol{\Omega}^{-1}(\boldsymbol{Y}-\boldsymbol{\alpha} - \boldsymbol{\mu} - \boldsymbol{X}\boldsymbol{\beta}_0))$. This posterior retains its band-matrix structure as in Figure~\ref{fig:precision_Kbeta}, which makes for fast computation with any sparse matrix routine.
To sample the state standard deviations $\boldsymbol{\sigma}$, we can rely again on standard regression results. Let $\boldsymbol{\tilde{X}}$ contain the stacked $\tilde{X}_q$ matrices across all quantiles, then the posterior for $\boldsymbol{\sigma}$ is normal:
\begin{equation}
    \boldsymbol{\sigma}|\boldsymbol{Y},\vartheta \propto \mvn\left(\overline{\boldsymbol{\sigma}},K_{\boldsymbol{\sigma}}^{-1}\right),
\end{equation}
where $K_{\boldsymbol{\sigma}} = (\boldsymbol{\tilde{X}}^{T}\boldsymbol{\Omega}^{-1}\boldsymbol{\tilde{X}} + \boldsymbol{\tilde{\Sigma}}^{-1})$ and $\overline{\boldsymbol{\sigma}} = K_{\boldsymbol{\sigma}}^{-1} (\boldsymbol{\tilde{X}}^{T}\boldsymbol{\Omega}^{-1}(\boldsymbol{Y} - \boldsymbol{\alpha} - \boldsymbol{\mu} - \boldsymbol{X}\boldsymbol{\beta}_0))$. 

Shift of the domain of $\sigma_{q,j}$ to the entire real line has advantages for computation but it introduces sign-unidentifiability. To aid mixing,  we randomly permute the signs of $(\boldsymbol{\sigma},\boldsymbol{\tilde{\beta}})$ as proposed in \citet{bitto2019achieving}.
%
\section{Post-processing the posteriors}\label{sec:savs}
Although horseshoe priors on the quantile invariant and difference vectors shrink toward sparsity, exact sparsity cannot be achieved in finite samples due to absolute continuity of the prior distributions \citep{carvalho_handling_2009}. We suggest post-processing the posterior with a thresholding algorithm motivated from decision theory \citep{lindley1968choice} to project the posterior onto a possibly sparse subset.\footnote{Simply calculating Bayesian p-values based on the posteriors for $\beta_0$ and $\boldsymbol{\sigma}$ might be misleading due the effect of correlation in the posteriors as well as potential multi-modality} On the one hand, this helps intuitively assessing the marginal importance of quantile invariant as well as quantile variant effects. \citet{piironen_sparsity_2017} show that post-estimation projection may also reduce variance in variable selection. On the other, it can sharpen inference of true zero effects, as well as lead to improved predictions \citep{huber2021inducing,kohns2025flexible}. This literature has renewed attention with \citet{hahn2015decoupling} for normal linear models which has been extended to Bayesian quantile regression in \citet{kohns2021decoupling}, and \citet{feldman2023bayesian}. 
%

Define the linear predictor of interest for the quantile invariant effect as $X\beta_0$, then a possibly sparse vector, $\xi^0$, can be found by solving:
\begin{equation} \label{eq:sparse_beta_0}
\overline{\xi}^0 = \underset{\xi^0}{\text{argmin}} \; \frac{1}{2}\norm{X\beta_0 - X\xi^0}^2_2 + \sum^{K}_{j=1}\upsilon^0_j|\xi^0_{j}|,
\end{equation}
where $\upsilon^0_j$ is an adaptive penalty factor akin to adaptive lasso \citep{zou2006adaptive}. Likewise, define the linear predictor that captures quantile variation as $\boldsymbol{\tilde{X}}\boldsymbol{\sigma}$, then a possibly sparse vector, $\boldsymbol{\xi^{\sigma}}$, can be found by solving:
\begin{equation} \label{eq:sparse_sigma}
    \overline{\xi}^{\boldsymbol{\sigma}} = \underset{\boldsymbol{\xi_{\sigma}}}{\text{argmin}} \; \frac{1}{2}\norm{\boldsymbol{\tilde{X}}\boldsymbol{\sigma} - \boldsymbol{\tilde{X}}\xi^{\boldsymbol{\sigma}}}^2_2 + \sum^{Q}_{q=1}\sum^{K}_{j=1}\upsilon^{\sigma}_{q,j}|\xi^{\boldsymbol{\sigma}}_{q,j}|.
\end{equation}
These projections can be solved for each MCMC draw $s = 1,\dotsc,S$ in order to retrieve pseudo model-average posteriors \citep{bhattacharya2016fast}. The level of sparsity in Equations~\ref{eq:sparse_beta_0} and~\ref{eq:sparse_sigma} is determined by the penalties $(\upsilon^0,\upsilon^{\sigma})$. For computational convenience, the penalty term is set inversely proportional to the posterior draw of the coefficient following \citet{ray2018signal}. We maintain the naming of the authors in describing this sparsification as algorithm as the $\mathrm{SAVS}$ algorithm. We refer to the $\NCQVP$ with $\mathrm{SAVS}$ algorithm applied as the $\NCQVP_\mathrm{SAVS}$ model.

\section{Shrinkage Properties of the $\mathrm{QVP}$} \label{sec:theoretical-properties}
\begin{figure}
    \centering
    \includegraphics[width=\textwidth]{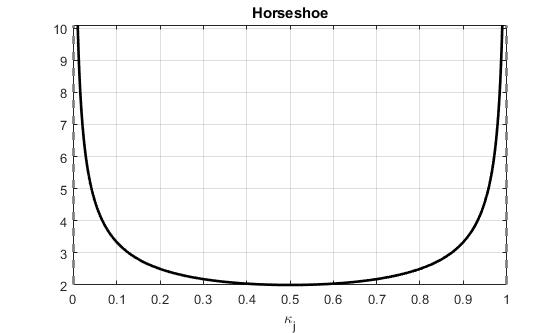}
    \caption{Distribution of $\kappa_{t,j}$, the shrinkage coefficient implied by the horseshoe prior.}
    \label{fig:ShrinkageCoefficients}
\end{figure}

In this section, we analyse the $\QVP$ prior in terms of its implications on the shrinkage scale space. It is common to analyse global-local priors in terms of their implied distribution on a scale that allows to gauge the shrinkage effect away from the maximum likelihood estimate \citep{polson_half-cauchy_2012}.  As is common in the literature, we analyse the shrinkage scale distribution for the normal model.\footnote{Derivations can be extended to the $\mathcal{ALD}$ using the shrinkage coefficient definitions in \citet{kohns2024horseshoe}} Applied to an isotropic normal observation model, we define fused shrinkage akin to the $\QVP$ in the following model structure: 
\begin{IEEEeqnarray}{rl}
    y_t & = x_t^T\beta_t + \epsilon_t, \quad \epsilon_t \sim \normal\left(0,\sigma_{y}^2\right) \\
    \beta_t & \sim \normal\left(\beta_{t-1},\nu^2\Lambda_t\right),\; \Lambda_t = \mathrm{diag}\left( \lambda_{t,1}^2,\dotsc, \lambda_{t,K}^2 \right) \\
    \beta_t\vert \beta_{1},\dotsc,\beta_{t-1},y,\vartheta \; & \propto \normal\left(\overline{\beta}_t,K_{\beta_t}^{-1}\right).
\end{IEEEeqnarray}
$K_{\beta_t} = (\nu^{-2}\Lambda_t^{-1} + \frac{1}{\sigma_y^2}x_t^Tx_t)$ and $\overline{\beta}_t = K^{-1}_{\beta_t}\left(\nu^{-2}\Lambda_t^{-1}\beta_{t-1} + \frac{1}{\sigma_y^2}x_t^Ty_t\right)$ where $\beta_t \in \mathbbm{R}^{K}$. Assume further that $x_t^Tx_t \approx \mathrm{diag}(1,\dotsc,1)$, then the posterior mean can be further decomposed as:
\begin{equation}
    \kappa_{t,j} \beta_{t-1,j} + (1-\kappa_{t,j})\beta_{\mathrm{ML},t,j},
\end{equation}
where $\kappa_{t,j} = \frac{1}{\nu^2\lambda_{t,j}^2\sigma^{-2}_y+1}$ and $\beta_{\mathrm{ML},t,j} = \left(x_{t,j}\right)^{-2}x_{t,j}y_t$ can be understood as a maximum likelihood esimate to the coefficient $\beta_{t,j}$.
This decomposition shows that the conditional posterior mean is a convex combination of the prior coefficient $\beta_{t-1}$ and a maximum likelihood estimate. Therefore, as $\lambda^2_{t,j}\rightarrow0$, then $\kappa_{t,j}\rightarrow1$ and $(1-\kappa_{t,j})\rightarrow 0$. Hence, with strong shrinkage toward the origin, there is no further updating implied from the data, and the posterior concentrates on the previous coefficient $\beta_{t-1,j}$. On the other-hand when $\lambda^2_{t,j}\rightarrow\infty$, then $\kappa_{t,j} \rightarrow0$ and $(1-\kappa_{t,j})\rightarrow1$, so updating the posterior only happens with the data information at observation $t$, and the previous coefficient has no further influence. 
\begin{definition}\label{eq:def-1}
    Suppose $\lambda_{t,j}\sim C_+(0,1)$, then the probability density function of the shrinkage coefficient $\kappa_{t,j}$, conditional on $(\nu,\sigma_y)$ can be shown to be
    \begin{equation}
        p(\kappa_{t,j}\vert \nu,\sigma_y) = \frac{1}{\pi}\frac{\nu\sigma^{-1}_y}{(\nu^2\sigma^{-2}_y-1)\kappa_{t,j}+1}\frac{1}{\sqrt{\kappa_{t,j}}\sqrt{1-\kappa_{t,j}}},
    \end{equation}
    which is proportional to $\betadist(1/2,1/2)$ when $\nu \sigma_y=1$.
\end{definition}
See Appendix~\ref{sec:shrinkage-properties} for further derivations. The implied prior distribution for $\kappa_{t,j}$ is shown in Figure~\ref{fig:ShrinkageCoefficients}. Hence, shrinkage properties that are well understood for the prior on levels of coefficients also transfer to difference shrinkage. The Cauchy prior on $\lambda_{t,j}$ results in a continuous approximation to variable selection type behaviour: probability density is highest on strong or very little shrinkage respectively.
%

\section{Simulation Study}\label{sec:simulation}
\subsection{Setup}
In this section we will test the performance of the proposed $\QVP$ priors via a simulation study following the data-generating processes of \citet{bondell2010noncrossing}. 
We generate data from the following location scale heteroscedastic error model:
\begin{equation} \label{eq:loc-scale}
    y_t=\alpha_0+\beta^Tx_t+(\eta_0+\varrho^T_t \odot \eta_1^Tx_t)\varepsilon_t,~x_{t,k}\sim U(0,1),~\varepsilon_t\sim N(0,1).
\end{equation}
In total we consider the following 5 DGPs: 
\begin{itemize}
    \item $\mathrm{DGP}$-1: 4 predictors, with the $\beta_{\mathrm{DGP}_1}=\mathbbm{1}_K$, $\eta_1=0.1\mathbbm{1}_K$, and $\varrho_t=\mathbbm{1}_K$.
    \item $\mathrm{DGP}$-2: 10 predictors, with the parameters $\beta_{\mathrm{DGP}_2}=(\mathbbm{1}_4^T,\textbf{0}_6^T)^T$, $\eta_1=(0.1\mathbbm{1}_4^T,\textbf{0}_6^T)^T$, and $\varrho_t=\mathbbm{1}_K$.
    \item $\mathrm{DGP}$-3: 7 predictors, with the parameters $\beta_{\mathrm{DGP}_3}=\mathbbm{1}_k$, $\eta_1=(0.1\mathbbm{1}_3^T,\textbf{0}_4^T)^T$, and $\varrho_t=\mathbbm{1}_K$.
    \item $\mathrm{DGP}$-4: 10 predictors, with the parameters $\beta_{\mathrm{DGP}_4}=(\mathbbm{1}_4^T,\textbf{0}_6^T)^T$, $\eta_1=(0.1\mathbbm{1}_8^T,\textbf{0}_2^T)^T$, and $\varrho_t=(\mathbbm{1}_4^T,\mathbbm{1}_4^T  [I(\epsilon_t> F^{-1}_\varepsilon(0.9))+I(\epsilon_t\leq F^{-1}_\varepsilon(0.1))],\textbf{0}_2^T)^T$.
    \footnote{Note that the $t$ subscript is needed since the presence of quantile variation will be dependent on the magnitude of $\varepsilon_t$.}
    \item $\mathrm{DGP}$-5: 4 predictors, with the parameters $\beta_{\mathrm{DGP}_5}=\mathbbm{1}_K$, $\eta_1=(0.1\mathbbm{1}_2^T,\mathbbm{1}_2^T)^T$, and $\varrho_t=\mathbbm{1}_K$.
\end{itemize}
%
%
$\mathrm{DGP}$-1 and $\mathrm{DGP}$-2 are identical to the simulation study in \citet{bondell2010noncrossing}. $\mathrm{DGP}$-1 features non-zero location effects of the covariates, with a shallow but continuous upward slope in the quantile coefficient profile. $\mathrm{DGP}$-2 adds sparsity to the location where true zero coefficients do not display quantile variation. $\mathrm{DGP}$-3 is defined by non-zero location effects of the covariates, where only the first three coefficients have a shallow quantile profile. $\mathrm{DGP}$-4 adds to $\mathrm{DGP}$-3 extra covariates with true zero location and quantile varying effects. Additionally, 4 covariates display what \citet{kohns2021decoupling} call quantile specific sparsity patterns: location and quantile variation are zero for central quantiles, with large quantile variation in the extreme quantiles $\tau_q\geq0.9$ and $0.1\leq\tau_q$. $\mathrm{DGP}$-5 is generated with non-zero location effects, where for two covariates there is large continuous quantile variation. For ease of discussion below, we refer to coefficients of covariates with only location effects as quantile constant coefficients, to those coefficients with only a shallow profile $({\eta_1}_i = 0.1)$ as quantile varying coefficients, and those quantile varying coefficients with big jumps or large variation as extreme varying coefficients. 
%

%

%
We simulate sample sizes $\mathcal{T}=\{100,300\}$, and estimate models for various number of total quantiles $\mathcal{Q}=\{9,19,49\}$. We do this to address the fact that the $\QVP$ prior's shrinkage depends on how many quantiles are being estimated and because the previous literature warns of increasing probability of crossing quantile curves with many estimated quantiles \citep{jiang2013interquantile}. We allow the design matrix to be correlated with constant correlation of $\varDelta \in \{0, 0.5, 0.9\}$.\footnote{For $\varDelta>0$, the design is sampled form a normal distribution and then transformed using the min-max transformation on a column-wise basis.} We generate $N_{\mathrm{sim}} = 500$ simulated data sets. For posterior inference we obtain per model 25000 MCMC draws for 4 chains in parallel, of which we discard the first 10000 each as burnin.  

Recovery of $\boldsymbol{\beta}$ is measured based on the root-mean-squared-error for the i\textsuperscript{th} DGP, by $\mathrm{RMSE}_i$:
\begin{equation} \label{eq:rmse_def}
    \mathrm{RMSE}_i =  \sqrt{\frac{1}{N_{\mathrm{sim}}\mathcal{Q}}\sum_{j = 1}^{N_{\mathrm{sim}}}\sum_{k=1}^{K_{\mathrm{DGP}_{i}}} \sum_{q=1}^{\mathcal{Q}} \left( {\beta_{k,q}} - {{\hat{\beta}_{j,k,q}}} \right)^2},
\end{equation}
where $\beta_{k,q}$ is the DGP's true coefficient at the $q$\textsuperscript{th} quantile and $\hat{\boldsymbol{\beta}}$ refers to the posterior mean defined as: 
\begin{equation}
    \begin{split}
        \hat{\boldsymbol{\beta}} = \E{\boldsymbol{\beta}|y} &  = \int_{\mathbbm{R}^{\mathcal{Q}K}}\boldsymbol{\beta} \; p(\boldsymbol{\beta}|y)\;\mathrm{d}\boldsymbol{\beta} \\
        & \approx \frac{1}{S}\sum_{s = 1}^S \boldsymbol{\beta}^{(s)}.
    \end{split}
\end{equation}
Denote the $q\textsuperscript{th}$ fitted quantile for observation $t$ based on the posterior mean as $\hat{Q}_{\tau_q,t}$. We evaluate predictions with the Quantile score ($\mathrm{QS}_{q,i}$) for the q\textsuperscript{th} quantile and i\textsuperscript{th} DGP:
\begin{equation}
    \mathrm{QS}_{\tau_q,i} =  \frac{1}{N_{\mathrm{sim}}}\sum_{j = 1}^{N_{\mathrm{sim}}} \frac{1}{T}\sum^\mathcal{T}_{t=1}\rho_{\tau_q}\left( {y_{j,t}} - \hat{Q}_{j,\tau_q,t} \right)
\end{equation}
The $\mathrm{QS}$ score belongs to the strictly proper scoring rules \citep{gneiting_strictly_2007} calculated for every simulation run based on 100 independently generated out of sample observations. The $\mathrm{QS}$ may be used to evaluate different combinations of quantiles to get an idea of the performance at different parts of the predictive density. To achieve this, we will follow \citet{gneiting2011comparing} in calculating the quantile weighted $\mathrm{QS}$:

\begin{equation}\label{eq:qwQS}
    \mathrm{qwQS}_i = w_{\tau_q} \mathrm{QS}_{i,\tau_q}
\end{equation}

\noindent where $w_{\tau_q}$ denotes a weighting scheme. We consider four different weighting schemes: (a) $w_{\tau_q}^1=\frac{1}{Q}$ places equal weight on all quantiles, which is equivalent to taking an average of the weighted residuals; (b) $w_{\tau_q}^2={\tau_q}(1-{\tau_q})$ places more weight on central quantiles; (c) $w_{\tau_q}^3=(1-{\tau_q})^2$ places more weight on the left tail; and (d) $w_{\tau_q}^4={\tau_q}^2$ places more weight on the right tail.  
%

Lastly, we measure the incidence of crossing. The crossing incidence is calculated by comparing the fitted quantiles with the sorted quantiles following the procedure of \citet{chernozhukov2010quantile}:

\begin{equation}\label{eq:cross-i}
    \mathrm{Cross}_{\tau_q,i}=\frac{1}{N_{\mathrm{sim}}} \sum_{j =1}^{N_{sim}} \frac{1}{\mathcal{Q}\mathcal{T}}\sum^{\mathcal{T}}_{t=1}\sum^{\mathcal{Q}}_{q=1}I[{\hat{Q}_{{j,\tau_q,t}}\neq\hat{Q}^{\mathrm{sort}}_{{j,\tau_q,t}}}],
\end{equation}

\noindent where $\hat{Q}^{\mathrm{sort}}_{{\tau_q,t}}$ is the sorted predicted quantile. The crossing incidence measures the proportion of quantiles that need sorting after estimation to adhere to the property of monotone quantile functions. The lower the $\mathrm{Cross}$ value, the less quantiles need to be rearranged after estimation.%

Next to the $\QVP$, $\NCQVP$, $\NCQVP_{\mathrm{SAVS}}$, we consider two approaches for independent Bayesian quantile regression, the $\BQR$ as presented in \citet{kozumi2011gibbs} with flat priors and the $\HSBQR$ of \citet{kohns2024horseshoe} which uses horseshoe priors on the quantile coefficients, respectively. 
Since both assume that the quantile functions are unrelated, we refer to these as independent quantile regression methods. We expect the independent quantile methods to do well for $\mathrm{DGP}$-4 where quantile specific sparsity is present. Additionally, to illustrate the benefit of estimating the quantile profile $\beta_q$ in addition to the composite quantile vector $\beta_0$, we also estimate a Bayesian composite quantile model which only models $\beta_0$, the $\mathrm{CQR}$ model. 
Lastly, since the QVP prior is motivated from the non-crossing quantile objective function of \citet{bondell2010noncrossing}, we include their model too for comparison, denoted $\BRW$.
\subsection{Results}
\subsubsection{Coefficient Recovery}
\begin{figure}
    \centering
    \includegraphics[width=\linewidth]{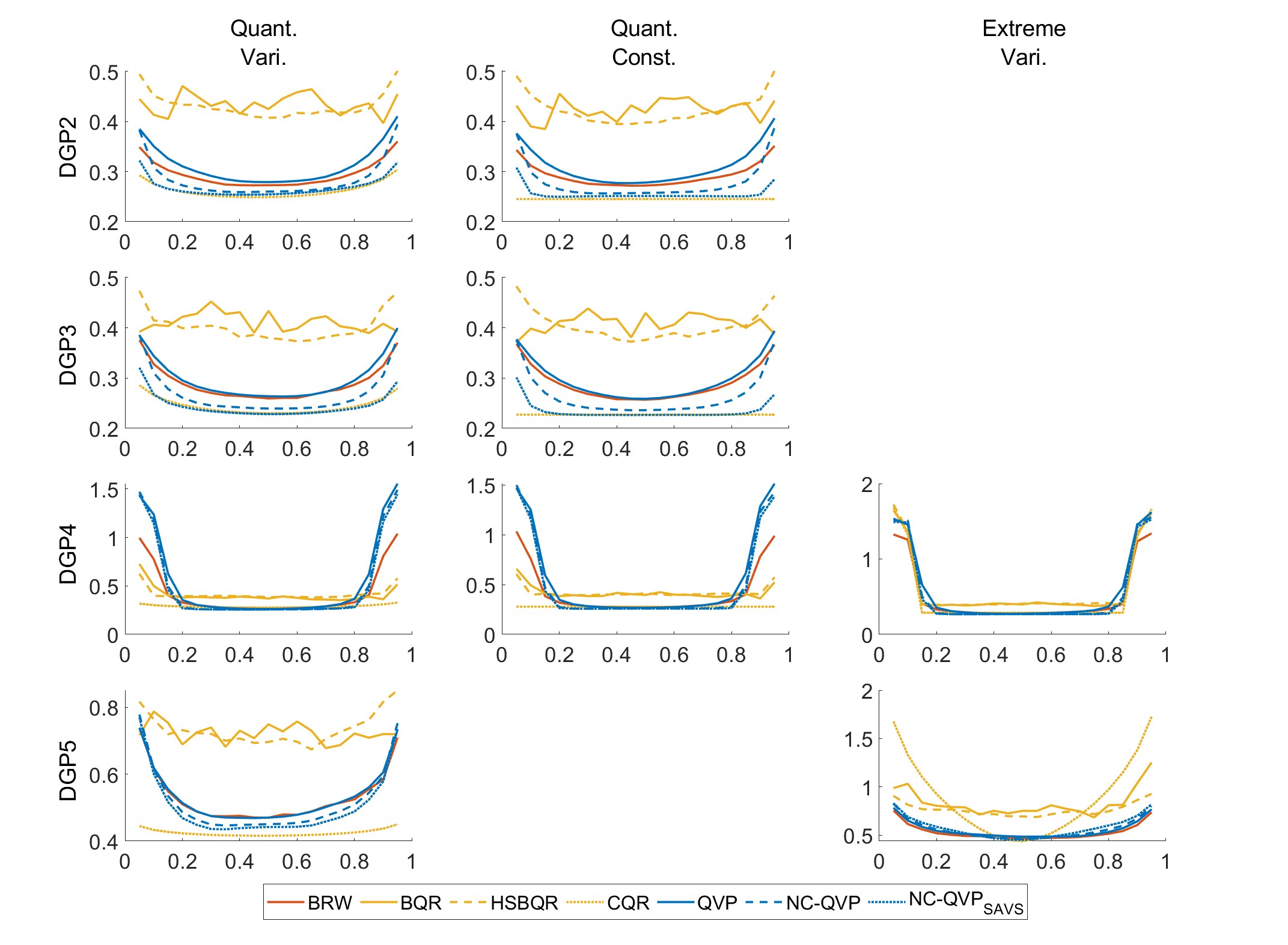}
    \caption{$\rmse$ profile of $\boldsymbol{\beta}$ across 19 quantiles $\tau_q \in \left\{0.05,\dotsc,0.95\right\}$ for $\mathcal{T}=300$ and $\varDelta$=0. Estimates are based on the posterior mean. }
    \label{fig:SpecCoeffBias}
\end{figure}
\begin{figure}
    \centering
    \includegraphics[width=\linewidth]{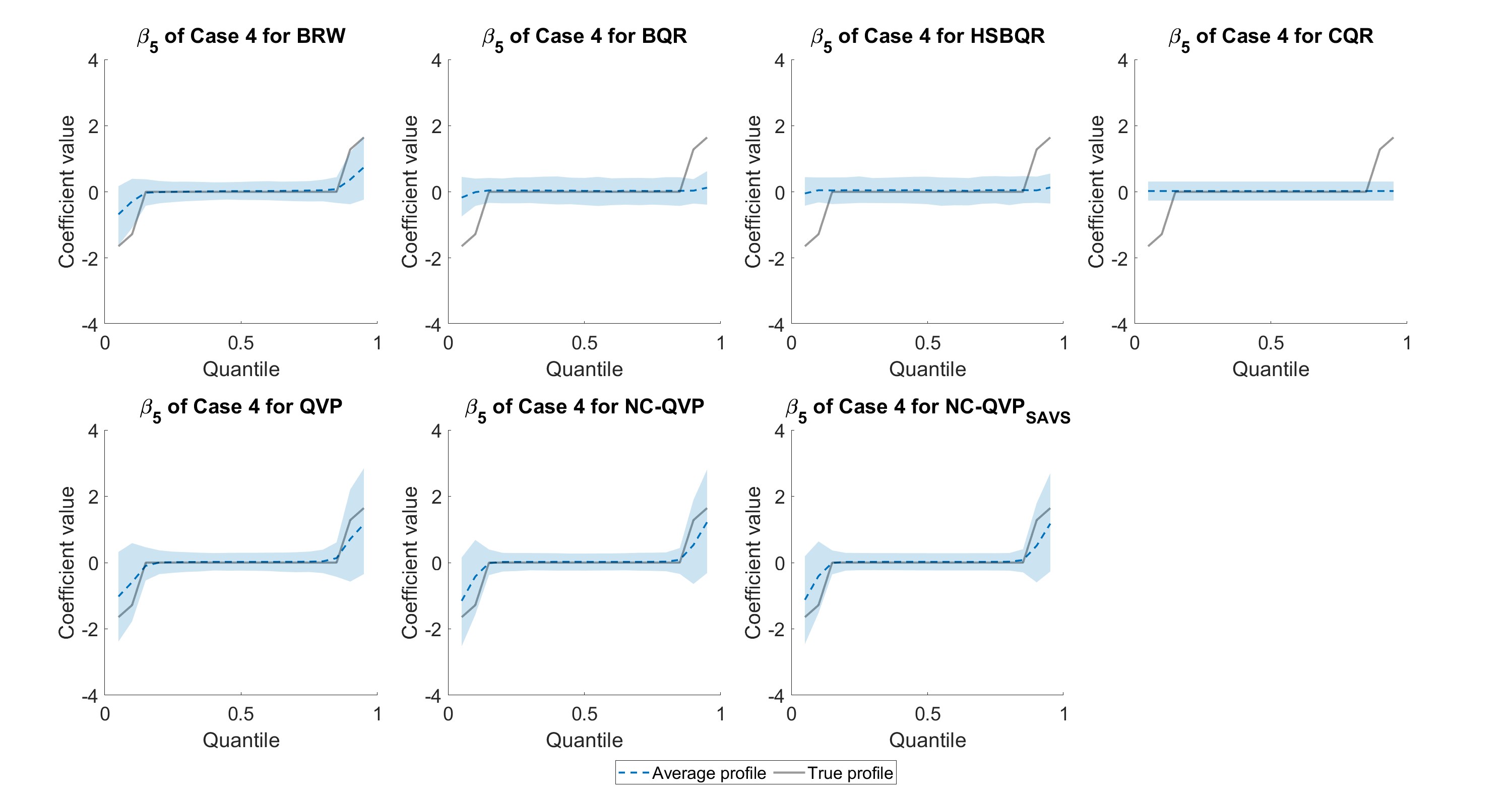}
    \caption{Coefficient posterior for $\mathrm{DGP}$-4, $\mathcal{T}=300$, $\varDelta$=0. Dotted lines show the average of the posterior means. The blue area reprents the central 95$\%$ interval of the posterior means.} 
    \label{fig:C4}
\end{figure}
\begin{figure}[h]
    \centering
    \includegraphics[width=\linewidth]{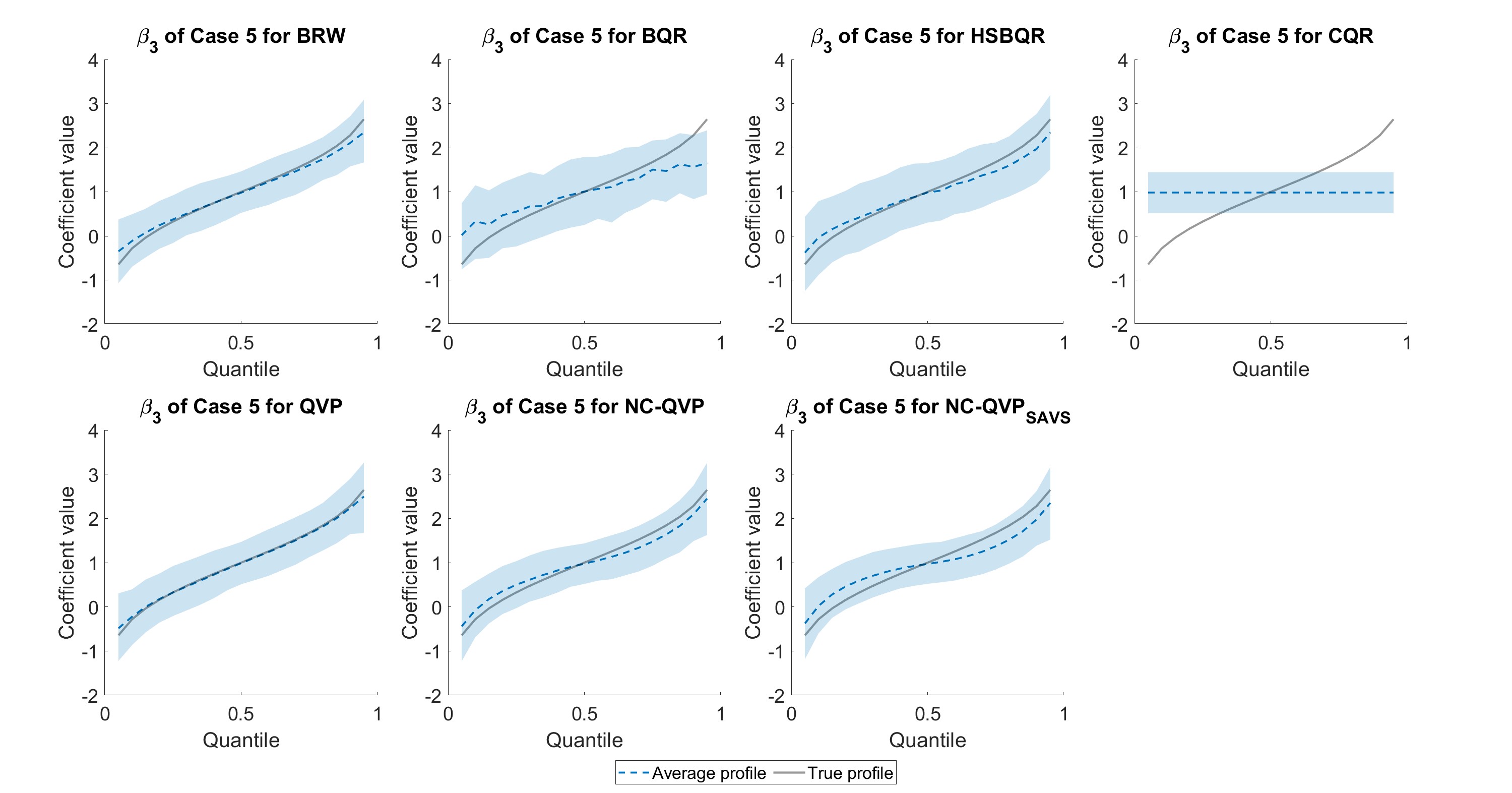}
    \caption{Coefficient posterior for $\mathrm{DGP}$-5, $\mathcal{T}=300$, $\varDelta$=0. Dotted lines show the average of the posterior means. The blue area reprents the central 95$\%$ interval of the posterior means.}
    \label{fig:C5}
\end{figure}

We summarise coefficient recovery in Figure~\ref{fig:SpecCoeffBias} for $\mathcal{T}=300$ and $\mathcal{Q}=19$ estimated quantiles. In Figures~\ref{fig:C4} and~\ref{fig:C5}, we additionally show estimated coefficient profiles for a variable with quantile specific sparsity for $\mathrm{DGP}$-4 and large quantile variation for $\mathrm{DGP}$-5 respectively.\footnote{See  Appendix~\ref{app:extra-sim-results} for figures also for coefficients the other DGPs.} See Appendix~\ref{app-sec:fig-posteriors} for average performance for the central and extreme quantiles for the different permutations of $(T,\varDelta)$. 

As expected, Figure~\ref{fig:SpecCoeffBias} shows that the $\NCQVP$ and $\CQR$ models perform generally best for coefficients with low quantile variation or no quantile vairation at all (columns 1 and 2), particularly visible for DGPs 2-3. The $\mathrm{CQR}$ does well for these DGPs because estimating the true shallow quantile profile to be zero only causes small increases in $\mathrm{RMSE}$. $\NCQVP$ and $\NCQVP_{\mathrm{SAVS}}$ shrink low quantile variation heavily to zero, and therefore perform for these DGPs well, too. In fact the SAVS variant is on par with the $\mathrm{CQR}$ model. 
The $\NCQVP$ offers not only a large improvement over methods that estimate the quantiles independently ($\BQR$,$\HSBQR$), but it also clearly outperforms $\mathrm{BRW}$. This is particularly visible for the quantile constant coefficients of the $\mathrm{DGP}$-3, in which the true constant coefficients are non-zero. In line with the motivation of the $\QVP$ prior's structure, we find that for $\mathrm{DGP}$-2-3, it generally performs on par with the $\mathrm{BRW}$. 

For $\mathrm{DGP}$-4 in which there is pronounced quantile specific sparsity, we see that that  models which do not explicitly model the difference in quantile coefficients ($\BQR$, $\HSBQR$, $\CQR$) may outperform the $\QVP$ models. However, Figure~\ref{fig:C4} shows that these models also falsely shrink away the true profile to zero in the extreme tails. The $\QVP$ models, in contrast, correctly recover the coefficient profile, albeit with larger variance in the tails. With increasing number of observations, we would expect the reduction in posterior variance to lead also to superior $\mathrm{RMSE}$ for the $\QVP$ models over the independent quantile methods.

For coefficients that showcase large continuous variation across all quantiles, $\mathrm{DGP}$-4 and $\mathrm{DGP}$-5 (column 3 of Figure~\ref{fig:SpecCoeffBias}), we can see that recovery of the $\mathrm{QVP}$ models is competitive in terms of RMSE (Figure~\ref{fig:SpecCoeffBias}) and estimated coefficient profile (Figure~\ref{fig:C5}). As expected, the $\mathrm{QVP}$ model tends to outperform the $\NCQVP$ here since less shrinkage on the difference between quantiles is exerted in the centred formulation. The uncertainty bands compared to independent estimation of the quantiles, indicates large efficiency benefits to joint estimation. 

These findings are robust to the number of data points (Appendix, Figure~\ref{fig:SpecCoeffBias_T100}), magnitude of correlation between covariates (Appendix, Figures~\ref{fig:SpecCoeffBias_rho05} and \ref{fig:SpecCoeffBias_rho09}), the number of quantiles estimated (Appendix, Figures~\ref{fig:SpecCoeffBias_Q9} and \ref{fig:SpecCoeffBias_Q39}) and from the perspective of predictive performance (Appendix, Section~\ref{app:sims-predictive-results}). In fact, with more quantiles estimated, we find that there are further performance gains with the $\QVP$ compared to the $\NCQVP$ in $\mathrm{DGP}$-5. Here, less shrinkage of the coefficient profile with the $\QVP$ over the $\NCQVP$ allows for improved modelling in the tails with the finer resolution offered by modelling more quantiles.  

Hence, in terms of parameter recovery, the $\QVP$ framework provides an excellent balance between the composite quantile model that only models location effects and the $\BRW$ model which strictly enforces non-crossing. The $\NCQVP$ model in particular benefits from stronger between-quantile shrinkage when little quantile variation is present, yet also does not overtly shrink true large variation in quantile profiles. To choose in practice between the centred and non-centred formulation, we recommend using the $\NCQVP$ as the baseline with which to test for the presence of no quantile variation with the $\mathrm{SAVS}$ algorithm. If significant quantile variation is present, then we recommend using the $\QVP$ model.
\footnote{We leave investigation in terms of model selection properties for future research.} Despite the relatively low dimensionality of the covariate set, the DGPs show that the QVP methods are always preferable to independent quantile models.

\subsubsection{Robustness}
The improvements in parameter recovery of the $\mathrm{QVP}$ models also translate to a lower incidence of crossing fitted quantiles as well as improved sampling efficiency. Table~\ref{tab:cross} shows that independent of the simulation design, the $\mathrm{QVP}$ models almost completely eliminate crossing. Hence, free estimation of the implied non-crossing penalty parameter on the differences across quantiles (compare Equation~\ref{eq:fused-objective-function}), is sufficient to regularise the posterior toward the desired area of non-crossing. Independent quantile models struggle comparatively more with $\mathrm{DGP}$~1 and $\mathrm{DGP}$~2, where separation of quantile varying and non-varying coefficients is complicated by the relatively low amount of cross-quantile variation. 

\begin{figure}
    \centering
    \includegraphics[width=\linewidth]{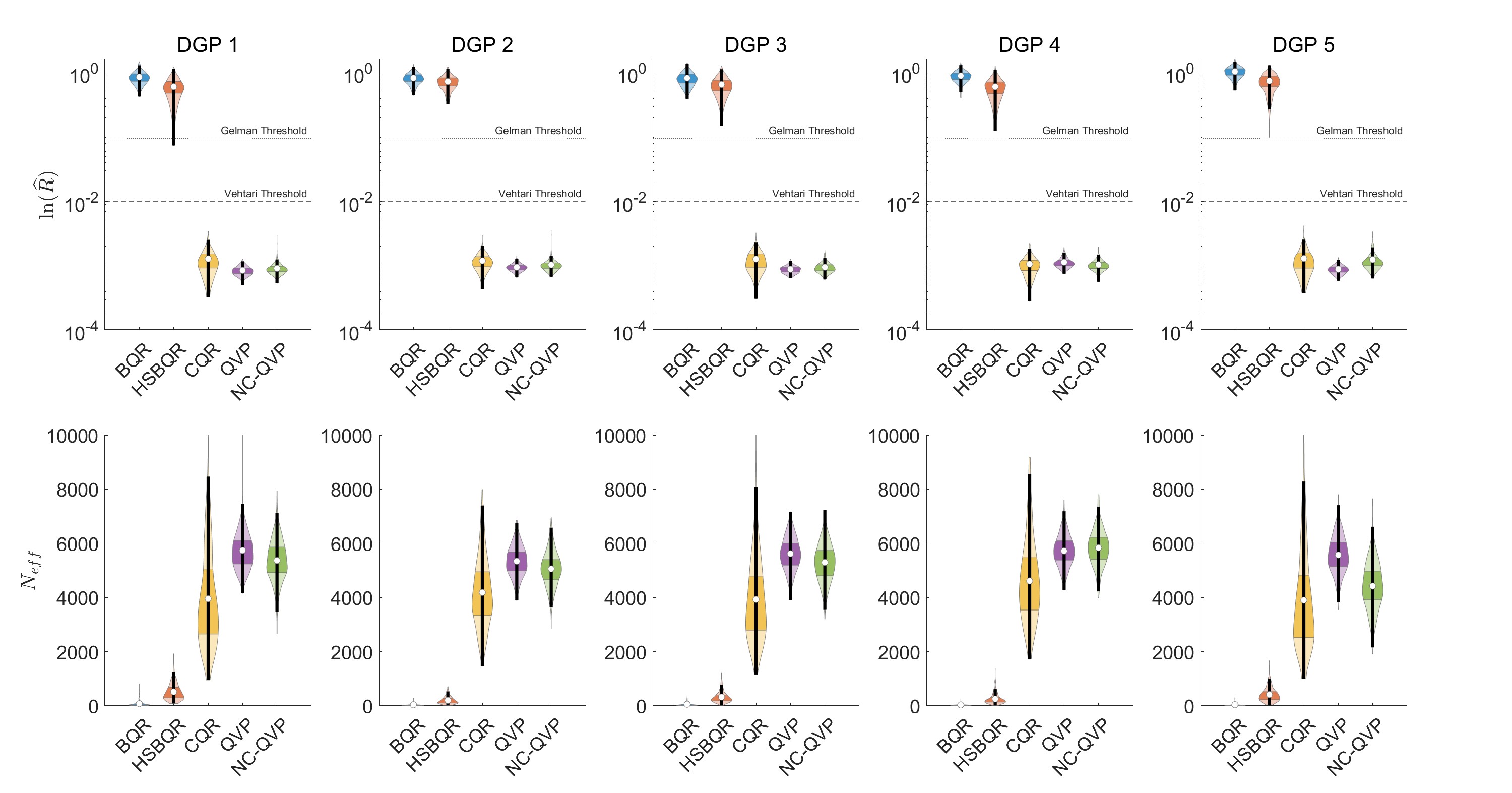}
    \caption{Robustness measures ($\mathcal{T}=300$,$\varDelta$=0) $\hat{R}$ and $N_{\mathrm{eff}}$ are calculated according to \citet{vehtari2021rank} from 4 $\mathrm{MCMC}$ chains with 10k burnin and 15k saved draws. We indicate $\hat{R}$ thresholds according to \citet{vehtari_practical_2017} as well as \citet{gelman_bayesian_1996}.}
    \label{fig:RobustMeas}
\end{figure}

Figure~\ref{fig:RobustMeas} shows that for all $\mathrm{DGP}$s, the $\QVP$ models' $\hat{R}$ and effective sample size, $\mathrm{N}_{\mathrm{eff}}$ of \citet{vehtari2021rank} indicate good mixing, as well as large efficiency gains of the $\mathrm{MCMC}$ sampler over the $\BQR$ and $\HSBQR$ models. As expected, sampling is more efficient for the $\mathrm{QVP}$ model for $\mathrm{DGP}$-5 due to the large quantile variation.

\vspace{-0.5cm}

\begin{table}[]
\centering
\begin{tabular}{l|ccccc}
   & C1 & C2 & C3 & C4 & C5 \\ \hline
$\BQR$ & 71.01\% & 75.17\% & 74.95\% & 72.86\% & 66.75\% \\
$\HSBQR$ & 34.24\% & 50.93\% & 44.06\% & 60.51\% & 35.72\% \\ \hdashline
$\QVP$ & 0.28\% & 0.11\% & 0.06\% & 1.22\% & 0.11\% \\
$\NCQVP$ & 0.06\% & 0.35\% & 0.17\% & 1.46\% & 0.28\% \\
$\NCQVP_{\mathrm{SAVS}}$ & 0.04\% & 0.11\% & 0.06\% & 1.35\% & 0.36\% \\ \hline
\end{tabular}%
\caption{Crossing incidence (see Equation~\ref{eq:cross-i}) for $\mathcal{T}=300$,$\varDelta$=0. Estimates are based on the posterior mean of the coefficients.}
\label{tab:cross}
\end{table}

\section{An Application to Quantile Vector-Autoregressions}\label{sec:application}
For a real word data application, we apply the $\QVP$ prior to the quantile vector-autoregressive ($\QVAR$) model, as presented in \citet{chavleishvili2024forecasting}. The $\QVAR$ generalises the autoregressive quantile model of \citet{koenker2006quantile} to vector valued targets and is commonly used to examine the interactions of endogenous variables across their respective conditional distributions. $\QVAR$s represent an important policy tool for conducting stress-tests on financial systems, and more recently to quantify probability of tail events in macroeconomic time-series \citep{chavleishvili2023quantifying,chavleishvili2023measuring}. The literature has proposed many solutions to the multiple quantile function estimation problem (see \citet{hallin2017multiple} for an overview). Ambiguity arises because, unlike the univariate case, no single universally accepted mathematical framework for defining  a quantile function in multiple dimensions is accepted. The approach proposed in \citet{wei2008approach} is particularly convenient for economic models since the statistical identification assumption is also often defensible from the standpoint of economic theory \citep{chavleishvili2024forecasting}.
The $\QVAR$ of order one, written $\QVAR(1)$, takes the following form for $t=2,\dotsc,\mathcal{T}$: 
\begin{equation} \label{eq:QVAR-qform}
    Q_{\tau_q}(Y_t \mid \Psi_t) = \mathcal{B}_{\tau_q} + \mathcal{A}_{0,{\tau_q}} Y_t + \mathcal{A}_{1,{\tau_q}} Y_{t-1},
\end{equation}
where $Y_t = (y_{1,t},\dotsc,y_{m,t})^T \in \mathbbm{R}^{m}$, $\mathcal{B}_{\tau_q} = ({b}_{1}, \dotsc, {b}_m) \in \mathbbm{R}^m $ is a vector of intercepts, $\mathcal{A}_{1,\tau_{q}} \in \mathbbm{R}^{m \times m}$ is the coefficient matrix on the lag vector, and $\mathcal{A}_{0,\tau_q} \in \mathbbm{R}^{m \times m}$ is a contemporaneous impact matrix. Denote the entry of the $i$\textsuperscript{ith} row and $j$\textsuperscript{th} column of $\mathcal{A}_{1,\tau_q}$ by $a_{i,j,1,q}$. \citet{wei2008approach} show that under the assumption of a lower triangular structure to $\mathcal{A}_0$, one obtains valid multivariate quantile function estimates by estimating the system one equation at a time. With this, $\Psi_t$ is the information set at time $t$, and differs for each variable due to the lower triangular structure of $\mathcal{A}_{0,\tau_q}$: $\Psi_{1,t} = \left\{ Y_{t-1} , Y_{t-2} ,\dotsc  \right\}$, $\Psi_{i,t} = \left\{ Y_{i-1,t},\Psi_{i-1,t} \right\}$ for $i = 2,\dotsc,m$. 

Thus, following the steps in Sections~\ref{sec:likelihood}-\ref{sec:qvp-prior}, the probabilistic representation of the model for the $i$\textsuperscript{th} variable and $q$\textsuperscript{th} quantile of the $\QVAR(1)$ becomes:
\begin{IEEEeqnarray}{rl}\label{eq:centred_qvp}
     y_{i,t} & = b_{i,q} + \tilde{z}_{i,t}^{T}\tilde{a}_{i,q} + \mu_{i,q,t} + \epsilon_{i,q,t}^{y},\; \epsilon^{y}_{i,q,t}\sim\normal\left(0,\theta_q^2\sigma^{y}_{i,q}\omega_{i,q,t}\right) \label{eq:observation-equation-centred-qvar} \\
     \tilde{a}_{i,q} & = \tilde{a}_{i,q-1} + \epsilon^{\tilde{a}}_{i,q},\; \epsilon^{\tilde{a}}_{i,q} \sim \mvn\left(0,\Sigma_{i,q}\right),  \quad q  = 1,\dotsc,\mathcal{Q} \label{eq:state-equation-centred-qvar} \\
     \tilde{a}_{i,0} & \sim \mvn\left(0,\Sigma_{i,0}\right) \label{eq:starting-condition-cented-qvar},
\end{IEEEeqnarray}
where $\tilde{z}_{i,t}$ refers to the $i^{\mathrm{th}}$ information set $\Psi_{i,t}$ in vectorised form, $\vect{\Psi_{i,t}}$, and similarly $\tilde{a}_{i,q}$ vectorises the $i^{\mathrm{th}}$ row of coefficients. Note that this setup differs from the recent literature on multi-variate modelling of multiple quantiles using the multivariate $\mathcal{ALD}$ ($\mathcal{MALD}$). This is not further considered here since the $\mathcal{MALD}$ does not, without further modification, prohibit crossing of quantiles. And inference on the quantile covariance matrix is complicated due to its non-standard conditional posterior \citep{iacopini2023money}.

Representing quantile function~\ref{eq:QVAR-qform} in the sample space of $Y_t$ is convenient for the subsequent forecasting and causal analysis. Define $U_t = (U_{1,t},\dotsc,U_{m,t}) \in (0,1)^m$, such that each element is distributed independently as a uniform distribution. \citet{wei2008approach} show that if the joint distribution of $Y_t$ is absolutely continuous, there exists a one-to-one mapping between the sample space of $Y$ and the hyper-cube $\left(0,1 \right)^m$\footnote{This is known as the Rosenblatt transformation.}: 
\begin{equation} \label{eq:QVARasUnif}
    Y_t = \mathcal{B}(U_t) + \mathcal{A}_0(U_t) Y_t + \mathcal{A}_1(U_t) Y_{t-1},
\end{equation}
where, one can obtain the standard $\VAR$ like representation of the model by making the right hand side, a  function of a constant and lags only:
\begin{equation}
    Y_t = v(U_t) + \mathcal{C}(U_t)Y_{t-1},
\end{equation}
where $v(U_t) \equiv \left( \mathbbm{I}_m - \mathcal{A}_0(U_t) \right)^{-1}\mathcal{B}(U_t)$ and $\mathcal{C}(U_t) \equiv \left( \mathbbm{I}_m - \mathcal{A}_0(U_t) \right)^{-1}\mathcal{A}_1(U_t)$. Set in Equation~\ref{eq:QVARasUnif} $U_t = \tau_q$ for $t=2,\dotsc,\mathcal{T}$.\footnote{These inverses can be shown to always exists whenever the $\QVAR$ coefficients imply a stationary process, which is equivalent to the stationary conditions on parameter matrices of standard $\VAR$ models.} 

\begin{figure}
    \centering
    \includegraphics[width=0.9\linewidth]{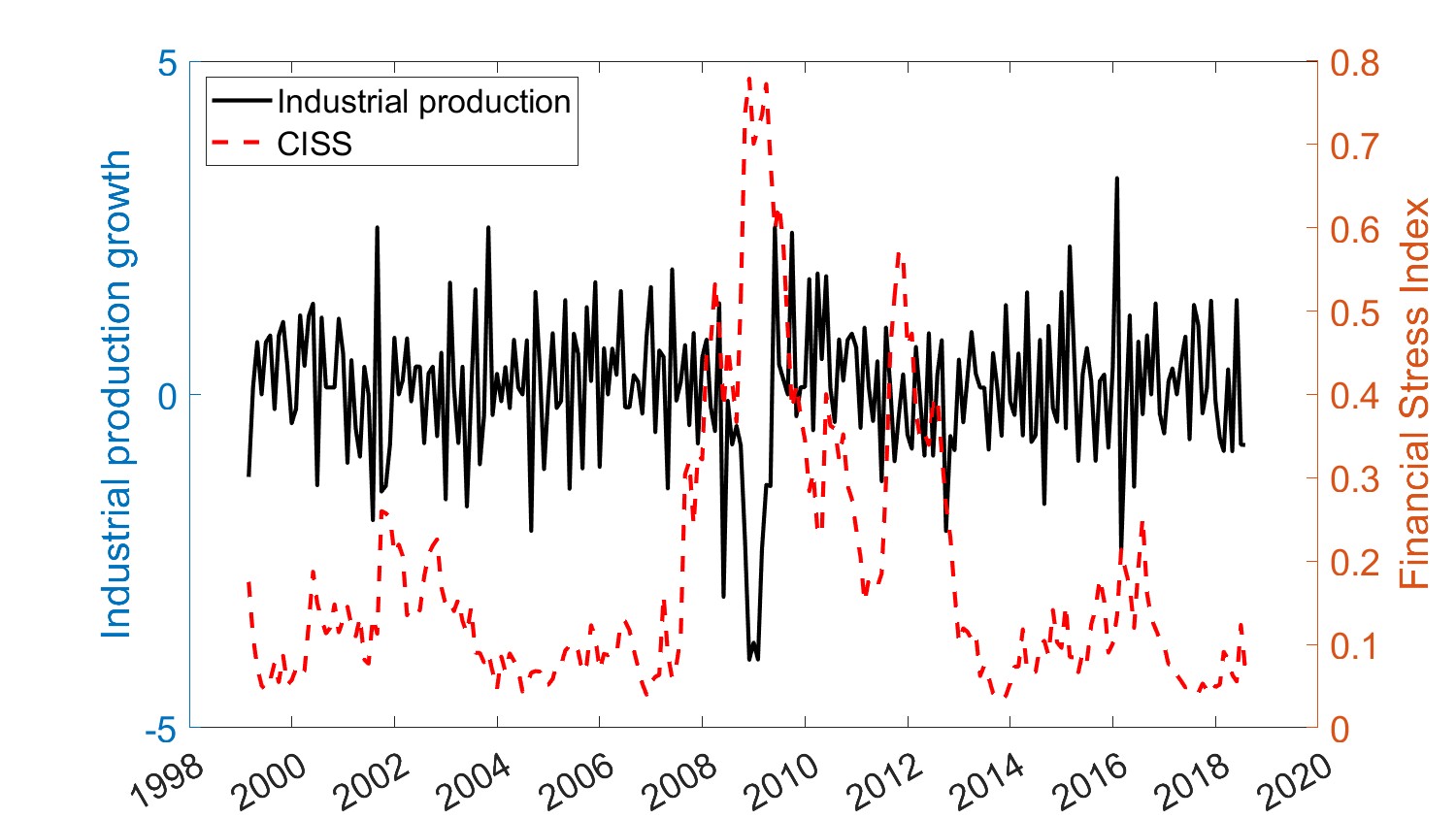}
    \caption{IP and CISS index for the available time-frame of January 1999 until July 2018}
    \label{fig:EmpAppData}
\end{figure}

The data set for this application, obtained from  \citet{chavleishvili2024forecasting} on Euro Area macrofinancial data, contains the industrial production growth-rate $(\IP)$, which measures real economic activity, and the composite indicator of systemic stress $(\CISS)$, representing a measure of financial health of the Euro Area.\footnote{For more examples of the $\CISS$ used in Growth-at-Risk models, see \citet{figueres2020vulnerable}, \citet{szendrei2023revisiting} or \citet{varga2025non} among others.} The time-series are plotted in Figure~\ref{fig:EmpAppData}. Data are monthly and available from January 1999 to July 2018. The goal of the study is to jointly forecast $\IP$ and $\CISS$ with the $\QVAR$ system and perform causal analysis of how the conditional distribution of $\IP$ responds to perturbations in the $\CISS$ indicator.
\citet{chavleishvili2024forecasting} give economic justification to a lower-triangular identification scheme where $\IP$ impacts $\CISS$ contemporaneously, but $\IP$ only impacts the $\CISS$ with a lag.\footnote{This identification assumption is implicit in \citet{adrian2019vulnerable} and many more studies that followed. Lower‐triangular identification entails that shocks are identified through how they propagate dynamically through the system of equations. This approach is widely used in macroeconomics because it yields a unique decomposition of structural shocks for location scale $\VAR$ models, often matching intuitive causal narratives \citep{sims1980macroeconomics}. } 


We estimate the $\QVAR$ with the same set of priors as in Section~\ref{sec:simulation}. We generate h-step-ahead quantile predictions $Y_{t+h}$ for $h = 1,3,6$, which are evaluated with the $\QS$ score presented in Section~\ref{sec:simulation}. Then, we perform causal analysis common to the $\VAR$ applications in which we estimate the impulse response function of $\IP$ to a one-standard-deviation shock in the $\CISS$. 

To retrieve quantile predictions, we follow the procedure given in \citet{chavleishvili2024forecasting}, however using the MCMC draws where appropriate. We summarise this in Algorithm~\ref{algorithm:forecasts}.

\begin{algorithm}[H] 
\caption{Algorithm for Multi-step Quantile Forecasts} \label{algorithm:forecasts}
\begin{algorithmic}[1]
\State \textbf{Input:} Data window: $\mathcal{T}_r$,  Data: $\left\{Y_t \right\}^{\mathcal{T}_{r }}_{t=1}$, posterior samples: $\left\{v(U)^{(s)}, \mathcal{C}(U)^{(s)}  \right\}_{s=1}^S$
\For{$n \in \left\{1,\dotsc,N\right\}$}
\For{$l \in {1,\dotsc,h}$}
    \State Randomly permute quantile level $U_{T_r+l}^{(n)}$ 
    \State Randomly draw from posterior sample $v(U_{T_r+l})^{(n)}, \mathcal{C}(U_{T_r+l})^{(n)}$
    \If{$l = 1$}
        \State Set $\hat{Y}_{T_r+1}^{(n)} = v(U_{T_r+1}^{(n)})^{(n)} + \mathcal{C}(U_{T_r+1}^{(n)})^{(n)} Y_{T_r}$
    \Else
        \State Set $Y_{T_r+l}^{(n)} = v(U_{T_r+l}^{(n)})^{(n)} + \mathcal{C}(U_{T_r+l}^{(n)})^{(n)} \hat{Y}_{T_r+l-1}^{(n)}$
    \EndIf
    \EndFor
\EndFor
\State \textbf{Output:} $ \boldsymbol{\hat{Y}} = \left\{\hat{Y}_{T_r+1}^{(n)},\dotsc, \hat{Y}_{T_r+h}^{(n)} \right\}_{n=1}^N$, where percentile $\tau_q$ of $\left\{ \hat{Y}^{(n)}_{T_r+l}\right\}_{n=1}^N$ yields the corresponding quantile prediction for period $t+l$.
\end{algorithmic}
\end{algorithm}
\subsection{Forecast Results}
Forecasts are produced on an expanding in-sample time-window $t = 1\dotsc,\mathcal{T}_r$, with initial in-sample period $\mathcal{T}_{\mathrm{start}}=96$ and $\mathcal{T}_{\mathrm{end}}=224$.
Quantile weighted $\QS$ scores for the $\QVAR$s are shown in Table~\ref{tab:ForcRes}. Similar to the results in Section~\ref{sec:simulation}, we show forecasting performance relative to the $\BQR$-$\QVAR$ model.
\begin{table}[]
\centering
\resizebox{\textwidth}{!}{%
\begin{tabular}{ll|cccc|cccc|cccc}
 &  & \multicolumn{4}{c|}{h=1} & \multicolumn{4}{c|}{h=3} & \multicolumn{4}{c}{h=6} \\
 &  & CRPS & Centre & Left & Right & CRPS & Centre & Left & Right & CRPS & Centre & Left & Right \\ \hline
\multicolumn{2}{l|}{Equation 1: IP} &  &  &  &  &  &  &  &  &  &  &  &  \\
& $\BQR$  & 0.378 & 0.069 & 0.121 & 0.119 & 0.397 & 0.072 & 0.130 & 0.123 & 0.416 & 0.076 & 0.142 & 0.124 \\ \hdashline
& $\HSBQR$  & 91.1\% & 95.2\% & 88.3\% & 89.0\% & 90.7\% & 95.0\% & 88.3\% & 88.2\% & 90.1\% & 94.4\% & 88.7\% & 86.5\% \\
& $\BRW$  & 87.3\% & 92.9\% & 83.6\% & 84.5\% & 86.4\% & 92.3\% & 82.8\% & 83.2\% & 85.2\% & 91.5\% & 80.3\% & 83.2\% \\
& $\QVP$  & 86.3\% & 92.1\% & 83.1\% & 82.9\% & 86.4\% & 92.6\% & 83.2\% & 82.6\% & 85.0\% & 91.4\% & 80.3\% & 82.6\% \\
& $\NCQVP$  & 87.0\% & 92.8\% & 83.6\% & 83.7\% & 86.4\% & 92.3\% & 82.8\% & 83.2\% & 85.8\% & 92.0\% & 81.0\% & 83.5\% \\
& $\NCQVP_{\mathrm{SAVS}}$  & 87.4\% & 92.9\% & 84.5\% & 84.0\% & 86.6\% & 92.4\% & 83.0\% & 83.6\% & 84.8\% & 90.9\% & 79.3\% & 83.7\% \\
 \hline
\multicolumn{2}{l|}{Equation 2: CISS} &  &  &  &  &  &  &  &  &  &  &  &  \\
& $\BQR$  & 0.019 & 0.003 & 0.005 & 0.007 & 0.034 & 0.006 & 0.009 & 0.013 & 0.052 & 0.009 & 0.013 & 0.020 \\ \hdashline
& $\HSBQR$  & 91.9\% & 94.7\% & 89.9\% & 90.7\% & 95.4\% & 98.3\% & 92.3\% & 94.6\% & 96.8\% & 99.3\% & 92.6\% & 97.0\% \\
& $\BRW$  & 86.2\% & 90.9\% & 84.7\% & 82.7\% & 89.1\% & 94.1\% & 88.2\% & 84.9\% & 91.6\% & 96.6\% & 92.8\% & 86.3\% \\
& $\QVP$  & 85.7\% & 90.5\% & 84.2\% & 82.0\% & 88.5\% & 93.7\% & 87.9\% & 84.1\% & 90.9\% & 96.3\% & 92.1\% & 85.2\% \\
& $\NCQVP$  & 86.0\% & 90.6\% & 84.4\% & 82.4\% & 88.8\% & 93.9\% & 88.1\% & 84.3\% & 90.8\% & 96.1\% & 91.8\% & 85.2\% \\
& $\NCQVP_{\mathrm{SAVS}}$  & 91.2\% & 95.3\% & 89.9\% & 88.0\% & 91.9\% & 96.1\% & 90.1\% & 89.2\% & 91.5\% & 95.4\% & 93.3\% & 86.7\% \\
 \hline
\multicolumn{2}{l|}{Overall} &  &  &  &  &  &  &  &  &  &  &  &  \\
& $\BQR$  & 0.198 & 0.036 & 0.063 & 0.063 & 0.215 & 0.039 & 0.069 & 0.068 & 0.234 & 0.042 & 0.077 & 0.072 \\ \hdashline
& $\HSBQR$  & 91.1\% & 95.2\% & 88.4\% & 89.1\% & 91.1\% & 95.3\% & 88.6\% & 88.8\% & 90.8\% & 94.9\% & 89.0\% & 88.0\% \\
& $\BRW$  & 87.2\% & 92.8\% & 83.6\% & 84.4\% & 86.6\% & 92.4\% & 83.2\% & 83.3\% & 85.9\% & 92.0\% & 81.3\% & 83.7\% \\
& $\QVP$  & 86.3\% & 92.0\% & 83.2\% & 82.8\% & 86.6\% & 92.6\% & 83.5\% & 82.8\% & 85.7\% & 91.9\% & 81.2\% & 83.0\% \\
& $\NCQVP$  & 86.9\% & 92.7\% & 83.6\% & 83.6\% & 86.6\% & 92.4\% & 83.2\% & 83.3\% & 86.3\% & 92.5\% & 81.9\% & 83.8\% \\
& $\NCQVP_{\mathrm{SAVS}}$  & 87.6\% & 93.0\% & 84.7\% & 84.2\% & 87.0\% & 92.6\% & 83.4\% & 84.2\% & 85.6\% & 91.4\% & 80.4\% & 84.2\% \\
 \hline
\end{tabular}%
}
\caption{Forecast performance as measured by the weighted quantile score, $\mathrm{qwQS}$, see Equation~\ref{eq:qwQS}. Weighting schemes are $w_{q} = 1/\mathcal{Q}$ ($\QS$) which is equal to the $\mathrm{CRPS}$. $w_{q} = \tau_q(1-\tau_q)$ (Centre) places more weight the central quantiles, $w_{q} = (1-\tau_q)^2$ (Left) places more weight on left tail quantiles, $w_{\tau_q} = \tau^2_q$ (Right) places more weight on right tail quantiles. Performance is shown relative to $\BQR$ whose absolute performance shown above the dotted lines respectively. For the $\NCQVP_{\mathrm{SAVS}}$ only, the posterior means were used for this forecast exercise.}
\label{tab:ForcRes}
\end{table}
Several clear patterns emerge. First, joint‐quantile models consistently outperform independent quantile models ($\BQR$, $\HSBQR$), confirming the benefit of `partially pooling' information across quantiles observed in Section~$\ref{sec:simulation}$. Second, while $\HSBQR$ often excels at central quantiles, its tail performance is much worse than the $\QVP$ models, especially at longer forecast horizons. This is particularly visible for the $\IP$ variable. Third, the $\QVP$ prior variants can even outperform the $\BRW$ model, particularly in the tails. Here, integration over the parameter space with the Bayesian models induces more smoothness of the quantile function of the coefficients (see Section~\ref{app-sec:fig-posteriors}) and therefore reduces variance in the tails. Finally, forecast accuracy gets worse as the horizon increases which is a consequence of model parsimony and accumulation of forecast uncertainty. However, the loss in forecasts accuracy is less pronounced for joint estimation frameworks, particularly for the $\IP$ equation. 

In terms of crossing of the estimated (in-sample) quantiles, we find, similar to the Section~\ref{sec:simulation}, that the $\QVP$ priors almost completely eradicate the issue of crossing quantile curves (see Table~\ref{tab:cross-in-sample}). 




\begin{table}[]
\centering
\begin{tabular}{r|ccc}
   & Eq. 1 & Eq. 2 & Overall \\ \hline
$\BQR$ & 23.06\% & 50.26\% & 36.66\% \\
$\HSBQR$ & 16.83\% & 15.04\% & 15.94\% \\ \hdashline
$\QVP$ & 0.05\% & 0.14\% & 0.09\% \\
$\NCQVP$ & 0.20\% & 0.09\% & 0.15\% \\
$\NCQVP_{\mathrm{SAVS}}$ & 0.16\% & 0.05\% & 0.10\% \\ \hline
\end{tabular}%
\caption{Crossing incidence (see Equation~\ref{eq:cross-i}) based on the entire sample. Estimates are based on the posterior mean of the coefficients.}
\label{tab:cross-in-sample}
\end{table}

\subsection{Quantile IRF}
$\QIRF$s measure the distributional causal effect of an unexpected change in $Y_{i,t}$ on $Y_{j,t}$ over some horizon $h = 1,\dotsc,W$. Such analyses are of interest to policy institutions like central banks, who tailor their policy instruments to the likelihood the potential paths real economic output take in relation to movements of the financial sector. Common to location-scale $\VAR$ analysis, the $\QVAR$ approach allows the analysis of such dynamic responses along particular points of the conditional distributions, in particular the tails.
%
%

Denote by $Y_t^{*}$ a hypothetical `shocked' vector $Y_t$, to which $\iota \in \mathbbm{R}^m$ is added. As in \citet{chavleishvili2024forecasting}, we define $\iota$ as a vector of zeros, except entry $i$ which is equal to the shock's magnitude. The quantile impulse-response function at time point $t$ is defined as $\delta_{t}(U_t) \equiv Y_t^{*} - Y_t$.  \citet{chavleishvili2024forecasting} show that under the triangular identification scheme from above, this reduces to $\mathcal{D}(U_t)\iota$, where $\mathcal{D}(U_t) = \left( \mathbbm{I}_m - \mathcal{A}_0(U_t) \right)^{-1}$. The impulse response function for $h$-steps ahead is then given by: $\delta_{t+h}\left( U_{t+h} \vert U_t,\dotsc, U_{t+h-1}\right) = \prod_{g=1}^h\mathcal{C}(U_{t+g})\mathcal{D}(U_t)\iota$. We define the quantile response levels of interest of $\IP$ to be equal to $U_{\IP,t+g} = \tau = (0.05,0.1,\dotsc,0.95) \forall g$ while that of $\CISS$ are fixed to the median level, $U_{\CISS,t+g} = 0.5 \forall g$. Note, however, that for estimation of the coefficient-posteriors, it remains that both equations are estimated for all $\tau$.\footnote{We keep the quantile levels for the $\QIRF$ constant respective to each equation of the $\QVAR$. In principle, the quantile indices must stay constant across the horizons of the $\QIRF$.} The initial shock $\iota$ is set equal to a one standard deviation of the residuals at the median quantile of the $\CISS$ equation \citep{chavleishvili2024forecasting}. The previous literature finds that shocks to financial conditions have a more pronounced negative impact on the left tail of real economic output \citep{adrian2019vulnerable,chavleishvili2024forecasting}. It is therefore expected that the $\QIRF$s show a pronounced negative impact at low quantile levels that even out to 0 at high quantile levels.

\begin{figure}
    \centering
    \includegraphics[width=\linewidth]{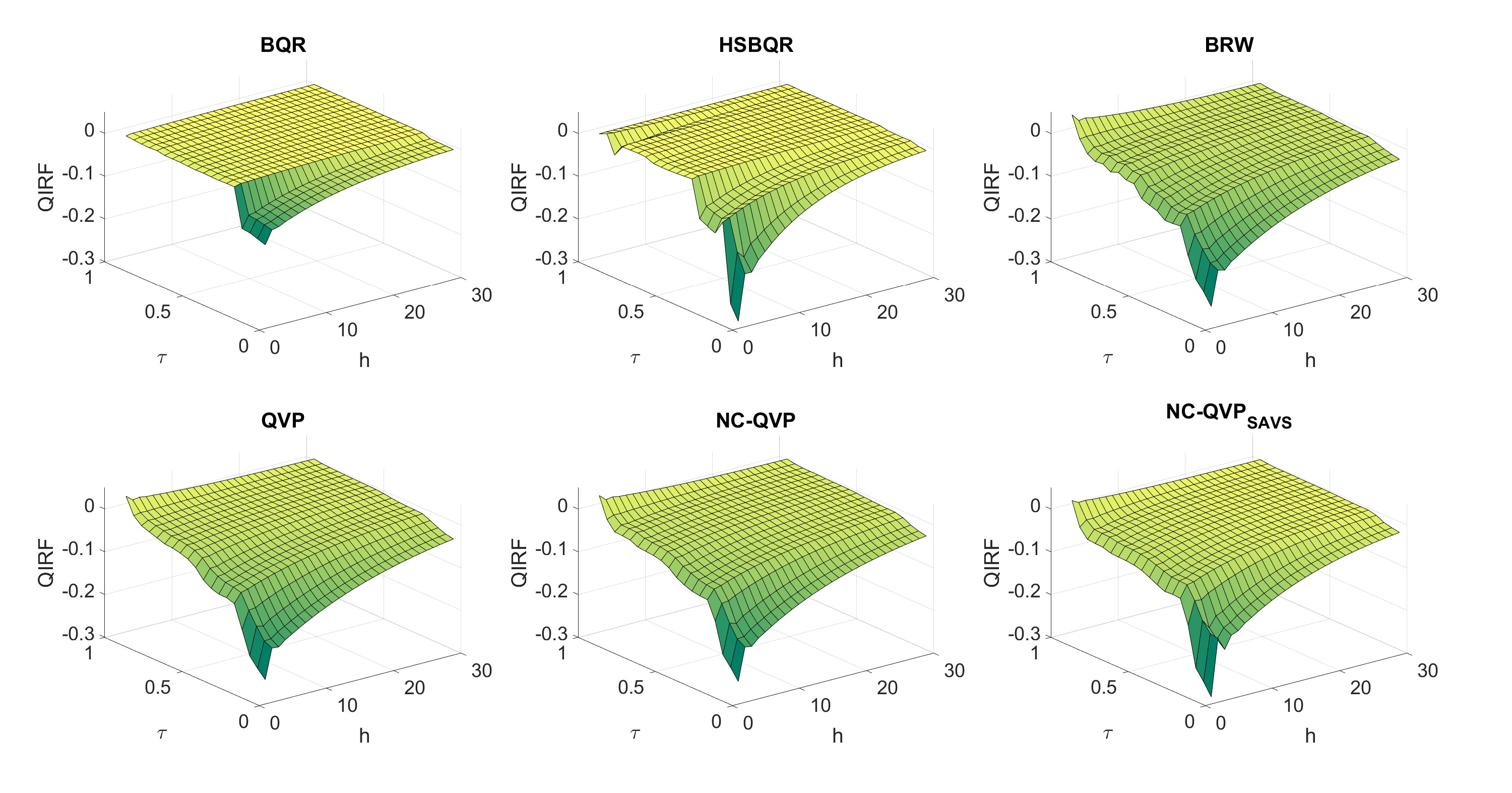}
    \caption{$\QIRF$ surface of $\IP$ shown in response to one standard deviation shock to the $\CISS$ variable. Estimates are  based on the posterior means for quantile levels $\tau_q \in \left\{0.05,\dotsc,0.95\right\}$.}
    \label{fig:QIRF}
\end{figure}

Figure~\ref{fig:QIRF} shows the impulse response functions for the various quantile models for $h=1,\dotsc,30$. 



\begin{figure}
    \centering
    \includegraphics[width=\linewidth]{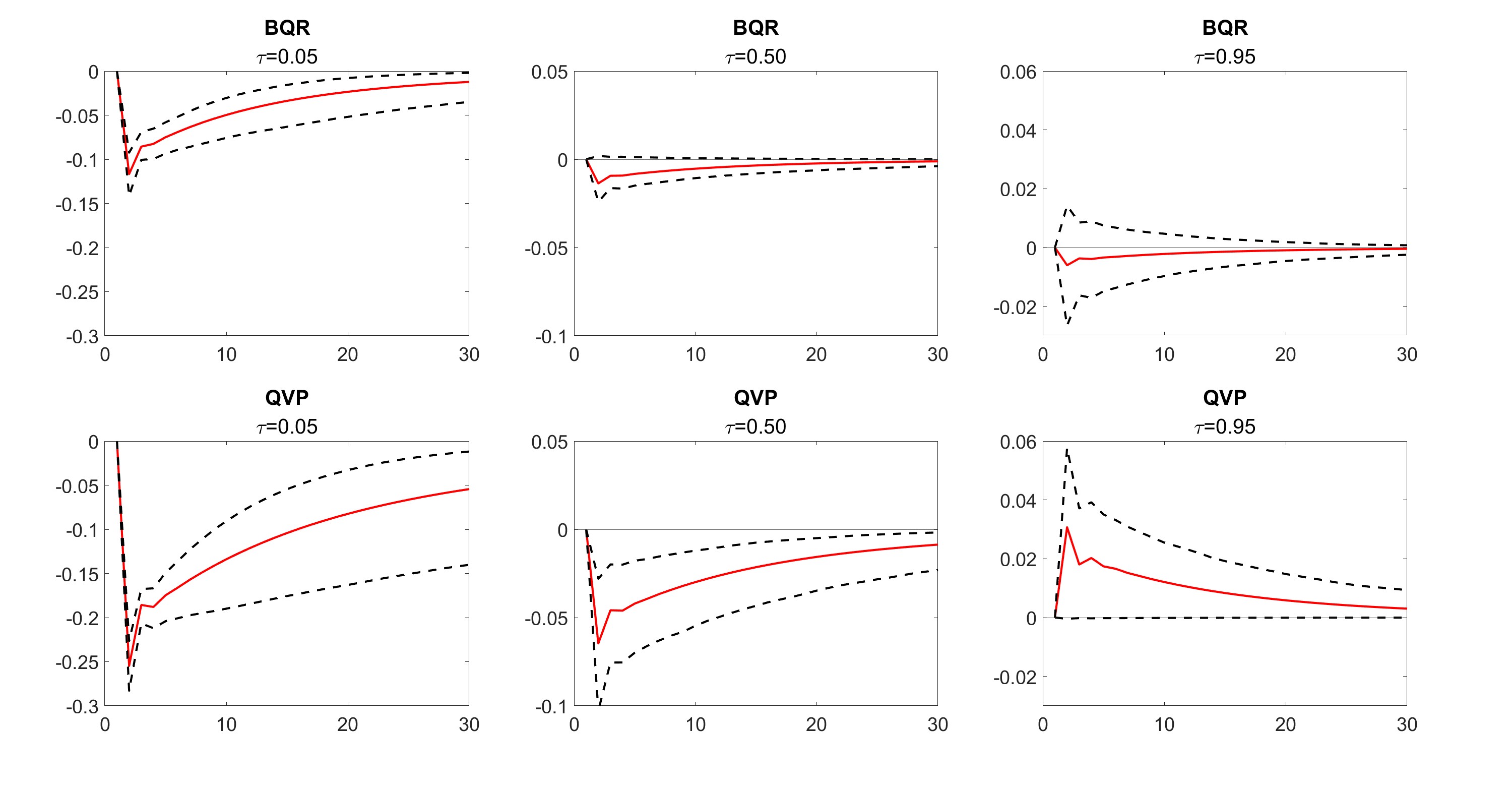}
    \caption{$\QIRF$ for $\BQR$ and $\QVP$ shown for selected quantiles $\tau_q \in \left\{0.05, 0.5, 0.95 \right\}$.}
    \label{fig:QIRFQuantSpec}
\end{figure}


%
As expected, Figure~\ref{fig:QIRF} shows generally that all models estimate a pronounced negative impact of a shock to $\CISS$ to the left tail of $\IP$ with impulse response functions petering out to zero as the quantile level increases.
While the $\BQR$ and $\HSBQR$ models exhibit a notably sharp dip in the lower quantile only, the $\QVP$ models estimate a more gentle slope along the quantile levels. This is due to the $\QVP$ priors leading to smoother posterior quantile coefficient profiles - even compared to the $\BRW$ model.\footnote{Posteriors of the coefficients are shown in the appendix in figures \ref{fig:CoeffsIntercept}, \ref{fig:CoeffsContemp}, and \ref{fig:CoeffsLag}.}


Compared to \citet{chavleishvili2024forecasting}, we find that joint estimation of quantiles leads to significant differences in the $\QIRF$s, particularly at the median. Figure~\ref{fig:QIRFQuantSpec} shows the $\QIRF$ at selected quantiles for better visibility (including posterior uncertainty intervals). While the $\BQR$ model, in line with \citet{chavleishvili2024forecasting}, predicts no significant impact of the $\CISS$ shock at the median, the $\QVP$ model predicts a persistent negative one. Taken together, we confirm the previous literature's finding that post shock distributions of $\IP$ exhibit negative skewness, however, we also observe a significant downward location shift identified by the $\QVP$ models.  This highlights how using joint estimation, along with a prior that regulates information `pooling', can have a significant influence on the inference drawn from these models.


\section{Conclusion}\label{sec:conclusion}

In this paper, we defined a prior for multiple quantile regression in which information across quantiles is shared via an adaptive joint shrinkage prior. The structure of the prior is motivated from the penalised non-crossing objective function from \citet{bondell2010noncrossing}, and is shown to imply a quantile state-space representation, named $\QVP$ model, where unknown states are equal to the quantile regression coefficients. This allows for the derivation of efficient sampling methods where the resultant triangular structure of the conditional posterior precision allows for fast computation. We extend the $\QVP$ framework to a non-centred formulation ($\NCQVP$) as well as a post-estimation sparsification algorithm that allow for stronger shrinkage on state variability and sparsity, respectively. With this method, we were able to tackle the issue of quantile crossing through a structured prior that regularises toward the desired parameter sub-space, rather than modifying the likelihood. 
     
A simulation exercise shows that the $\QVP$ priors result in far superior predictive performance and parameter recovery compared to Bayesian methods that estimate quantiles independently. Additionally, crossing is nearly completely eliminated with this approach. For low true quantile variation, the $\NCQVP$ models can also offer large gains over frequentist methods that strictly enforce non-crossing.
    
In the empirical application of a $\QVAR$ on the Euro Area, following \citet{chavleishvili2024forecasting}, we show the practical advantages of the $\QVP$ prior in modelling complex macroeconometric dynamics. We produce quantile forecasts as well as conduct a causal study of the effect of financial shocks to the distribution of industrial production, $\IP$. $\QVP$ models produce very competitive forecasts, often outperforming all models under comparison.
    
For the causal study, we generate impulse response functions ($\QIRF$) of $\IP$ in reaction to shocks to worsening financial conditions. We verify the finding that financial shocks exert markedly asymmetric and persistent effects across the conditional distribution of $\IP$. The $\QVP$ priors produce smoother $\QIRF$s with larger negative effects at the lower tails which are more persistent.

Despite these advantages, there are several avenues by which the method can be improved. First, the paper focuses on implementing the framework to linear quantile regression models. However, the method can be extended to nonlinear settings as well. This extension can enhance the applicability of the $\QVP$ prior framework. Second, we have exclusively focused on the horseshoe prior for modelling the differences across quantiles. The $\QVP$ framework can be used on various other types of shrinkage priors, such as the GIGG prior (see for example \citet{kohns2025flexible}).

\subsection*{Acknowledgments}
We acknowledge the computational resources provided by the Aalto Science-IT project, and the support of the Research Council of Finland Flagship programme: Finnish Center for Artificial Intelligence, Research Council of Finland project (340721), and the Finnish Foundation for Technology Promotion. We thank Gary Koop, Niko Hauzenberger, Ping Wu, Aristeidis Raftapostolos, and all the participants of the 2024 CFE conference, 2025 RSS conference for their feedback. We further thank the Scotland national rugby union teams for their continued effort both on and off the field.
\bibliography{main.bbl}

\begin{thebibliography}{}

\bibitem[Adrian et~al., 2019]{adrian2019vulnerable}
Adrian, T., Boyarchenko, N., and Giannone, D. (2019).
\newblock Vulnerable growth.
\newblock {\em American Economic Review}, 109(4):1263--1289.

\bibitem[Alhamzawi, 2015]{alhamzawi2015model}
Alhamzawi, R. (2015).
\newblock Model selection in quantile regression models.
\newblock {\em Journal of Applied Statistics}, 42(2):445--458.

\bibitem[Bhattacharya et~al., 2016]{bhattacharya2016fast}
Bhattacharya, A., Chakraborty, A., and Mallick, B.~K. (2016).
\newblock Fast sampling with gaussian scale mixture priors in high-dimensional regression.
\newblock {\em Biometrika}, page asw042.

\bibitem[Bissiri et~al., 2016]{bissiri2016general}
Bissiri, P.~G., Holmes, C.~C., and Walker, S.~G. (2016).
\newblock A general framework for updating belief distributions.
\newblock {\em Journal of the Royal Statistical Society: Series B (Statistical Methodology)}, 78(5):1103--1130.

\bibitem[Bitto and Fr{\"u}hwirth-Schnatter, 2019]{bitto2019achieving}
Bitto, A. and Fr{\"u}hwirth-Schnatter, S. (2019).
\newblock Achieving shrinkage in a time-varying parameter model framework.
\newblock {\em Journal of Econometrics}, 210(1):75--97.

\bibitem[Bondell et~al., 2010]{bondell2010noncrossing}
Bondell, H.~D., Reich, B.~J., and Wang, H. (2010).
\newblock Noncrossing quantile regression curve estimation.
\newblock {\em Biometrika}, 97(4):825--838.

\bibitem[Cadonna et~al., 2020]{cadonna2020triple}
Cadonna, A., Fr{\"u}hwirth-Schnatter, S., and Knaus, P. (2020).
\newblock Triple the gamma—a unifying shrinkage prior for variance and variable selection in sparse state space and tvp models.
\newblock {\em Econometrics}, 8(2):20.

\bibitem[Carpenter et~al., 2017]{carpenter_stan_2017}
Carpenter, B., Gelman, A., Hoffman, M.~D., Lee, D., Goodrich, B., Betancourt, M., Brubaker, M., Guo, J., Li, P., and Riddell, A. (2017).
\newblock \textit{{Stan}} : {A} {Probabilistic} {Programming} {Language}.
\newblock {\em Journal of Statistical Software}, 76(1).

\bibitem[Carvalho et~al., 2009]{carvalho_handling_2009}
Carvalho, C.~M., Polson, N.~G., and Scott, J.~G. (2009).
\newblock Handling {Sparsity} via the {Horseshoe}.
\newblock In {\em Proceedings of the {Twelth} {International} {Conference} on {Artificial} {Intelligence} and {Statistics}}, pages 73--80. PMLR.

\bibitem[Chavleishvili and Kremer, 2023]{chavleishvili2023measuring}
Chavleishvili, S. and Kremer, M. (2023).
\newblock Measuring systemic financial stress and its risks for growth.
\newblock {\em ECB Working Paper}.

\bibitem[Chavleishvili et~al., 2023]{chavleishvili2023quantifying}
Chavleishvili, S., Kremer, M., and Lund-Thomsen, F. (2023).
\newblock Quantifying financial stability trade-offs for monetary policy: a quantile var approach.
\newblock {\em ECB Working Paper}, (2833).

\bibitem[Chavleishvili and Manganelli, 2024]{chavleishvili2024forecasting}
Chavleishvili, S. and Manganelli, S. (2024).
\newblock Forecasting and stress testing with quantile vector autoregression.
\newblock {\em Journal of Applied Econometrics}, 39(1):66--85.

\bibitem[Chernozhukov et~al., 2010]{chernozhukov2010quantile}
Chernozhukov, V., Fern{\'a}ndez-Val, I., and Galichon, A. (2010).
\newblock Quantile and probability curves without crossing.
\newblock {\em Econometrica}, 78(3):1093--1125.

\bibitem[Chernozhukov and Hong, 2003]{chernozhukov2003mcmc}
Chernozhukov, V. and Hong, H. (2003).
\newblock An mcmc approach to classical estimation.
\newblock {\em Journal of econometrics}, 115(2):293--346.

\bibitem[Devroye, 2014]{devroye2014random}
Devroye, L. (2014).
\newblock Random variate generation for the generalized inverse gaussian distribution.
\newblock {\em Statistics and Computing}, 24(2):239--246.

\bibitem[Duan et~al., 2020]{duan2020bayesian}
Duan, L.~L., Young, A.~L., Nishimura, A., and Dunson, D.~B. (2020).
\newblock Bayesian constraint relaxation.
\newblock {\em Biometrika}, 107(1):191--204.

\bibitem[Feldman and Kowal, 2023]{feldman2023bayesian}
Feldman, J. and Kowal, D. (2023).
\newblock Bayesian quantile regression with subset selection: A posterior summarization perspective.
\newblock {\em arXiv preprint arXiv:2311.02043}.

\bibitem[Figueres and Jaroci{\'n}ski, 2020]{figueres2020vulnerable}
Figueres, J.~M. and Jaroci{\'n}ski, M. (2020).
\newblock Vulnerable growth in the euro area: Measuring the financial conditions.
\newblock {\em Economics Letters}, 191:109126.

\bibitem[Fr{\"u}hwirth-Schnatter and Wagner, 2010]{fruhwirth2010stochastic}
Fr{\"u}hwirth-Schnatter, S. and Wagner, H. (2010).
\newblock Stochastic model specification search for gaussian and partial non-gaussian state space models.
\newblock {\em Journal of Econometrics}, 154(1):85--100.

\bibitem[Gelman, 1996]{gelman_bayesian_1996}
Gelman, A. (1996).
\newblock Bayesian {Model}-{Building} {By} {Pure} {Thought}: {Some} {Principles} and {Examples}.
\newblock {\em Statistica Sinica}, 6(1).

\bibitem[Gneiting and Raftery, 2007]{gneiting_strictly_2007}
Gneiting, T. and Raftery, A.~E. (2007).
\newblock Strictly {Proper} {Scoring} {Rules}, {Prediction}, and {Estimation}.
\newblock {\em Journal of the American Statistical Association}, 102(477):359--378.

\bibitem[Gneiting and Ranjan, 2011]{gneiting2011comparing}
Gneiting, T. and Ranjan, R. (2011).
\newblock Comparing density forecasts using threshold-and quantile-weighted scoring rules.
\newblock {\em Journal of Business \& Economic Statistics}, 29(3):411--422.

\bibitem[Hahn and Carvalho, 2015]{hahn2015decoupling}
Hahn, P.~R. and Carvalho, C.~M. (2015).
\newblock Decoupling shrinkage and selection in bayesian linear models: a posterior summary perspective.
\newblock {\em Journal of the American Statistical Association}, 110(509):435--448.

\bibitem[Hallin and \v{S}iman, 2017]{hallin2017multiple}
Hallin, M. and \v{S}iman, M. (2017).
\newblock Multiple-output quantile regression.
\newblock In Koenker, R., Chernozhukov, V., He, X., and Peng, L., editors, {\em Handbook of Quantile Regression}, pages 185--207. Chapman and Hall/CRC.

\bibitem[H{\"o}rmann and Leydold, 2014]{hormann2014generating}
H{\"o}rmann, W. and Leydold, J. (2014).
\newblock Generating generalized inverse gaussian random variates.
\newblock {\em Statistics and Computing}, 24:547--557.

\bibitem[Huber et~al., 2021]{huber2021inducing}
Huber, F., Koop, G., and Onorante, L. (2021).
\newblock Inducing sparsity and shrinkage in time-varying parameter models.
\newblock {\em Journal of Business \& Economic Statistics}, 39(3):669--683.

\bibitem[Iacopini et~al., 2023]{iacopini2023money}
Iacopini, M., Poon, A., Rossini, L., and Zhu, D. (2023).
\newblock Money growth and inflation: A quantile sensitivity approach.
\newblock {\em arXiv preprint arXiv:2308.05486}.

\bibitem[Jiang et~al., 2013]{jiang2013interquantile}
Jiang, L., Wang, H.~J., and Bondell, H.~D. (2013).
\newblock Interquantile shrinkage in regression models.
\newblock {\em Journal of Computational and Graphical statistics}, 22(4):970--986.

\bibitem[Koenker, 2005]{koenker2005}
Koenker, R. (2005).
\newblock {\em Quantile regression}.
\newblock New York: Cambridge University Press.

\bibitem[Koenker and Xiao, 2006]{koenker2006quantile}
Koenker, R. and Xiao, Z. (2006).
\newblock Quantile autoregression.
\newblock {\em Journal of the American statistical association}, 101(475):980--990.

\bibitem[Kohns and Potjagailo, 2025]{kohns2025flexible}
Kohns, D. and Potjagailo, G. (2025).
\newblock Flexible bayesian midas: time-variation, group-shrinkage and sparsity.
\newblock {\em Journal of Business \& Economic Statistics}, pages 1--17.

\bibitem[Kohns and Szendrei, 2021]{kohns2021decoupling}
Kohns, D. and Szendrei, T. (2021).
\newblock Decoupling shrinkage and selection for the bayesian quantile regression.
\newblock {\em arXiv preprint arXiv:2107.08498}.

\bibitem[Kohns and Szendrei, 2024]{kohns2024horseshoe}
Kohns, D. and Szendrei, T. (2024).
\newblock Horseshoe prior bayesian quantile regression.
\newblock {\em Journal of the Royal Statistical Society Series C: Applied Statistics}, 73(1):193--220.

\bibitem[Kottas and Gelfand, 2001]{kottas2001bayesian}
Kottas, A. and Gelfand, A.~E. (2001).
\newblock Bayesian semiparametric median regression modeling.
\newblock {\em Journal of the American Statistical Association}, 96(456):1458--1468.

\bibitem[Kotz et~al., 2001]{kotz2001asymmetric}
Kotz, S., Kozubowski, T.~J., Podg{\'o}rski, K., Kotz, S., Kozubowski, T.~J., and Podg{\'o}rski, K. (2001).
\newblock Asymmetric laplace distributions.
\newblock {\em The Laplace distribution and generalizations: a revisit with applications to communications, economics, engineering, and finance}, pages 133--178.

\bibitem[Kozumi and Kobayashi, 2011]{kozumi2011gibbs}
Kozumi, H. and Kobayashi, G. (2011).
\newblock Gibbs sampling methods for bayesian quantile regression.
\newblock {\em Journal of statistical computation and simulation}, 81(11):1565--1578.

\bibitem[Lancaster and Jae~Jun, 2010]{lancaster2010bayesian}
Lancaster, T. and Jae~Jun, S. (2010).
\newblock Bayesian quantile regression methods.
\newblock {\em Journal of Applied Econometrics}, 25(2):287--307.

\bibitem[Li et~al., 2010]{li2010bayesian}
Li, Q., Lin, N., and Xi, R. (2010).
\newblock Bayesian regularized quantile regression.

\bibitem[Lindley, 1968]{lindley1968choice}
Lindley, D.~V. (1968).
\newblock The choice of variables in multiple regression.
\newblock {\em Journal of the Royal Statistical Society: Series B (Methodological)}, 30(1):31--53.

\bibitem[Makalic and Schmidt, 2015]{makalic2015simple}
Makalic, E. and Schmidt, D.~F. (2015).
\newblock A simple sampler for the horseshoe estimator.
\newblock {\em IEEE Signal Processing Letters}, 23(1):179--182.

\bibitem[McLatchie et~al., 2025]{mclatchie2025predictive}
McLatchie, Y., Fong, E., Frazier, D.~T., and Knoblauch, J. (2025).
\newblock Predictive performance of power posteriors.
\newblock {\em Biometrika}, page asaf034.

\bibitem[Park and Casella, 2008]{park2008bayesian}
Park, T. and Casella, G. (2008).
\newblock The bayesian lasso.
\newblock {\em Journal of the american statistical association}, 103(482):681--686.

\bibitem[Piironen and Vehtari, 2017a]{piironen_hyperprior_2017}
Piironen, J. and Vehtari, A. (2017a).
\newblock On the {Hyperprior} {Choice} for the {Global} {Shrinkage} {Parameter} in the {Horseshoe} {Prior}.
\newblock arXiv:1610.05559 [stat].

\bibitem[Piironen and Vehtari, 2017b]{piironen_sparsity_2017}
Piironen, J. and Vehtari, A. (2017b).
\newblock Sparsity information and regularization in the horseshoe and other shrinkage priors.
\newblock {\em Electronic Journal of Statistics}, 11(2).
\newblock arXiv:1707.01694 [stat].

\bibitem[Polson and Scott, 2010]{polson2010shrink}
Polson, N.~G. and Scott, J.~G. (2010).
\newblock Shrink globally, act locally: Sparse bayesian regularization and prediction.
\newblock {\em Bayesian statistics}, 9(501-538):105.

\bibitem[Polson and Scott, 2012]{polson_half-cauchy_2012}
Polson, N.~G. and Scott, J.~G. (2012).
\newblock On the {Half}-{Cauchy} {Prior} for a {Global} {Scale} {Parameter}.
\newblock {\em Bayesian Analysis}, 7(4).

\bibitem[Ray and Bhattacharya, 2018]{ray2018signal}
Ray, P. and Bhattacharya, A. (2018).
\newblock Signal adaptive variable selector for the horseshoe prior.
\newblock {\em arXiv preprint arXiv:1810.09004}.

\bibitem[Reich, 2012]{reich2012spatiotemporal}
Reich, B.~J. (2012).
\newblock Spatiotemporal quantile regression for detecting distributional changes in environmental processes.
\newblock {\em Journal of the Royal Statistical Society Series C: Applied Statistics}, 61(4):535--553.

\bibitem[Reich et~al., 2011]{reich2011bayesian}
Reich, B.~J., Fuentes, M., and Dunson, D.~B. (2011).
\newblock Bayesian spatial quantile regression.
\newblock {\em Journal of the American Statistical Association}, 106(493):6--20.

\bibitem[Reich and Smith, 2013]{reich2013bayesian}
Reich, B.~J. and Smith, L.~B. (2013).
\newblock Bayesian quantile regression for censored data.
\newblock {\em Biometrics}, 69(3):651--660.

\bibitem[Rodrigues and Fan, 2017]{rodrigues2017regression}
Rodrigues, T. and Fan, Y. (2017).
\newblock Regression adjustment for noncrossing bayesian quantile regression.
\newblock {\em Journal of Computational and Graphical Statistics}, 26(2):275--284.

\bibitem[Scaccia and Green, 2003]{scaccia2003bayesian}
Scaccia, L. and Green, P.~J. (2003).
\newblock Bayesian growth curves using normal mixtures with nonparametric weights.
\newblock {\em Journal of Computational and Graphical Statistics}, 12(2):308--331.

\bibitem[Simpson et~al., 2017]{simpson2017penalising}
Simpson, D., Rue, H., Riebler, A., Martins, T.~G., and S{\o}rbye, S.~H. (2017).
\newblock Penalising model component complexity: A principled, practical approach to constructing priors.

\bibitem[Sims, 1980]{sims1980macroeconomics}
Sims, C.~A. (1980).
\newblock Macroeconomics and reality.
\newblock {\em Econometrica: journal of the Econometric Society}, pages 1--48.

\bibitem[Sriram et~al., 2013]{sriram2013posterior}
Sriram, K., Ramamoorthi, R., and Ghosh, P. (2013).
\newblock Posterior consistency of bayesian quantile regression based on the misspecified asymmetric laplace density.

\bibitem[Szendrei et~al., 2024]{szendrei2023fused}
Szendrei, T., Bhattacharjee, A., and Schaffer, M.~E. (2024).
\newblock Fused {LASSO} as non-crossing quantile regression.
\newblock {\em arXiv preprint arXiv:2403.14036}.

\bibitem[Szendrei and Varga, 2023]{szendrei2023revisiting}
Szendrei, T. and Varga, K. (2023).
\newblock Revisiting vulnerable growth in the euro area: Identifying the role of financial conditions in the distribution.
\newblock {\em Economics Letters}, 223:110990.

\bibitem[Taddy and Kottas, 2010]{taddy2010bayesian}
Taddy, M.~A. and Kottas, A. (2010).
\newblock A bayesian nonparametric approach to inference for quantile regression.
\newblock {\em Journal of Business \& Economic Statistics}, 28(3):357--369.

\bibitem[Varga and Szendrei, 2025]{varga2025non}
Varga, K. and Szendrei, T. (2025).
\newblock Non-stationary financial risk factors and macroeconomic vulnerability for the uk.
\newblock {\em International Review of Financial Analysis}, 97:103866.

\bibitem[Vehtari et~al., 2017]{vehtari_practical_2017}
Vehtari, A., Gelman, A., and Gabry, J. (2017).
\newblock Practical {Bayesian} model evaluation using leave-one-out cross-validation and {WAIC}.
\newblock {\em Statistics and Computing}, 27(5):1413--1432.

\bibitem[Vehtari et~al., 2021]{vehtari2021rank}
Vehtari, A., Gelman, A., Simpson, D., Carpenter, B., and B{\"u}rkner, P.-C. (2021).
\newblock Rank-normalization, folding, and localization: An improved r-hat for assessing convergence of mcmc (with discussion).
\newblock {\em Bayesian analysis}, 16(2):667--718.

\bibitem[Wang and Cai, 2024]{wang2024composite}
Wang, Q. and Cai, Z. (2024).
\newblock A composite bayesian approach for quantile curve fitting with non-crossing constraints.
\newblock {\em Communications in Statistics-Theory and Methods}, 53(20):7119--7143.

\bibitem[Wei, 2008]{wei2008approach}
Wei, Y. (2008).
\newblock An approach to multivariate covariate-dependent quantile contours with application to bivariate conditional growth charts.
\newblock {\em Journal of the American Statistical Association}, 103(481):397--409.

\bibitem[Wu and Narisetty, 2021]{wu2021bayesian}
Wu, T. and Narisetty, N.~N. (2021).
\newblock Bayesian multiple quantile regression for linear models using a score likelihood.
\newblock {\em Bayesian Analysis}, 16(3):875--903.

\bibitem[Yang and He, 2012]{yang2012bayesian}
Yang, Y. and He, X. (2012).
\newblock Bayesian empirical likelihood for quantile regression.

\bibitem[Yang and He, 2015]{yang2015quantile}
Yang, Y. and He, X. (2015).
\newblock Quantile regression for spatially correlated data: An empirical likelihood approach.
\newblock {\em Statistica Sinica}, pages 261--274.

\bibitem[Yang and Tokdar, 2017]{yang2017joint}
Yang, Y. and Tokdar, S.~T. (2017).
\newblock Joint estimation of quantile planes over arbitrary predictor spaces.
\newblock {\em Journal of the American Statistical Association}, 112(519):1107--1120.

\bibitem[Yu and Moyeed, 2001]{yu2001bayesian}
Yu, K. and Moyeed, R.~A. (2001).
\newblock Bayesian quantile regression.
\newblock {\em Statistics \& Probability Letters}, 54(4):437--447.

\bibitem[Zou, 2006]{zou2006adaptive}
Zou, H. (2006).
\newblock The adaptive lasso and its oracle properties.
\newblock {\em Journal of the American statistical association}, 101(476):1418--1429.

\bibitem[Zou and Yuan, 2008]{zou2008composite}
Zou, H. and Yuan, M. (2008).
\newblock Composite quantile regression and the oracle model selection theory.

\end{thebibliography}
\onecolumn
\begin{appendices}
\section{Outline of the Appendix}
In the appendix, we list the posteriors and remaining sampling steps for of the $\MCMC$ sampling algorithm for the $\QVP$ and $\NCQVP$ models discussed in the paper in Section~\ref{app:qvp-posteriors}. We additionally give sampling steps for a MCMC-sampler in Section~\ref{app:asis} which interweaves the posterior sampling steps for the centred and non-centred parameterisation in the sense of \citet{bitto2019achieving}. In Section~\ref{sec:shrinkage-properties}, we give further details on the derivation of the implicit prior on the shrinkage coefficient, introduced in Section~\ref{sec:theoretical-properties}. In Section~\ref{app:note} we provide some discussion on how the prior can be tuned to reach an appropriate level of prior probability of crossing quantile functions. In Section~\ref{app:extra-sim-results}, we give extra results on the simulation exercise of the paper, Section~\ref{sec:simulation}. In Section~\ref{app:extra-results-application}, we give further results regarding the empirical application of the paper. In Section~\ref{app:stress-testing}, we show how the $\QVAR$ model in combination with the priors presented can be used to conduct stress-test scenarios, relevant for risk management of central banks \citep{chavleishvili2023measuring}.

\section{Posteriors}\label{app:qvp-posteriors}
\subsection{Centred Model: $\QVP$}
Here we give further details to the posterior derivations presented in Definition~\ref{def:qvp_centred}.
\subsubsection{$\alpha$}
The conditional posterior for $\boldsymbol{\alpha}$ is normal:
\begin{equation}
    \boldsymbol{\alpha}\vert \boldsymbol{Y},\vartheta \propto \mvn(\overline{\boldsymbol{\alpha}}, K^{-1}_{\boldsymbol{\alpha}}),
\end{equation}
where $K_{\alpha} = \boldsymbol{X}_{\alpha}^T\boldsymbol{\Omega}^{-1}\boldsymbol{X}_{\alpha}$ and $\overline{\boldsymbol{\alpha}} = \boldsymbol{X}_{\alpha}^T\boldsymbol{\Omega}^{-1}(\boldsymbol{Y} - \boldsymbol{\mu} - \boldsymbol{X}\boldsymbol{\beta})$. Define $\boldsymbol{X}_{\alpha} \in \mathbbm{R}^{\mathcal{T}\mathcal{Q} \times \mathcal{Q} }$ as a matrix where the first rows of the first column of 1, the $\mathcal{T}+1$ until  $2\mathcal{T}$ are 1 for the second column, and so on, while all other entries are 0.
\subsubsection{$\omega$}
The conditional posterior for $\omega_{q,t}$ is distributed according to a generalised inverse Gaussian, $\GIG{a,b,c}$, where $a \in \mathbbm{R}$, $b>0$ and $c>0$:
\begin{equation}
    \omega_{q,t} \vert \boldsymbol{Y}, \vartheta \propto \GIG{1/2,2/\sigma^y_q + \theta_q/(\zeta^2_q \sigma^y_q),(y_t - \alpha_q - x_t^T\beta_q)/(\sigma^y_q\zeta^2_q)},
\end{equation}
for which we use the sampling algorithm from \citet{devroye2014random}.
\subsubsection{$\sigma^y_q$}
The conditional posterior for $\sigma^y_q$ is distributed according to a standard inverse Gamma, $\IG{a,b}$, with shape $a$ and scale $b$. We set the prior with relatively weakly informative priors $\sigma^y_q \sim \IG{\underline{a}=0.1,\underline{b}=0.1}$:
\begin{equation}
    \IG{\underline{a} + 3\mathcal{T}/2, \underline{b} + \sum_{t=1}^{\mathcal{T}} \omega_{q,t} + \sum_{t=1}^{\mathcal{T}} (y_t - \alpha_q - \mu_{q,t} - x^{T}_t \beta_q)/(2\zeta^2_q\omega_{q,t})}
\end{equation}
\subsubsection{$\beta_0$}
The conditional posterior for $\beta_0$ is normal, where the likelihood contribution is given by $p(\beta_1\vert\beta_0,\Sigma_1) = \mvn(\beta_0,\Sigma_1)$, and the prior $p(\beta_0|0,\Sigma_0) = \mvn(0,\Sigma_0)$:
\begin{equation}
    \beta_0 \vert \boldsymbol{Y}, \vartheta \propto \mvn(\overline{\beta}_0,K^{-1}_{\beta_0}),
\end{equation}
where $K = \Sigma_1^{-1} + \Sigma^{-1}_0$ and $\overline{\beta_0} = K^{-1}_{\beta_0}\left( \Sigma^{-1}_1\beta_1 \right) $
\subsubsection{$\{\nu,\lambda\}$}
The posteriors for $\nu_q$ as well as $\lambda_{q,j}$ for $q=1,\dotsc,\mathcal{Q}$ and $j = 1,\dotsc,K$ are also Cauchy distributions. In this paper, we make use of the mixture of inverse Gammas representation of the Cauchy. Namely, if $x^2\vert a \sim \IG{1/2,1/a}$ and $a \sim \IG{1/2,1/A^2}$, then $x \sim C_+(0,A)$,  following \citet{makalic2015simple}. Therefore
\begin{equation}
\begin{split}
    \nu^2_q \sim \IG{1/2,1/\xi^{\nu_q}_{q}},\; & \xi^{\nu_q}_{q} \sim \IG{1/2,T}, \; \mathrm{for} \; q = 1,\dotsc,\mathcal{Q} \\
    \lambda^{2}_{q,j} \sim \IG{1/2,1/\xi_{q,j}^{\lambda_{q,j}}},\; &  \xi_{q,j}^{\lambda_{q,j}} \sim \IG{1/2,1} \mathrm{for} \; j = 1,\dotsc,K,
 \end{split}
\end{equation}
and
\begin{equation}
    \nu^2_0 \sim \IG{1/2,1/\xi^{\nu_0}_{0}},\; \xi^{\nu_0}_{0} \sim \IG{1/2,T\mathcal{Q}}.
\end{equation}
Then, the posteriors are also all distributed according to a $\IG{}$ distribution: 
\begin{equation}
\begin{split}
    \nu^2_q \vert \boldsymbol{Y},\cdot & \sim \IG{(K+1)/2,1/\xi^{\nu_q}_{q} + \sum_{j=1}^K (\beta_{q,j} - \beta_{q-1,j})^2/(2 \lambda^2_{q,j}) } \\
    \xi^{\nu_q}_{q} \vert \boldsymbol{Y},\cdot & \sim \IG{1,\mathcal{T} + 1/\nu^2_q} \\ 
    \lambda^{2}_{q,j} \vert \boldsymbol{Y},\cdot & \sim \IG{1,1/\xi_{q,j}^{\lambda_{q,j}} + (\beta_{q,j} - \beta_{q-1,j})^2/\nu^2_q} \\
    \xi_{q,j}^{\lambda_{q,j}} \vert \boldsymbol{Y}, \cdot & \sim \IG{1, 1 + 1/\lambda^2_{q,j}},
 \end{split}
\end{equation}
for $j=1,\dotsc,K$ and $q = 1,\dotsc,\mathcal{Q}$. The posterior for the vector $\beta_0$ is found analogously. 
\subsection{Non-centred Model: $\NCQVP$}
Here we give further details to the posterior derivations presented in Section~\ref{sec:alt-parameterisation}.
\subsubsection{$\beta_0$}
The conditional posterior for $\beta_0$ is normal:
\begin{equation}
    \beta_0\vert\boldsymbol{Y},\vartheta \propto \mvn(\overline{\beta}_0,K^{-1}_{\beta_0}),
\end{equation}
where $K_{\beta_0} = (\sum_{q=1}^{\mathcal{Q}}X^T\boldsymbol{\Omega}_qX) + \Sigma_0^{-1}$ and $\overline{\beta}_0  = K^{-1}_{\beta_0}(\sum_{q=1}^{\mathcal{Q}}X^T\Omega_q^{-1}(y - \alpha_q - \mu_q - \tilde{X}_q\sigma_q))$. Denote by $\tilde{y}_q = y - \alpha_q - \mu_q - \tilde{X}_q\sigma_q.$
That is the following structure emerges from conditionally updating one quantile at a time:
\begin{equation}\label{eq:second-moment-beta0-NC}
  K_{\beta_0,q} =
  \begin{cases}
    \left( X^T\Omega_q^{-1}X + \Sigma_0^{-1} \right) & \text{for $q = 1$ } \\
    \left( X^T\Omega_q^{-1}X + K_{\beta_0,q-1}  \right) & \text{for $q = 2,\dotsc,\mathcal{Q}$}
  \end{cases}
\end{equation}
and 
\begin{equation}\label{eq:first-moment-beta0-NC}
  \overline{\beta}_{0,q} =
  \begin{cases}
    K^{-1}_{\beta_0,q} \left( X^T\Omega_q^{-1}\tilde{y}_q \right) & \text{for $q = 1$ } \\
    K^{-1}_{\beta,q} \left( X^T\Omega_q^{-1}\tilde{y}_q + K_{\beta_0,q-1}\overline{\beta}_{0,q-1} \right) & \text{for $q = 2,\dotsc,\mathcal{Q}$},
  \end{cases}
\end{equation}
which by substituting in Equation~\ref{eq:second-moment-beta0-NC} into Equation~\ref{eq:first-moment-beta0-NC} allows representation of the posterior updated with all quantiles as in Equation~\ref{eq:NC-beta0-posterior}.

The rest of the sampling steps are standard and equivalent to the exposition of the $\QVP$ model, see Section~\ref{app:qvp-posteriors}.

\subsection{Interweaving QVP and NC-QVP for Sampling Efficiency}\label{app:asis}
Instead of using the $\NCQVP$ or the $\QVP$ prior, one may combine both by interweaving the sampling step of the non-centred model with that of the centred model. This is akin to the methods used for state-space modelling, presented in \citet{bitto2019achieving}. 
The updated sampling algorithm looks as follows:
\begin{enumerate}
    \item ${{\beta}_0}^{(s)} \sim ({\beta}_0\mid \boldsymbol{y},\boldsymbol{\tilde{\beta}}^{(s-1)},\boldsymbol{\sigma}^{(s-1)},\boldsymbol{\tilde{\Sigma}}^{(s-1)},\Sigma_0^{(s-1)},\boldsymbol{\alpha}^{(s-1)},\boldsymbol{\mu}^{(s-1)},\Omega^{(s-1)} )$, 
    \item ${\boldsymbol{\tilde{\beta}}}^{(s)} \sim ({\boldsymbol{\tilde{\beta}}}\mid \boldsymbol{y},\boldsymbol{\sigma}^{(s-1)},\boldsymbol{\tilde{\Sigma}}^{(s-1)},\boldsymbol{{\beta}}_0^{(s)},\Sigma_0^{(s)},\boldsymbol{\alpha}^{(s-1)},\boldsymbol{\mu}^{(s-1)},\Omega^{(s-1)} )$, 
    \item $\boldsymbol{\sigma}^{(s)} \sim (\boldsymbol{\sigma} | \boldsymbol{y},\boldsymbol{\tilde{\beta}}^{(s)},\boldsymbol{\tilde{\Sigma}}^{(s-1)},\boldsymbol{{\beta}}_0^{(s)},\Sigma_0^{(s)},\boldsymbol{\alpha}^{(s-1)},\boldsymbol{\mu}^{(s-1)},\Omega^{(s-1)} )$, 
    \item Interweaving Steps
    \begin{enumerate}
        \item Transform to centred parameterisation $\boldsymbol{\beta}^{(s)} = \mathbbm{1}_{Q}  \otimes \beta_0^{(s)} + \boldsymbol{\tilde{\beta}}^{(s)}\text{diag}(\boldsymbol{\sigma}^{(s)})$ and save sign of $\boldsymbol{\sigma}^{(s)}$
        \item ${\boldsymbol{\sigma^2}}^{(s)} \sim (\boldsymbol{\sigma^2} | \boldsymbol{y},\boldsymbol{\beta}^{(s)},\boldsymbol{\tilde{\Sigma}}^{(s-1)},{{\beta}}_0^{(s)},\Sigma_0^{(s)},\boldsymbol{\alpha}^{(s-1)},\boldsymbol{\mu}^{(s-1)},\Omega^{(s-1)} )$
        \item ${\boldsymbol{\beta}_0}^{C,(s)} \sim (\boldsymbol{\beta}_0^{C}\mid \boldsymbol{y},\boldsymbol{{\beta}}^{(s)},{\boldsymbol{\sigma^2}}^{(s)},\boldsymbol{\tilde{\Sigma}}^{(s-1)},\Sigma_0^{(s-1)},\boldsymbol{\alpha}^{(s-1)},\boldsymbol{\mu}^{(s-1)},\Omega^{(s-1)} )$ and set $\beta_0^{(s)} = {{\beta}_0}^{C,(s)}$,  
        \item Set $\boldsymbol{\sigma}^{(s)} = \sqrt{{\boldsymbol{\sigma^2}}^{(s)}}$ and adjust for the saved sign in (a). Set $\tilde{\beta}_{q,j}^{(s)} = \frac{\beta_{q,j}^{(s)}-\beta_{0,j}^{C,(s)}}{{\sigma_{q,j}}^{C,(s)}}$
    \end{enumerate}
    \item $\Sigma^{(s)}_0 \sim (\Sigma_0\mid \boldsymbol{y},\boldsymbol{\tilde{\beta}}^{(s-1)},\boldsymbol{\sigma}^{(s-1)},\boldsymbol{\tilde{\Sigma}}^{(s-1)},\boldsymbol{\beta}_0^{(s)},\boldsymbol{\alpha}^{(s-1)},\boldsymbol{\mu}^{(s-1)}, \Omega^{(s-1)})$, 
    \item $\boldsymbol{\tilde{\Sigma}}^{(s)} \sim (\boldsymbol{\tilde{\Sigma}}\mid \boldsymbol{y},\boldsymbol{\tilde{\beta}}^{(s)},\boldsymbol{\sigma}^{(s)},\boldsymbol{{\beta}}_0^{(s)},\Sigma_0^{(s)},\boldsymbol{\alpha}^{(s-1)},\boldsymbol{\mu}^{(s-1)}, \Omega^{(s-1)})$, 
    \item $\boldsymbol{\alpha}^{(s)} \sim (\boldsymbol{\alpha} \mid \boldsymbol{y},\boldsymbol{\tilde{\beta}}^{(s)},\boldsymbol{\sigma}^{(s)},\boldsymbol{\tilde{\Sigma}}^{(s)},\boldsymbol{{\beta}}_0^{(s)},\Sigma_0^{(s)},\boldsymbol{\mu}^{(s-1)},\Omega^{(s-1)} )$, 
    \item $\boldsymbol{\mu}^{(s)} \sim (\boldsymbol{\mu} \mid \boldsymbol{y},\boldsymbol{\tilde{\beta}}^{(s)},\boldsymbol{\sigma}^{(s)},\boldsymbol{\tilde{\Sigma}}^{(s)},\boldsymbol{{\beta}}_0^{(s)},\Sigma_0^{(s)},\boldsymbol{\alpha}^{(s),},\Omega^{(s-1)})$, 
    \item $\Omega^{(s)} \sim (\Omega\mid \boldsymbol{y},\boldsymbol{\tilde{\beta}}^{(s)},\boldsymbol{\sigma}^{(s)},\boldsymbol{\tilde{\Sigma}}^{(s)},\boldsymbol{{\beta}}_0^{(s)},\Sigma_0^{(s)},\boldsymbol{\alpha}^{(s)},\boldsymbol{\mu}^{(s)} )$,
\end{enumerate}
for $s = (1,\dotsc,S)$ until convergence.
Steps 1.-3. and 5.-9. remain the same to the non-centred sampling algorithm, so we focus on steps 4.(b) to 4.(c) here. 
By standard transformation of variables, $x|\zeta \sim \normal(0,\zeta^2) \implies x^|\zeta \sim \text{G}(\frac{1}{2},\frac{1}{2\zeta^2})$, where 
$\text{G}()$ stands for the Gamma distribution. Hence, $\sigma^2_{q,j}|\nu^2\lambda^2_{q,j} \sim \text{G}(\frac{1}{2},\frac{1}{2\nu^2\lambda_{q,j}^2})$ for $q = 1,\dotsc,Q$ and $j = 1,\dotsc,K$. The relevant conditional likelihood and prior contributions for posterior inference are:
\begin{equation}
    p(\sigma^2_{q,j}| \boldsymbol{y},\theta) \propto p(\sigma^2_{q,j}|\nu^2\lambda^2_{q,j})\times p(\beta_{q,j}|\beta_{q,j},\sigma^2_{q,j},\theta)
\end{equation}
which can be shown to be proportional to a Generalised-inverse-Gaussian distribution (GiG)\footnote{$\text{GiG}(a,b,c)$ stands for the generalised inverse Gaussian distribution with $a\in \mathbbm{R}, b>0,c>0$}, $\sigma^2_{q,j}| \boldsymbol{y},\theta \sim \text{GiG}(-\frac{1}{2},\frac{1}{\nu^2\lambda_{q,j}^2},(\beta_{q,j}-\beta_{q-1,j})^2)$ which we can be sample efficiently for each $q$ and $j$ \footnote{We make use of the algorithm by \citet{hormann2014generating}}.
In order to sample for om the posterior in  4.(c), notice that the relevant conditional and prior likelihood contributions are yield a normal posterior distribution for $\beta_0$:
\begin{equation}
    p(\beta_0|\boldsymbol{y},\theta) \propto p(\beta_1|\beta_0,\Sigma_1) \times p(\beta_0 | \Sigma_0) = \mvn(\overline{\beta}_0,K^{-1}_0),
\end{equation}
where $K_0 = (\Sigma^{-1}_1 + \Sigma^{-1}_0)$ and $\overline{\beta}_0 = K^{-1}_0(\Sigma^{-1}_1\beta_1)$.
\citet{bitto2019achieving} discuss the possibility that sampling efficiency may also depend on whether one first samples from the non-centred to move to the centred distribution, or vice versa, however, we have found from our experiments that performance is similar regardless. 
%

\section{Further Derivations on Shrinkage properties}\label{sec:shrinkage-properties}
From Definition~\ref{eq:def-1}), $\kappa_{t,j} = \frac{1}{1 + \nu^2\lambda_{t,j}^2 \sigma^{-2}_y}$. Now apply change of variables $\lambda_{t,j} \mapsto \kappa_{t,j}$, where the probability density function for $\lambda_{t,j}$ is $\frac{1}{\pi} \frac{1}{1 + \lambda_{t,j}^2}$ for $\lambda_{t,j} > 0$: 

\begin{equation} \label{kappa_dist_raw}
    \frac{1}{a} \times \frac{\frac{1}{\kappa_{t,j}} + \frac{1- \kappa_{t,j}}{\kappa_{t,j}} }{\sqrt{\frac{1- \kappa_{t,j}}{\kappa_{t,j}}}} \times \frac{1}{\pi} \frac{1}{1 + \sqrt{\frac{1- \kappa_{t,j}}{\kappa_{t,j}}} \frac{1}{a^2}},
\end{equation}
where $a = \nu \sigma^{-1}_y$. Then Equation (\ref{kappa_dist_raw}) simplifies to that in Definition~\ref{eq:def-1}. \qed{}

\section{A Note on Non-crossing}\label{app:note}
Although the quantile differences are penalised, it is not necessarily guaranteed that the methodology outlined above will lead to exact non-crossing. Since the coefficient process is centred on the quantile invariant vector $\beta_0$ and the priors shrink $\epsilon_q^{\beta}$ toward zero, the priors will shrink \textit{towards} parallel quantile curves. As such, the framework provides continuous approximation to the non-crossing constraint of \citet{bondell2010noncrossing}. 

Our approach to constrained parameter spaces follows the view taken by \citet{simpson2017penalising} who advocate that strict constraints are often ill supported by the data, assuming that the model is only an approximation of the true data-generating process. Theoretical justification for continuous penalisation toward the desired space can be found in \citet{duan2020bayesian}. Nevertheless, it is always possible to construct an importance sampling correction to the posteriors to enforce exact non-crossing.\footnote{This can be achieved by simply discarding $\mathrm{MCMC}$ draws that lead to crossing quantile functions.}
It is however, instructive to further investigate the implied prior probability that the constraint for non-crossing is adhered to. We follow the notation of \citet{bondell2010noncrossing}. Namely we denote by $\gamma_{0,q}=\alpha_q-\alpha_{q-1}$ the vector of differences of the intercept, and by $\gamma_{j,q}=\beta_{j,q}-\beta_{j,q-1}$ the difference of the $j$\textsuperscript{th} covariate. Then, under the transformation of $x_t \in [-1,1]^K$,  the non-crossing constraints in Equation~\ref{eq:NC-constraint-facevalue} becomes:
\begin{equation}
    \gamma_{0,q} - \sum_{j=1}^K \gamma_{j,q} \geq 0,~\forall q=2\dots,Q.
\end{equation}
%
Then under the QVP prior, conditional on hyper-parameters, $(\sum_{q=1}^Q\sum_{j=1}^K \epsilon_{q,j}^{\beta}) \sim \normal(0,\Phi)$, where $\Phi$ is some covariance matrix. Hence, in order to investigate the prior tendencies of the model to lead to crossing, one can calibrate the prior to high probability of $p((\sum_{q=2}^{\mathcal{Q}}\alpha_q-\sum_{q=2}^Q\sum_{j=1}^K \epsilon_{q,j}^{\beta}) = \normal(0,\Phi) \geq 0)$ to be of a reasonable level.


\section{Simulation Experiments}\label{app:extra-sim-results}

\begin{figure}[H]
    \centering
    \includegraphics[width=\linewidth]{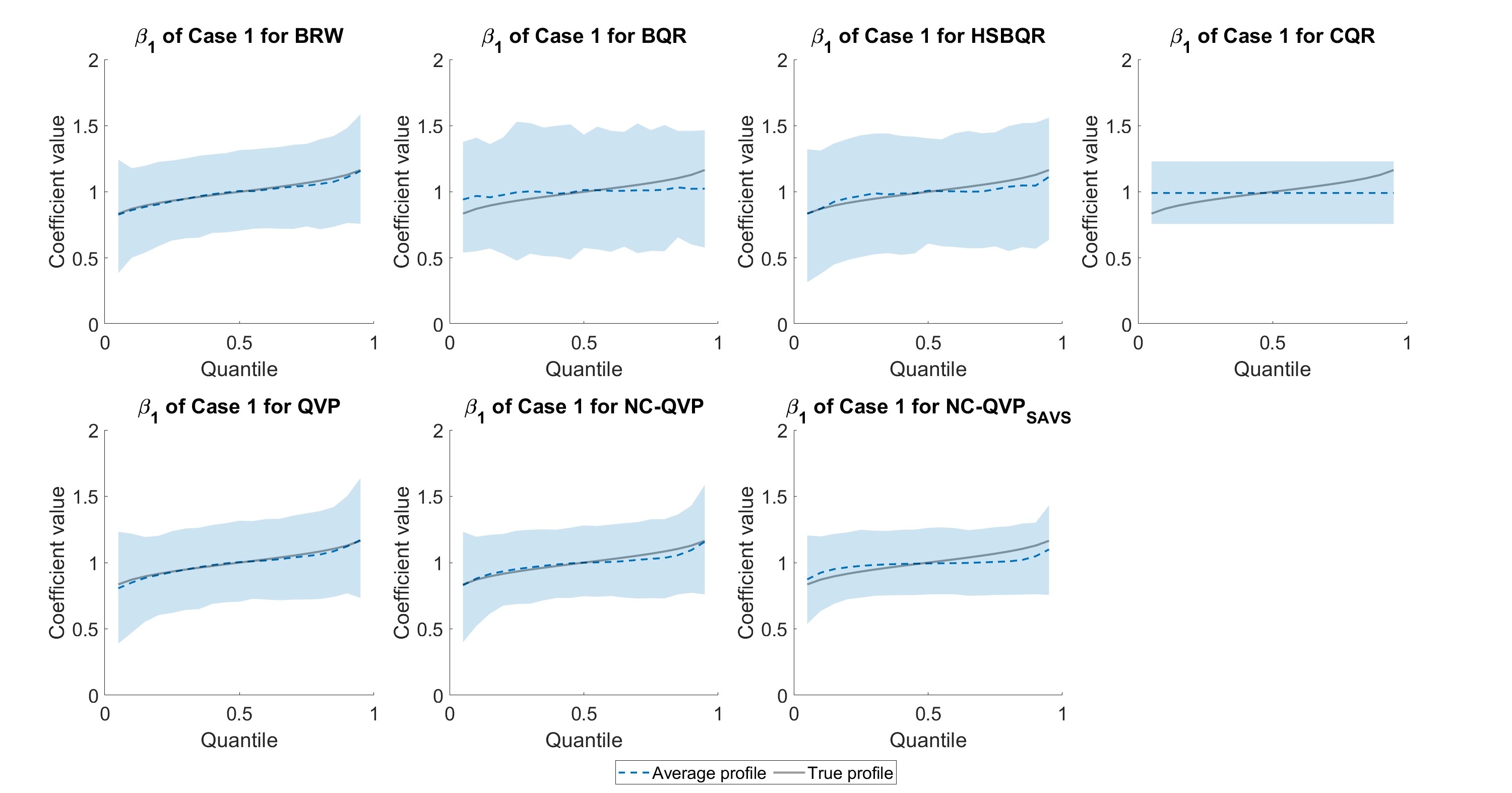}
    \caption{Quantile varying variable for Case=1, T=300, rho=0.0}
    \label{fig:DGP1}
\end{figure}

\begin{figure}[H]
    \centering
    \includegraphics[width=\linewidth]{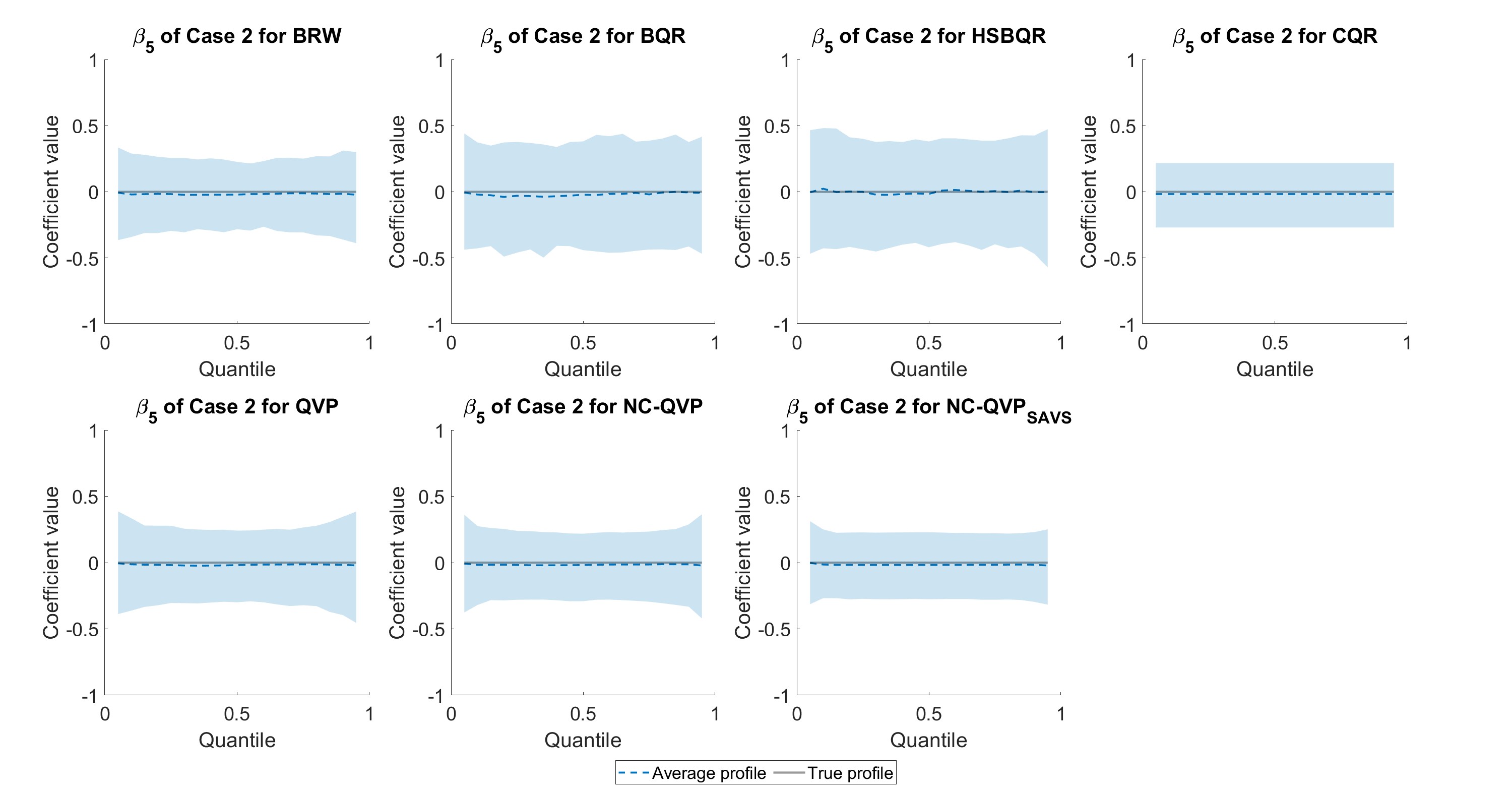}
    \caption{Quantile constant variable for Case=2, T=300, rho=0.0}
    \label{fig:DGP2}
\end{figure}

\begin{figure}[H]
    \centering
    \includegraphics[width=0.9\linewidth]{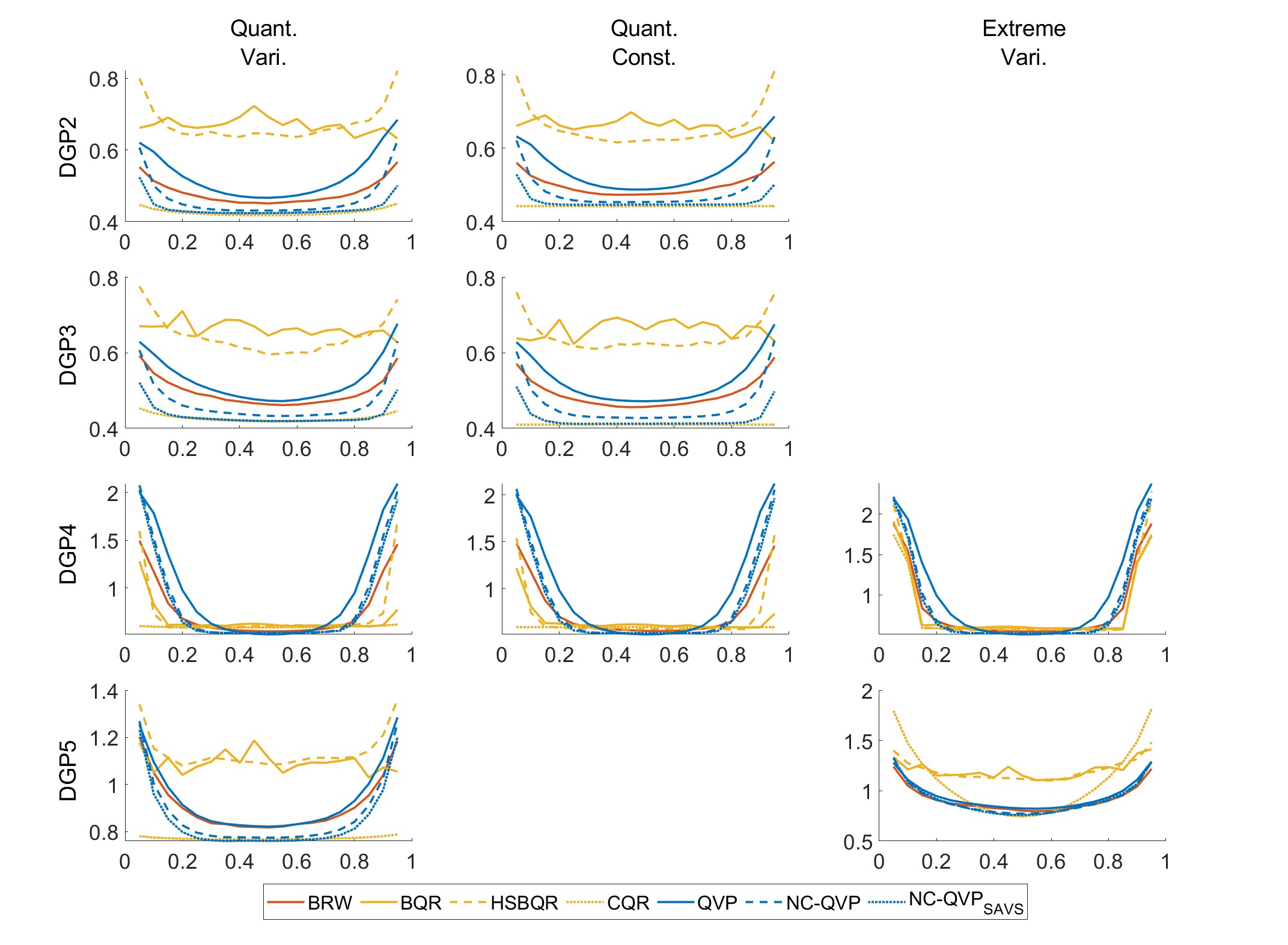}
    \caption{$\rmse$ profile of $\boldsymbol{\beta}$ across 19 quantiles $\tau_q \in \left\{0.05,\dotsc,0.95\right\}$ for $\mathcal{T}=100$ and $\varDelta$=0. Estimates are based on the posterior mean.}
    \label{fig:SpecCoeffBias_T100}
\end{figure}

\begin{figure}[H]
    \centering
    \includegraphics[width=0.9\linewidth]{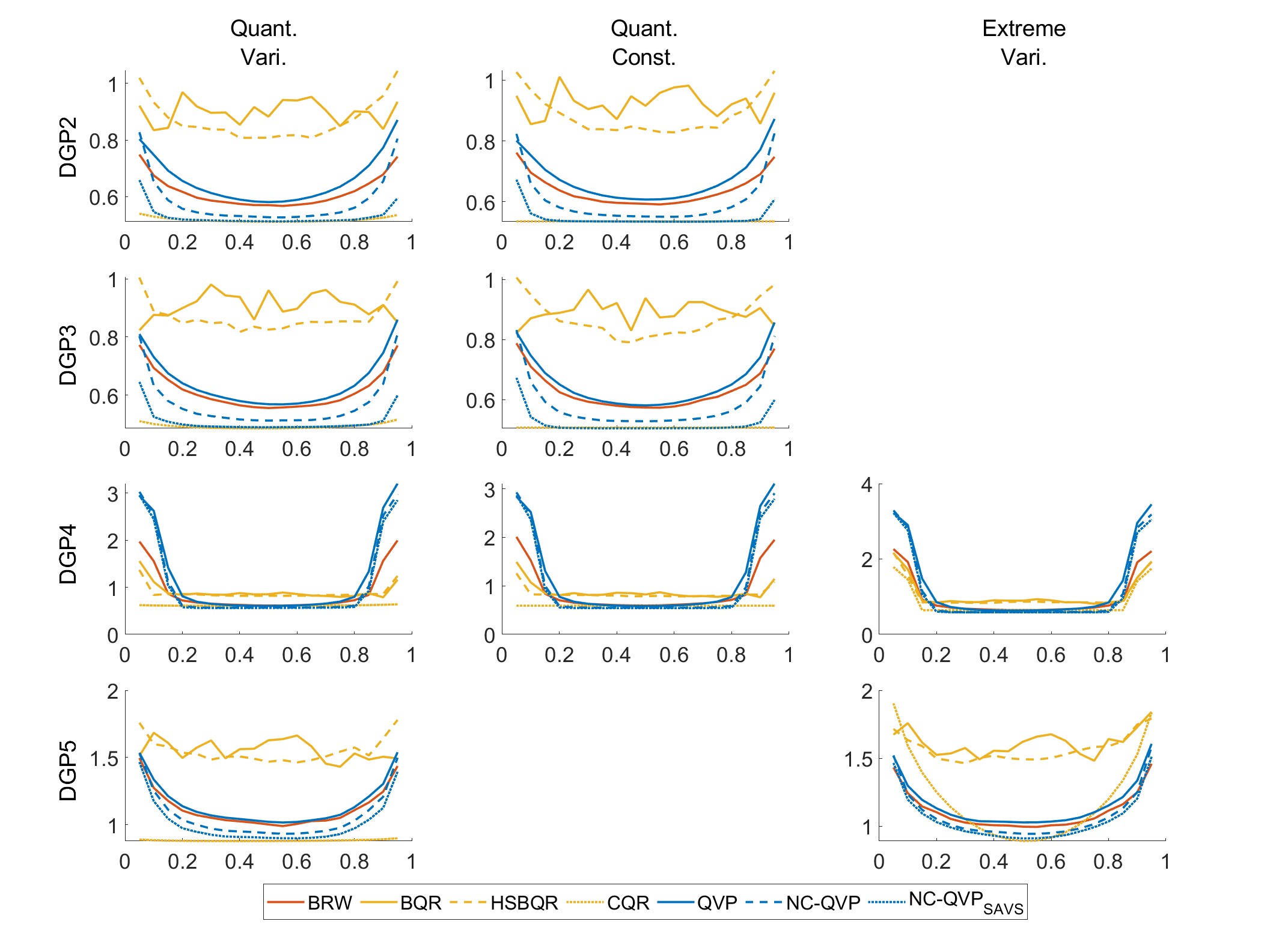}
    \caption{$\rmse$ profile of $\boldsymbol{\beta}$ across 19 quantiles $\tau_q \in \left\{0.05,\dotsc,0.95\right\}$ for $\mathcal{T}=300$ and $\varDelta$=0.5. Estimates are based on the posterior mean.}
    \label{fig:SpecCoeffBias_rho05}
\end{figure}

\begin{figure}[H]
    \centering
    \includegraphics[width=0.9\linewidth]{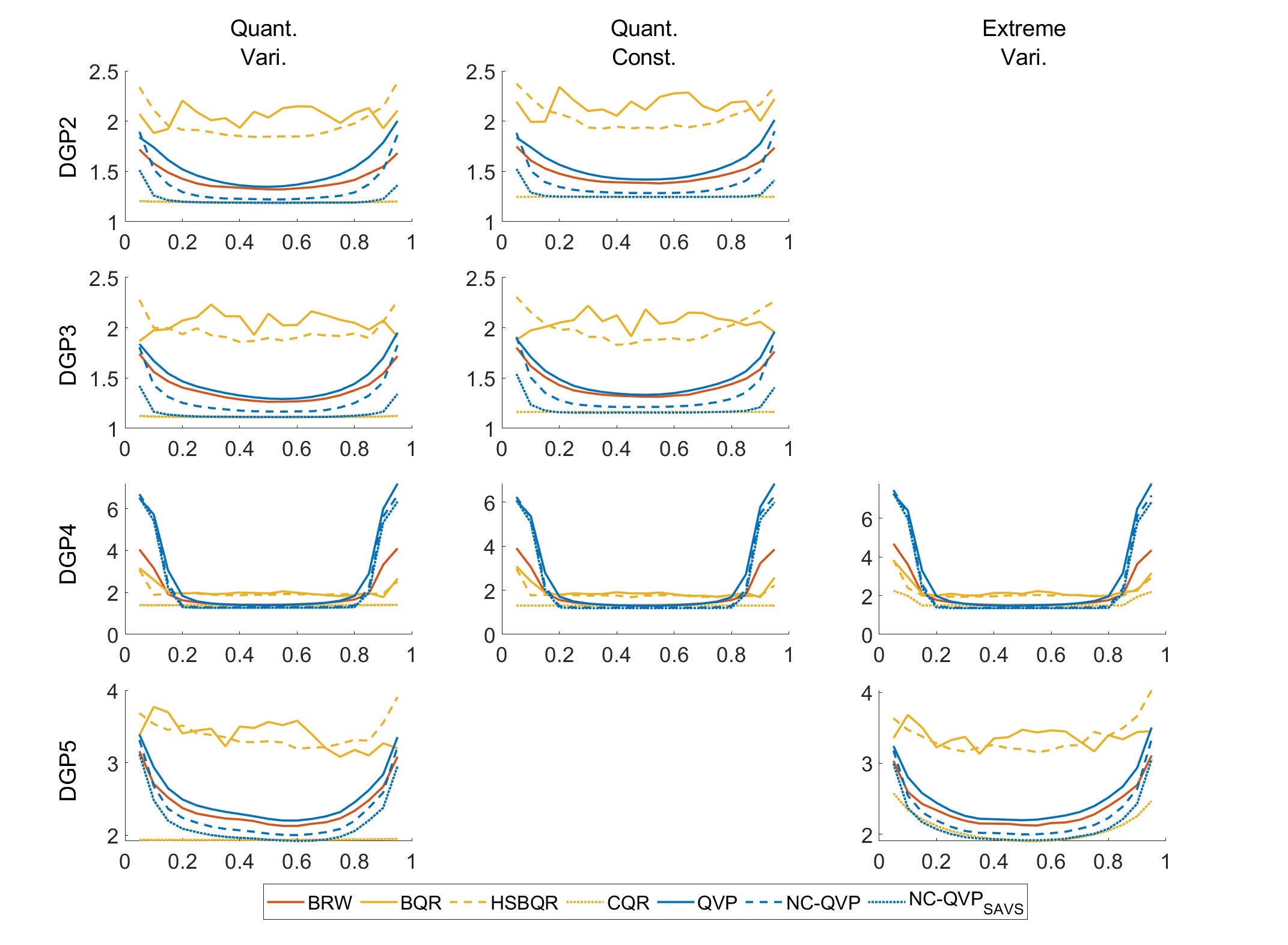}
    \caption{$\rmse$ profile of $\boldsymbol{\beta}$ across 19 quantiles $\tau_q \in \left\{0.05,\dotsc,0.95\right\}$ for $\mathcal{T}=300$ and $\varDelta$=0.9. Estimates are based on the posterior mean.}
    \label{fig:SpecCoeffBias_rho09}
\end{figure}

\begin{figure}[H]
    \centering
    \includegraphics[width=0.9\linewidth]{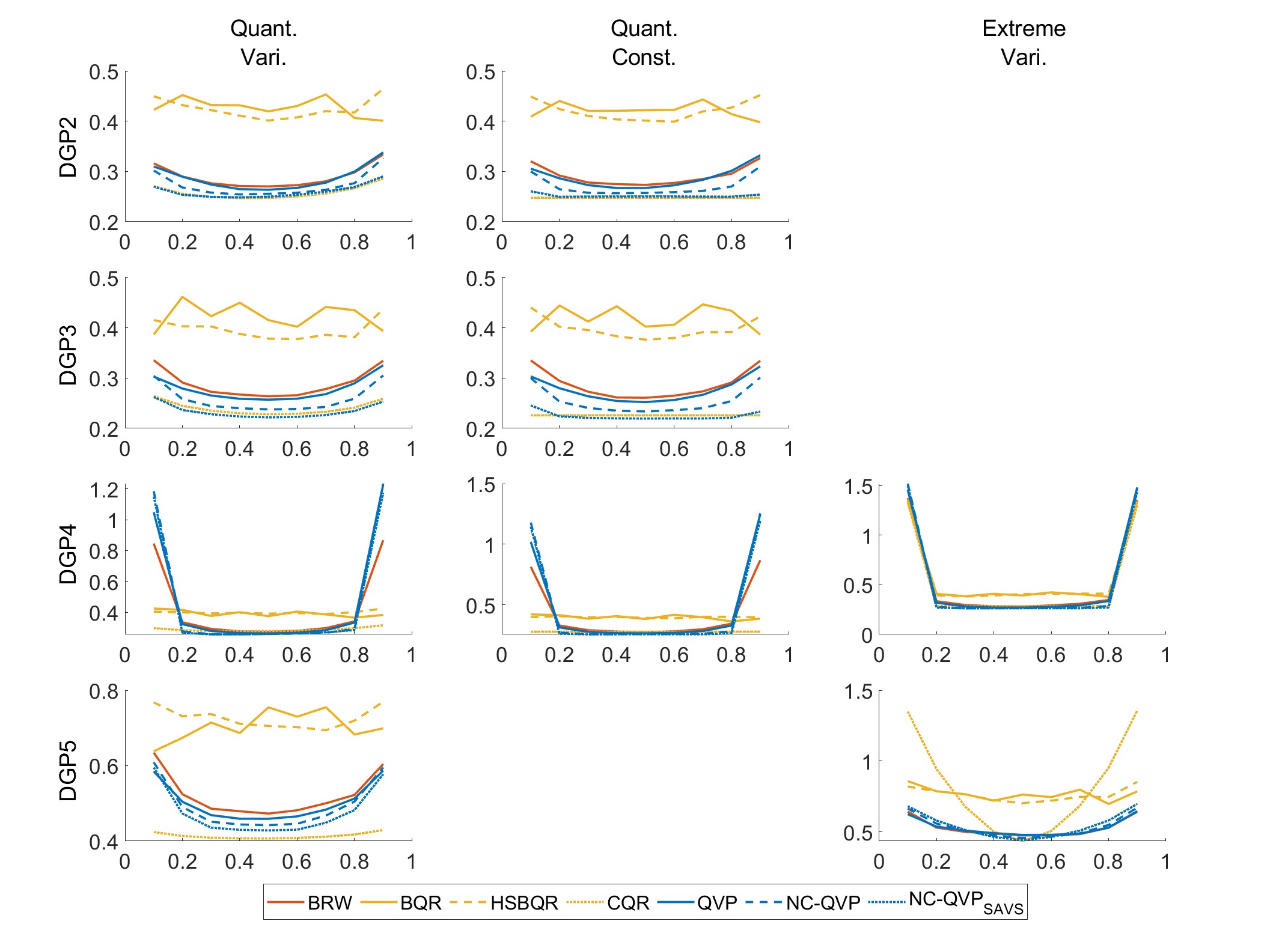}
    \caption{$\rmse$ profile of $\boldsymbol{\beta}$ across 9 quantiles $\tau_q \in \left\{0.1,\dotsc,0.9\right\}$ for $\mathcal{T}=300$ and $\varDelta$=0. Estimates are based on the posterior mean.}
    \label{fig:SpecCoeffBias_Q9}
\end{figure}

\begin{figure}[H]
    \centering
    \includegraphics[width=0.9\linewidth]{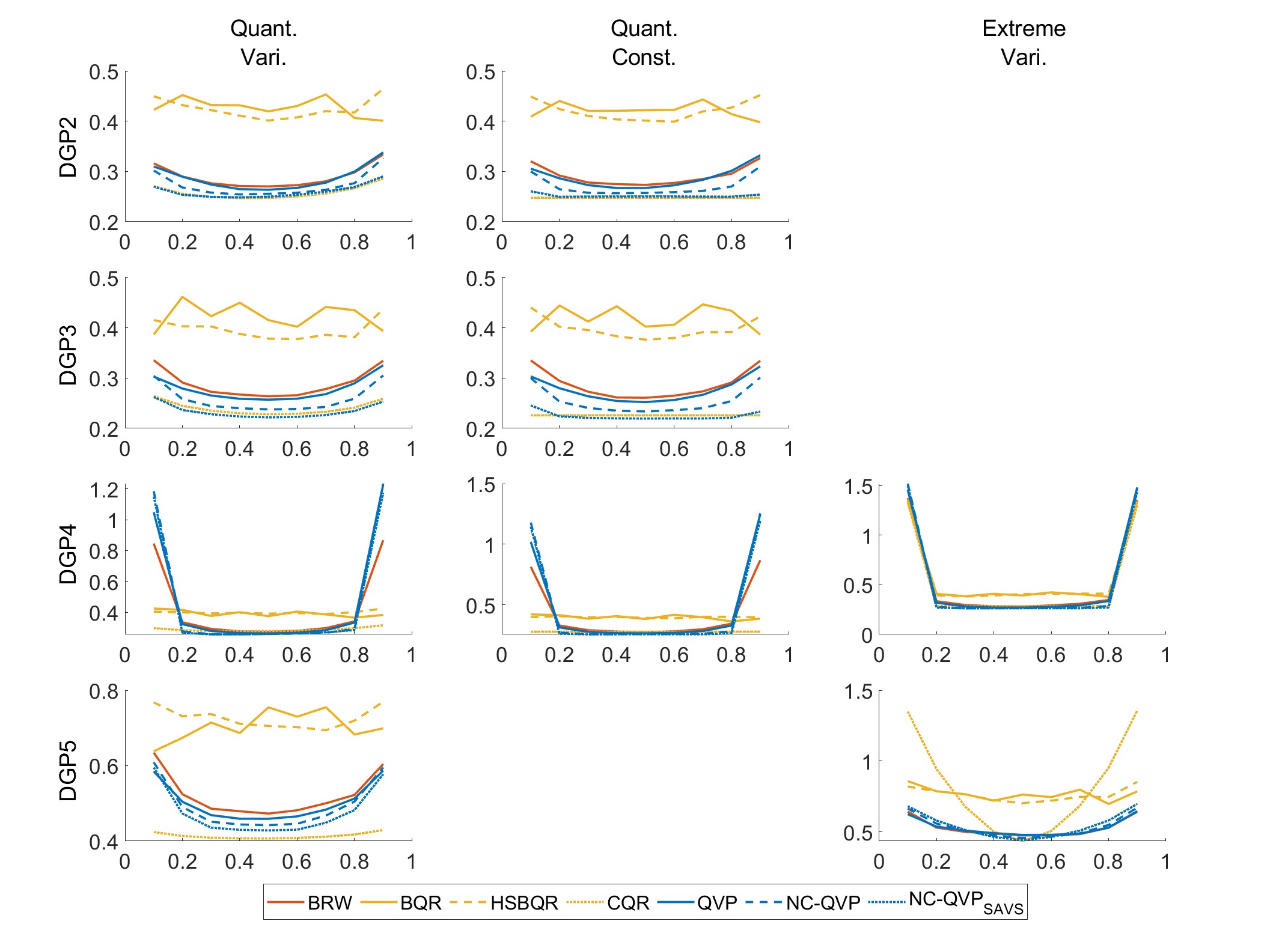}
    \caption{$\rmse$ profile of $\boldsymbol{\beta}$ across 39 quantiles $\tau_q \in \left\{0.025,\dotsc,0.975\right\}$ for $\mathcal{T}=300$ and $\varDelta$=0. Estimates are based on the posterior mean.}
    \label{fig:SpecCoeffBias_Q39}
\end{figure}

\subsection{Predictive Results}\label{app:sims-predictive-results}
\clearpage

\begin{table}[H]
\centering
\resizebox{0.9\textwidth}{!}{%
\begin{tabular}{ll|cccc|cccc}
 &  & \multicolumn{4}{c|}{$\varDelta$=0.5} & \multicolumn{4}{c}{$\varDelta$=0.9} \\
 &  & $w_1$ & $w_2$ & $w_3$ & $w_4$ & $w_1$ & $w_2$ & $w_3$ & $w_4$ \\ \hline
\multicolumn{2}{l|}{$\mathrm{DGP}$-1} &  &  &  &  &  &  \\
 & $\BQR$   & 0.421 & 0.078 & 0.132 & 0.133 & 0.420 & 0.078 & 0.132 & 0.133 \\ \hdashline
 & $\HSBQR$ & 0.911 & 0.950 & 0.892 & 0.883 & 0.912 & 0.949 & 0.892 & 0.881 \\
 & $\BRW$   & 0.855 & 0.906 & 0.828 & 0.821 & 0.856 & 0.905 & 0.827 & 0.82 \\
 & $\CQR$   & 0.852 & 0.904 & 0.824 & 0.818 & 0.853 & 0.903 & 0.823 & 0.817 \\
 & $\QVP$   & 0.854 & 0.906 & 0.828 & 0.820 & 0.856 & 0.905 & 0.827 & 0.819 \\
 & $\NCQVP$ & 0.853 & 0.905 & 0.826 & 0.819 & 0.854 & 0.904 & 0.825 & 0.818 \\
 & $\NCQVP_{\mathrm{SAVS}}$ & 0.852 & 0.904 & 0.825 & 0.818 & 0.853 & 0.903 & 0.823 & 0.817 \\ \hline
\multicolumn{2}{l|}{$\mathrm{DGP}$-2} &  &  &  &  &  &  \\
 & $\BQR$   & 0.433 & 0.080 & 0.134 & 0.138 & 0.433 & 0.08 & 0.134 & 0.138 \\ \hdashline
 & $\HSBQR$ & 0.908 & 0.949 & 0.897 & 0.878 & 0.907 & 0.948 & 0.896 & 0.877 \\
 & $\BRW$   & 0.842 & 0.895 & 0.825 & 0.802 & 0.841 & 0.895 & 0.824 & 0.801 \\
 & $\CQR$   & 0.837 & 0.891 & 0.819 & 0.797 & 0.836 & 0.890 & 0.819 & 0.797 \\
 & $\QVP$   & 0.843 & 0.897 & 0.826 & 0.805 & 0.843 & 0.896 & 0.826 & 0.804 \\
 & $\NCQVP$ & 0.839 & 0.892 & 0.822 & 0.800 & 0.839 & 0.892 & 0.822 & 0.799 \\
 & $\NCQVP_{\mathrm{SAVS}}$ & 0.837 & 0.891 & 0.820 & 0.797 & 0.836 & 0.890 & 0.820 & 0.797 \\ \hline
\multicolumn{2}{l|}{$\mathrm{DGP}$-3} &  &  &  &  &  &  \\
 & $\BQR$   & 0.412 & 0.076 & 0.132 & 0.128 & 0.412 & 0.076 & 0.132 & 0.128 \\ \hdashline
 & $\HSBQR$ & 0.906 & 0.948 & 0.867 & 0.895 & 0.906 & 0.949 & 0.867 & 0.896 \\
 & $\BRW$   & 0.844 & 0.899 & 0.800 & 0.823 & 0.844 & 0.899 & 0.800 & 0.824 \\
 & $\CQR$   & 0.840 & 0.895 & 0.796 & 0.819 & 0.840 & 0.896 & 0.796 & 0.819 \\
 & $\QVP$   & 0.844 & 0.899 & 0.800 & 0.824 & 0.844 & 0.900 & 0.801 & 0.824 \\
 & $\NCQVP$ & 0.841 & 0.896 & 0.798 & 0.821 & 0.842 & 0.897 & 0.798 & 0.821 \\
 & $\NCQVP_{\mathrm{SAVS}}$ & 0.840 & 0.895 & 0.796 & 0.819 & 0.840 & 0.895 & 0.796 & 0.819 \\ \hline
\multicolumn{2}{l|}{$\mathrm{DGP}$-4} &  &  &  &  &  &  \\
 & $\BQR$   & 0.779 & 0.141 & 0.245 & 0.251 & 0.779 & 0.141 & 0.245 & 0.251 \\ \hdashline
 & $\HSBQR$ & 0.996 & 0.998 & 1.006 & 0.987 & 0.995 & 0.997 & 1.005 & 0.986 \\
 & $\BRW$   & 0.903 & 0.941 & 0.893 & 0.873 & 0.902 & 0.941 & 0.892 & 0.872 \\
 & $\CQR$   & 0.901 & 0.940 & 0.891 & 0.871 & 0.902 & 0.940 & 0.892 & 0.871 \\
 & $\QVP$   & 0.906 & 0.943 & 0.897 & 0.878 & 0.906 & 0.943 & 0.897 & 0.877 \\
 & $\NCQVP$ & 0.903 & 0.940 & 0.894 & 0.874 & 0.903 & 0.939 & 0.894 & 0.873 \\
 & $\NCQVP_{\mathrm{SAVS}}$ & 0.903 & 0.940 & 0.894 & 0.873 & 0.902 & 0.939 & 0.893 & 0.872 \\ \hline
\multicolumn{2}{l|}{$\mathrm{DGP}$-5} &  &  &  &  &  &  \\
 & $\BQR$   & 0.720 & 0.134 & 0.221 & 0.230 & 0.717 & 0.134 & 0.220 & 0.229 \\ \hdashline
 & $\HSBQR$ & 0.935 & 0.971 & 0.933 & 0.899 & 0.936 & 0.968 & 0.934 & 0.900 \\
 & $\BRW$   & 0.879 & 0.927 & 0.868 & 0.836 & 0.880 & 0.925 & 0.870 & 0.837 \\
 & $\CQR$   & 0.880 & 0.928 & 0.871 & 0.837 & 0.883 & 0.927 & 0.874 & 0.841 \\
 & $\QVP$   & 0.879 & 0.928 & 0.868 & 0.836 & 0.880 & 0.925 & 0.870 & 0.837 \\
 & $\NCQVP$ & 0.878 & 0.927 & 0.868 & 0.835 & 0.879 & 0.924 & 0.869 & 0.836 \\
 & $\NCQVP_{\mathrm{SAVS}}$ & 0.878 & 0.926 & 0.867 & 0.835 & 0.879 & 0.924 & 0.869 & 0.836 \\ \hline
\end{tabular}%
}
\caption{Out-of-sample predictive performance for $\mathcal{T}=300$ as measured by the weighted quantile score, $\mathrm{qwQS}$, see Equation~\ref{eq:qwQS}. Weighting schemes are $w_{q} = 1/\mathcal{Q}$ ($\QS$) which is equal to the $\mathrm{CRPS}$. $w_{q} = \tau_q(1-\tau_q)$ (Centre) places more weight the central quantiles, $w_{q} = (1-\tau_q)^2$ (Left) places more weight on left tail quantiles, $w_{\tau_q} = \tau^2_q$ (Right) places more weight on right tail quantiles. Performance is shown relative to $\BQR$ whose absolute performance shown above the dotted lines respectively.}
\label{tab:overallbias_main_correlation}
\end{table}
\begin{table}[H]
\centering
\resizebox{0.9\textwidth}{!}{%
\begin{tabular}{ll|cccc|cccc}
 &  & \multicolumn{4}{c|}{$\mathcal{Q}=9$} & \multicolumn{4}{c}{$\mathcal{Q}=39$} \\
 &  & $w_1$ & $w_2$ & $w_3$ & $w_4$ & $w_1$ & $w_2$ & $w_3$ & $w_4$ \\ \hline
\multicolumn{2}{l|}{$\mathrm{DGP}$-1} &  &  &  &  &  &  \\
 & $\BQR$   & 0.446 & 0.084 & 0.136 & 0.142 & 0.418 & 0.076 & 0.131 & 0.135 \\ \hdashline
 & $\HSBQR$ & 0.901 & 0.937 & 0.900 & 0.859 & 0.904 & 0.957 & 0.888 & 0.859 \\
 & $\BRW$   & 0.847 & 0.893 & 0.837 & 0.801 & 0.846 & 0.912 & 0.821 & 0.797 \\
 & $\CQR$   & 0.843 & 0.890 & 0.833 & 0.798 & 0.843 & 0.909 & 0.818 & 0.793 \\
 & $\QVP$   & 0.846 & 0.892 & 0.835 & 0.800 & 0.846 & 0.912 & 0.821 & 0.797 \\
 & $\NCQVP$ & 0.845 & 0.891 & 0.835 & 0.799 & 0.844 & 0.910 & 0.820 & 0.795 \\
 & $\NCQVP_{\mathrm{SAVS}}$ & 0.843 & 0.890 & 0.833 & 0.798 & 0.843 & 0.909 & 0.818 & 0.793 \\ \hline
\multicolumn{2}{l|}{$\mathrm{DGP}$-2} &  &  &  &  &  &  \\
 & $\BQR$   & 0.439 & 0.084 & 0.136 & 0.135 & 0.421 & 0.077 & 0.134 & 0.133 \\ \hdashline
 & $\HSBQR$ & 0.932 & 0.953 & 0.913 & 0.926 & 0.915 & 0.961 & 0.881 & 0.895 \\
 & $\BRW$   & 0.865 & 0.898 & 0.840 & 0.849 & 0.844 & 0.904 & 0.806 & 0.814 \\
 & $\CQR$   & 0.860 & 0.894 & 0.835 & 0.844 & 0.840 & 0.900 & 0.800 & 0.809 \\
 & $\QVP$   & 0.864 & 0.897 & 0.839 & 0.848 & 0.848 & 0.907 & 0.810 & 0.818 \\
 & $\NCQVP$ & 0.862 & 0.895 & 0.837 & 0.846 & 0.843 & 0.902 & 0.804 & 0.813 \\
 & $\NCQVP_{\mathrm{SAVS}}$ & 0.860 & 0.894 & 0.835 & 0.844 & 0.841 & 0.901 & 0.801 & 0.810 \\ \hline
\multicolumn{2}{l|}{$\mathrm{DGP}$-3} &  &  &  &  &  &  \\
 & $\BQR$   & 0.420 & 0.080 & 0.126 & 0.134 & 0.409 & 0.074 & 0.131 & 0.130 \\ \hdashline
 & $\HSBQR$ & 0.924 & 0.950 & 0.937 & 0.882 & 0.892 & 0.949 & 0.858 & 0.863 \\
 & $\BRW$   & 0.864 & 0.901 & 0.868 & 0.815 & 0.831 & 0.899 & 0.79 & 0.794 \\
 & $\CQR$   & 0.859 & 0.896 & 0.863 & 0.810 & 0.827 & 0.896 & 0.786 & 0.790 \\
 & $\QVP$   & 0.862 & 0.899 & 0.867 & 0.814 & 0.832 & 0.900 & 0.792 & 0.796 \\
 & $\NCQVP$ & 0.860 & 0.897 & 0.865 & 0.812 & 0.829 & 0.897 & 0.788 & 0.792 \\
 & $\NCQVP_{\mathrm{SAVS}}$ & 0.858 & 0.896 & 0.863 & 0.810 & 0.827 & 0.896 & 0.786 & 0.790 \\ \hline
\multicolumn{2}{l|}{$\mathrm{DGP}$-4} &  &  &  &  &  &  \\
 & $\BQR$   & 0.790 & 0.148 & 0.244 & 0.249 & 0.770 & 0.138 & 0.245 & 0.249 \\ \hdashline
 & $\HSBQR$ & 0.997 & 1.000 & 1.004 & 0.992 & 0.986 & 0.993 & 0.986 & 0.979 \\
 & $\BRW$   & 0.929 & 0.948 & 0.927 & 0.912 & 0.895 & 0.937 & 0.876 & 0.866 \\
 & $\CQR$   & 0.927 & 0.947 & 0.924 & 0.909 & 0.893 & 0.937 & 0.875 & 0.863 \\
 & $\QVP$   & 0.929 & 0.948 & 0.926 & 0.913 & 0.902 & 0.941 & 0.885 & 0.875 \\
 & $\NCQVP$ & 0.928 & 0.946 & 0.926 & 0.911 & 0.897 & 0.937 & 0.88 & 0.869 \\
 & $\NCQVP_{\mathrm{SAVS}}$ & 0.927 & 0.946 & 0.925 & 0.910 & 0.896 & 0.937 & 0.879 & 0.868 \\ \hline
\multicolumn{2}{l|}{$\mathrm{DGP}$-5} &  &  &  &  &  &  \\
 & $\BQR$   & 0.709 & 0.138 & 0.218 & 0.214 & 0.708 & 0.130 & 0.222 & 0.226 \\ \hdashline
 & $\HSBQR$ & 0.985 & 0.990 & 0.971 & 0.996 & 0.926 & 0.971 & 0.906 & 0.894 \\
 & $\BRW$   & 0.927 & 0.947 & 0.909 & 0.926 & 0.869 & 0.927 & 0.843 & 0.828 \\
 & $\CQR$   & 0.931 & 0.949 & 0.914 & 0.931 & 0.874 & 0.930 & 0.849 & 0.834 \\
 & $\QVP$   & 0.927 & 0.946 & 0.908 & 0.925 & 0.870 & 0.928 & 0.844 & 0.829 \\
 & $\NCQVP$ & 0.926 & 0.946 & 0.908 & 0.925 & 0.869 & 0.927 & 0.843 & 0.828 \\
 & $\NCQVP_{\mathrm{SAVS}}$ & 0.926 & 0.945 & 0.908 & 0.925 & 0.869 & 0.927 & 0.843 & 0.828 \\ \hline
\end{tabular}%
}
\caption{Out-of-sample predictive performance for $\mathcal{T}=300$ as measured by the weighted quantile score, $\mathrm{qwQS}$, see Equation~\ref{eq:qwQS}. Weighting schemes are $w_{q} = 1/\mathcal{Q}$ ($\QS$) which is equal to the $\mathrm{CRPS}$. $w_{q} = \tau_q(1-\tau_q)$ (Centre) places more weight the central quantiles, $w_{q} = (1-\tau_q)^2$ (Left) places more weight on left tail quantiles, $w_{\tau_q} = \tau^2_q$ (Right) places more weight on right tail quantiles. Performance is shown relative to $\BQR$ whose absolute performance shown above the dotted lines respectively.}
\label{tab:overallbias_quant}
\end{table}

\begin{table}[H]
\centering
\resizebox{0.9\textwidth}{!}{%
\begin{tabular}{ll|cccc|cccc}
 &  & \multicolumn{4}{c|}{$\mathcal{T}$=300} & \multicolumn{4}{c}{$\mathcal{T}$=100} \\
 &  & $w_1$ & $w_2$ & $w_3$ & $w_4$ & $w_1$ & $w_2$ & $w_3$ & $w_4$ \\ \hline
\multicolumn{2}{l|}{$\mathrm{DGP}$-1} &  &  &  &  &  &  \\
 & $\BQR$ & 0.424 & 0.078 & 0.133 & 0.134 & 0.438 & 0.081 & 0.137 & 0.138 \\ \hdashline
 & $\HSBQR$ & 0.911 & 0.957 & 0.892 & 0.882 & 0.920 & 0.957 & 0.903 & 0.898 \\
 & $\BRW$ & 0.853 & 0.912 & 0.826 & 0.819 & 0.845 & 0.897 & 0.821 & 0.815 \\
 & $\CQR$ & 0.850 & 0.908 & 0.822 & 0.815 & 0.836 & 0.888 & 0.811 & 0.805 \\
 & $\QVP$ & 0.853 & 0.911 & 0.825 & 0.819 & 0.845 & 0.896 & 0.821 & 0.815 \\
 & $\NCQVP$ & 0.851 & 0.910 & 0.824 & 0.817 & 0.840 & 0.891 & 0.816 & 0.810 \\
 & $\NCQVP_{\mathrm{SAVS}}$ & 0.850 & 0.908 & 0.822 & 0.815 & 0.837 & 0.888 & 0.812 & 0.806 \\ \hline
\multicolumn{2}{l|}{$\mathrm{DGP}$-2} &  &  &  &  &  &  \\
 & $\BQR$ & 0.433 & 0.08 & 0.135 & 0.138 & 0.471 & 0.087 & 0.149 & 0.149 \\ \hdashline
 & $\HSBQR$ & 0.909 & 0.950 & 0.893 & 0.878 & 0.910 & 0.945 & 0.885 & 0.888 \\
 & $\BRW$ & 0.841 & 0.894 & 0.817 & 0.802 & 0.816 & 0.866 & 0.784 & 0.785 \\
 & $\CQR$ & 0.836 & 0.890 & 0.812 & 0.797 & 0.806 & 0.857 & 0.773 & 0.773 \\
 & $\QVP$ & 0.842 & 0.895 & 0.819 & 0.804 & 0.822 & 0.871 & 0.790 & 0.793 \\
 & $\NCQVP$ & 0.839 & 0.892 & 0.815 & 0.800 & 0.811 & 0.860 & 0.779 & 0.78 \\
 & $\NCQVP_{\mathrm{SAVS}}$ & 0.837 & 0.891 & 0.813 & 0.798 & 0.806 & 0.857 & 0.774 & 0.774 \\ \hline
\multicolumn{2}{l|}{$\mathrm{DGP}$-3} &  &  &  &  &  &  \\
 & $\BQR$ & 0.413 & 0.076 & 0.133 & 0.128 & 0.437 & 0.081 & 0.133 & 0.143 \\ \hdashline
 & $\HSBQR$ & 0.906 & 0.950 & 0.864 & 0.896 & 0.905 & 0.938 & 0.915 & 0.851 \\
 & $\BRW$ & 0.843 & 0.900 & 0.796 & 0.825 & 0.823 & 0.870 & 0.823 & 0.763 \\
 & $\CQR$ & 0.839 & 0.897 & 0.791 & 0.821 & 0.812 & 0.861 & 0.811 & 0.753 \\
 & $\QVP$ & 0.844 & 0.901 & 0.796 & 0.825 & 0.826 & 0.872 & 0.826 & 0.767 \\
 & $\NCQVP$ & 0.841 & 0.898 & 0.793 & 0.822 & 0.816 & 0.864 & 0.816 & 0.757 \\
 & $\NCQVP_{\mathrm{SAVS}}$ & 0.839 & 0.896 & 0.791 & 0.821 & 0.812 & 0.860 & 0.811 & 0.752 \\ \hline
\multicolumn{2}{l|}{$\mathrm{DGP}$-4} &  &  &  &  &  &  \\
 & $\BQR$ & 0.779 & 0.142 & 0.244 & 0.252 & 0.811 & 0.147 & 0.252 & 0.265 \\ \hdashline
 & $\HSBQR$ & 0.997 & 0.992 & 1.009 & 0.986 & 0.986 & 0.992 & 1.003 & 0.964 \\
 & $\BRW$ & 0.904 & 0.936 & 0.898 & 0.872 & 0.905 & 0.941 & 0.907 & 0.863 \\
 & $\CQR$ & 0.903 & 0.935 & 0.895 & 0.87 & 0.901 & 0.940 & 0.900 & 0.858 \\
 & $\QVP$ & 0.909 & 0.938 & 0.903 & 0.878 & 0.920 & 0.950 & 0.926 & 0.882 \\
 & $\NCQVP$ & 0.906 & 0.935 & 0.900 & 0.874 & 0.906 & 0.939 & 0.910 & 0.867 \\
 & $\NCQVP_{\mathrm{SAVS}}$ & 0.905 & 0.935 & 0.899 & 0.873 & 0.904 & 0.938 & 0.908 & 0.864 \\ \hline
\multicolumn{2}{l|}{$\mathrm{DGP}$-5} &  &  &  &  &  &  \\
 & $\BQR$ & 0.719 & 0.134 & 0.221 & 0.230 & 0.731 & 0.138 & 0.221 & 0.234 \\ \hdashline
 & $\HSBQR$ & 0.935 & 0.970 & 0.931 & 0.898 & 0.962 & 0.982 & 0.973 & 0.929 \\
 & $\BRW$ & 0.879 & 0.927 & 0.868 & 0.833 & 0.886 & 0.921 & 0.891 & 0.839 \\
 & $\CQR$ & 0.884 & 0.930 & 0.873 & 0.84 & 0.886 & 0.920 & 0.893 & 0.840 \\
 & $\QVP$ & 0.879 & 0.927 & 0.868 & 0.833 & 0.886 & 0.921 & 0.892 & 0.840 \\
 & $\NCQVP$ & 0.878 & 0.926 & 0.867 & 0.833 & 0.883 & 0.918 & 0.889 & 0.837 \\
 & $\NCQVP_{\mathrm{SAVS}}$ & 0.878 & 0.926 & 0.867 & 0.833 & 0.883 & 0.917 & 0.888 & 0.836 \\ \hline
\end{tabular}%
}
\caption{Out-of-sample predictive performance as measured by the weighted quantile score, $\mathrm{qwQS}$, see Equation~\ref{eq:qwQS}. Performance is shown relative to $\BQR$ whose absolute performance shown above the dotted lines respectively, $\varDelta$=0.0.}
\label{tab:prediction-increasing-T}
\end{table}

\clearpage

\section{Performance with prior on $\alpha_q-\alpha_{q-1}$}\label{subsubsec:alpha-predictions}

For the below, we refer to the $\QVP$ model with inclusion of the difference prior implied by Objective function~\ref{eq:fused-objective-function} as $\QVP_{\alpha}$.

\begin{figure}[H]
    \centering
    \includegraphics[width=0.9\linewidth]{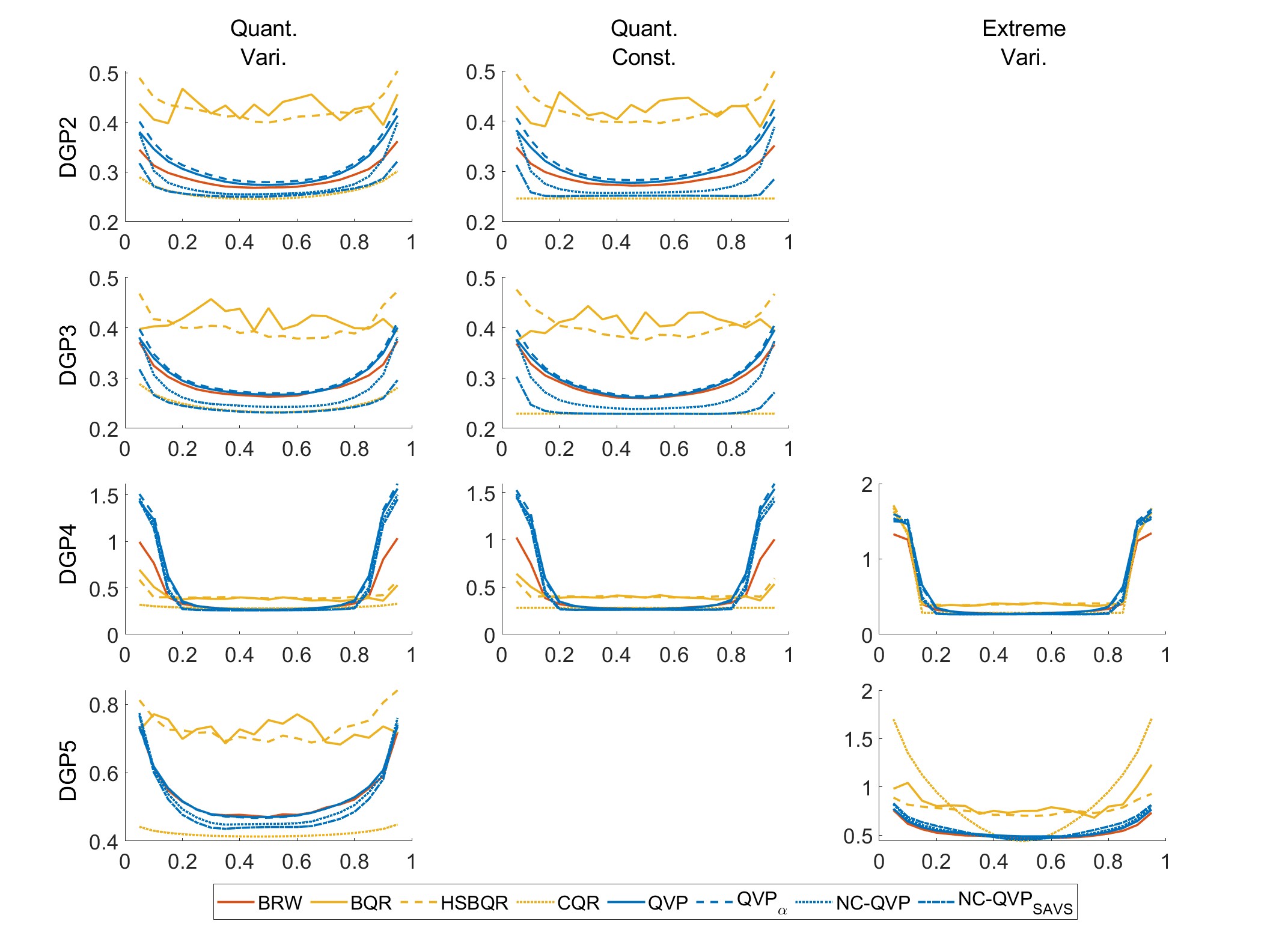}
    \caption{$\rmse$ profile of $\boldsymbol{\beta}$ across 19 quantiles $\tau_q \in \left\{0.05,\dotsc,0.95\right\}$ for $\mathcal{T}=300$ and $\varDelta$=0. Includes $\QVP_\alpha$. Estimates are based on the posterior mean. }
    \label{fig:AlphaCoeffBias}
\end{figure}

\begin{table}[h!]
\centering
\label{tab:QVP_a}
\begin{tabular}{lr|c:cc}
 &  & $\BQR$ & $\QVP$ & $\QVP_{\alpha}$ \\ \hline
\multicolumn{2}{l|}{DGP-1} &  &  &  \\
 & $w_1$ & 0.424 & 0.853 & 0.853 \\
 & $w_2$ & 0.078 & 0.906 & 0.906 \\
 & $w_3$ & 0.133 & 0.825 & 0.825 \\
 & $w_4$ & 0.134 & 0.818 & 0.818 \\ \hline
\multicolumn{2}{l|}{DGP-2} &  &  &  \\
 & $w_1$ & 0.432 & 0.843 & 0.845 \\
 & $w_2$ & 0.080 & 0.895 & 0.896 \\
 & $w_3$ & 0.135 & 0.821 & 0.823 \\
 & $w_4$ & 0.138 & 0.804 & 0.806 \\ \hline
\multicolumn{2}{l|}{DGP-3} &  &  &  \\
 & $w_1$ & 0.412 & 0.844 & 0.845 \\
 & $w_2$ & 0.076 & 0.897 & 0.898 \\
 & $w_3$ & 0.132 & 0.799 & 0.800 \\
 & $w_4$ & 0.128 & 0.826 & 0.827 \\ \hline
\multicolumn{2}{l|}{DGP-4} &  &  &  \\
 & $w_1$ & 0.779 & 0.909 & 0.909 \\
 & $w_2$ & 0.141 & 0.941 & 0.941 \\
 & $w_3$ & 0.245 & 0.900 & 0.901 \\
 & $w_4$ & 0.251 & 0.881 & 0.882 \\ \hline
\multicolumn{2}{l|}{DGP-5} &  &  &  \\
 & $w_1$ & 0.716 & 0.880 & 0.880 \\
 & $w_2$ & 0.134 & 0.926 & 0.926 \\
 & $w_3$ & 0.220 & 0.870 & 0.870 \\
 & $w_4$ & 0.228 & 0.837 & 0.837 \\ \hline
\end{tabular}%
\caption{Out-of-sample predictive performance of $\QVP$ vs. $\QVP_{\alpha}$ as measured by the weighted quantile score, $\mathrm{qwQS}$, see Equation~\ref{eq:qwQS}. Performance is shown relative to $\BQR$ whose absolute performance shown above the dotted lines respectively, $\varDelta$=0.0, $\mathcal{T}=300$, $\mathcal{Q}=19$.}
\end{table}

\clearpage

\section{Further Analysis to the Application}\label{app:extra-results-application}

\subsection{Stress Testing}\label{app:stress-testing}
\citet{chavleishvili2024forecasting} outline a method for stress-testing, using a simulation-based procedure for forecasting under stress scenarios using the QVAR framework. Importantly, in the forecasting setup described above and in \citet{chavleishvili2024forecasting} one can also restrict the quantiles to be of a specific sequence for the different variables thereby creating stress-test scenarios. For example, one can impose a series of consecutive tail quantile shocks (e.g., 10\% quantile events) over a designated forecast horizon. This approach leverages the model’s recursive structure to propagate the effect of these extreme shocks forward in time, producing a simulated forecast distribution that reflects potential stress conditions.

This type of analysis is useful because it offers an intuitive depiction of the range of possible outcomes under stress, thereby highlighting the potential magnitude and duration of adverse effects. This is an essential tool in stress testing because they not only illustrate the central forecast under baseline conditions but also reveal the dispersion and asymmetry in the forecast distribution when the system is subjected to extreme shocks. By comparing the stressed forecast paths with baseline forecasts, analysts and policymakers can gain insights into the robustness of the economy and identify periods of heightened vulnerability. This simulation-based approach, rooted in the QVAR framework, thus provides a more comprehensive view of risk by capturing the entire conditional distribution rather than focusing solely on the mean outcome.

\begin{figure}
    \centering
    \includegraphics[width=\linewidth]{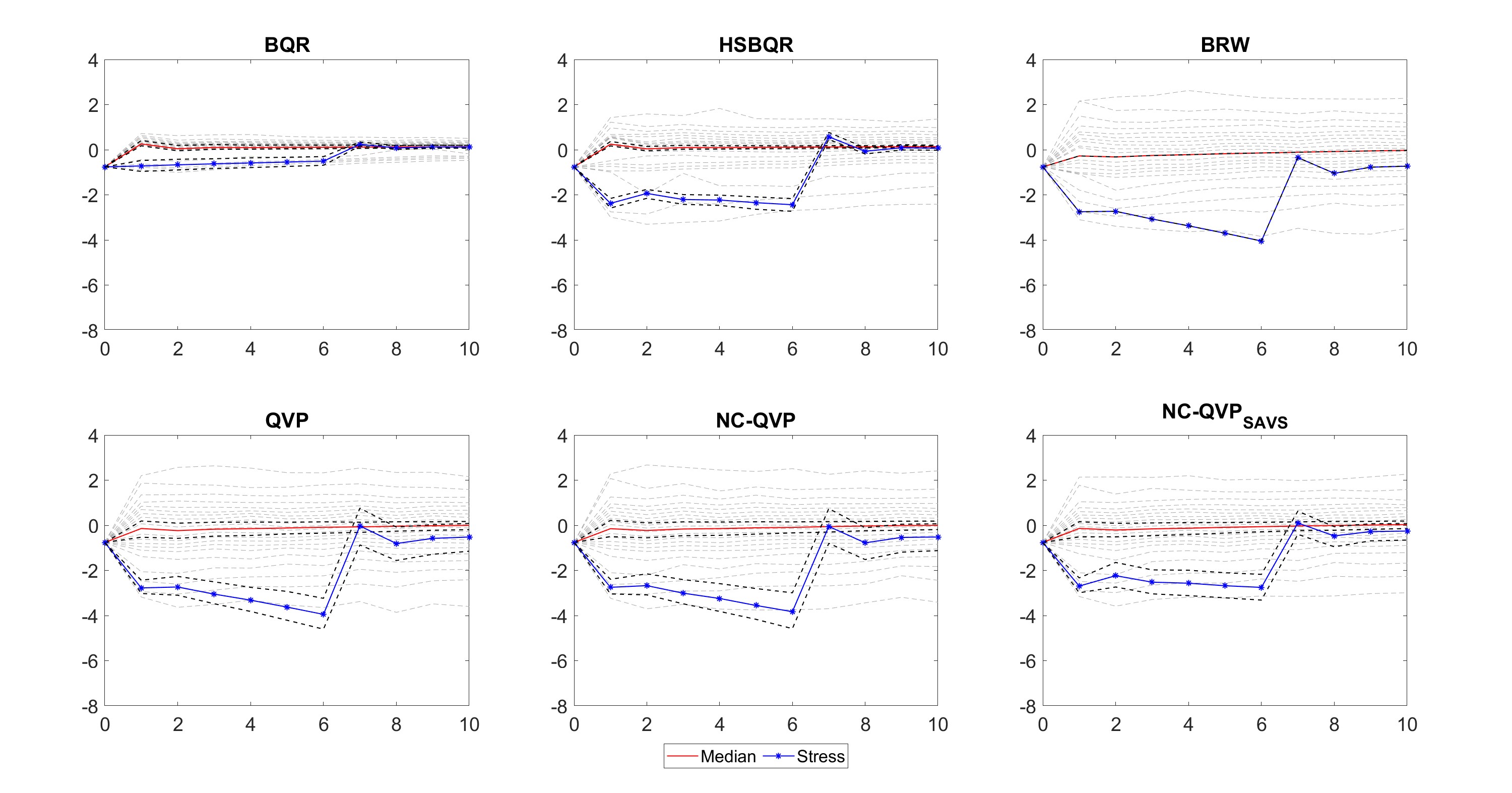}
    \caption{Stress-testing}
    \label{fig:StressTest}
\end{figure}

Here we note, that on account of the Bayesian framework, we can generate confidence bands around the stress-test scenarios by drawing randomly from the posterior draws. In this way we deviate from \citet{chavleishvili2024forecasting}, and draw randomly from the posterior draws as well to obtain a confidence band around the scenarios. Beyond this difference, everything else is the same as in \citet{chavleishvili2024forecasting}. Namely, the stress scenario is constructed by assuming that the CISS index sees a 0.9 quantile realisation for 6 months followed by median realisations, while the median scenario assumes that both GDP and CISS observe median outcomes across the forecast horizon. The results for IP for the stress-testing scenarios are shown in figure (\ref{fig:StressTest}).\footnote{No bands are constructed for the BRW. This was done because a different procedure would be neccessary to construct these bands than the Bayesian estimators.}

The figure reveals some key differences. First, the BQR produces much lower stress scenarios with narrower spreads, than the other methods. Second, the BRW, QVP, and NCQVP produces very similar profiles, with a sharp drop in IP in $h=1$, followed by a slower decrease in IP growth until $h=6$. Third, the SAVS variant showcase similar profiles as well, but they differ from their non-SAVS counterparts by showing slow recovery from $h=2$ to $h=6$. Finally, while the HSBQR produces a similar stress profile as the joint estimation methods, the stress scenario is not as negative as the joint estimation methods. These highlights how the choice of modelling quantiles jointly (with fused-shrinkage) has a noticeable influence on the outcomes. 

From a policy standpoint, these differences in tail forecasts can significantly influence decisions regarding capital buffers, liquidity requirements, or macroprudential measures. A method that systematically yields tighter stress estimates might understate risks, leading policymakers to adopt insufficient safeguards. Conversely, overly conservative forecasts could prompt more stringent policies, potentially dampening economic activity. By comparing a range of models and stress outcomes, policymakers can better triangulate the severity of risks, ensuring that policy interventions strike a balance between prudence and efficiency.

\subsection{Coefficient Posteriors}\label{app-sec:fig-posteriors}{\color{white}.}

\begin{figure}[H]
     \centering
     \begin{subfigure}[b]{\textwidth}
         \centering
         \includegraphics[width=\textwidth]{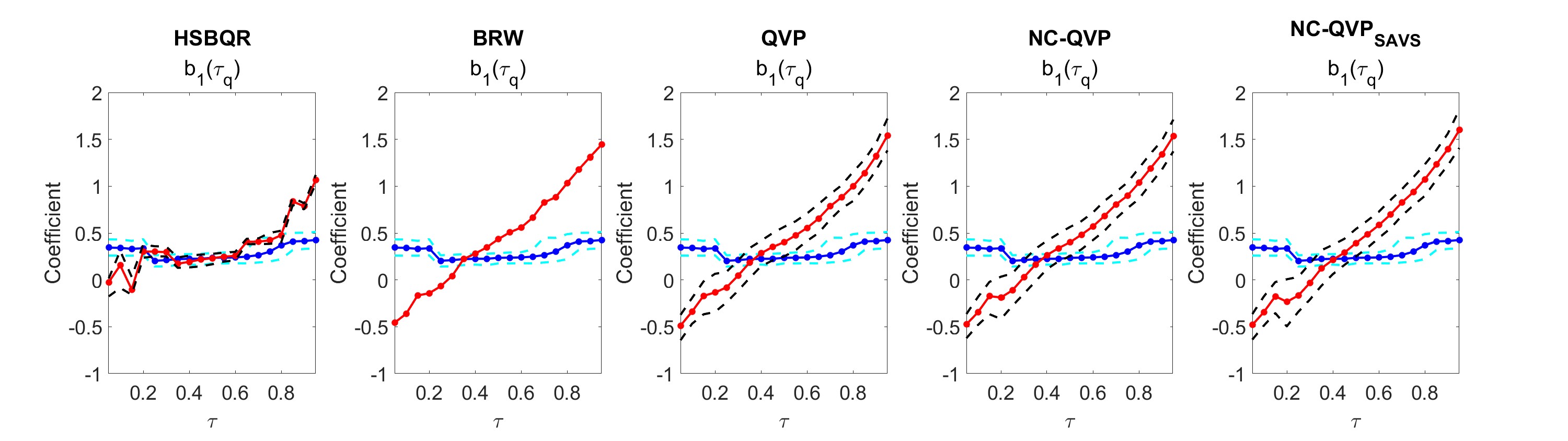}
     \end{subfigure}
     \vfill
     \begin{subfigure}[b]{\textwidth}
         \centering
         \includegraphics[width=\textwidth]{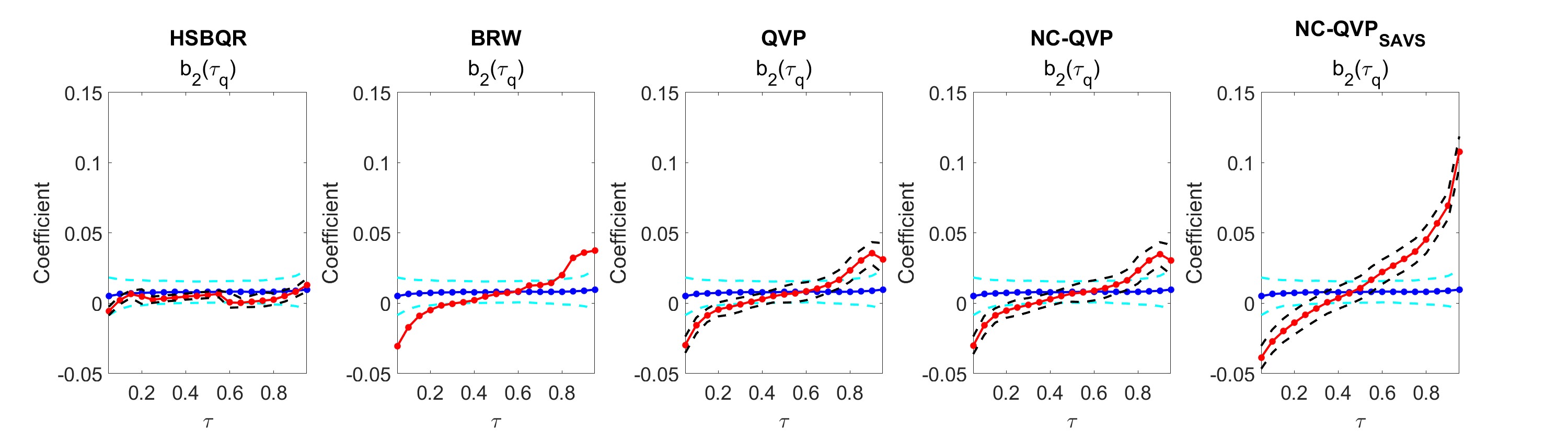}
     \end{subfigure}
        \caption{Intercept posteriors}
        \label{fig:CoeffsIntercept}
\end{figure}

\begin{figure}[H]
    \centering
    \includegraphics[width=\linewidth]{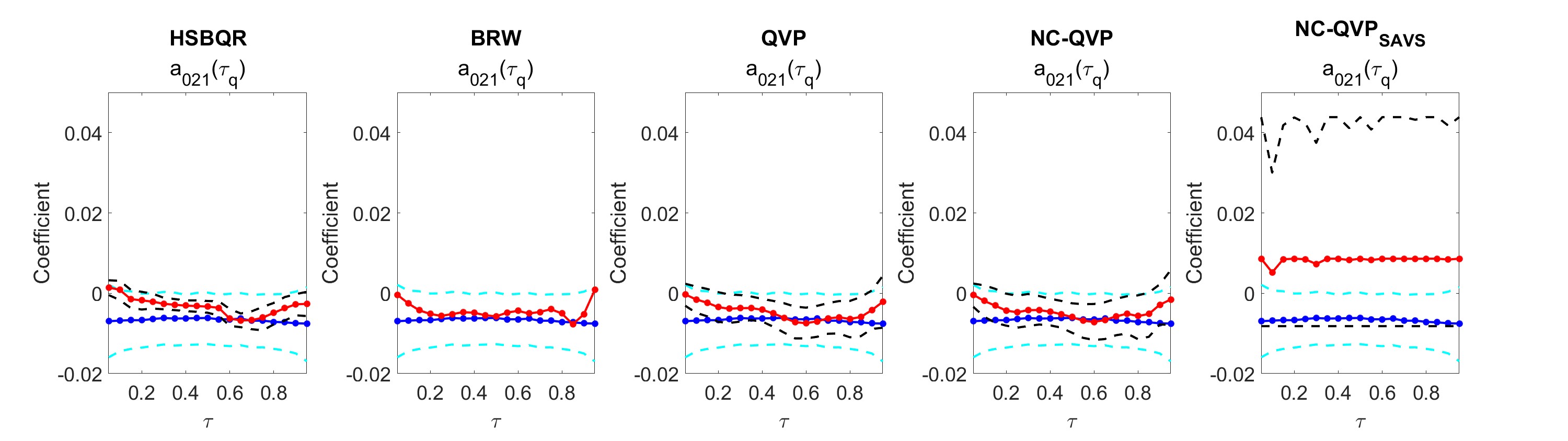}
    \caption{Contemporaneous effect posteriors}
    \label{fig:CoeffsContemp}
\end{figure}

\begin{figure}[H]
     \centering
     \begin{subfigure}[b]{\textwidth}
         \centering
         \includegraphics[width=\textwidth]{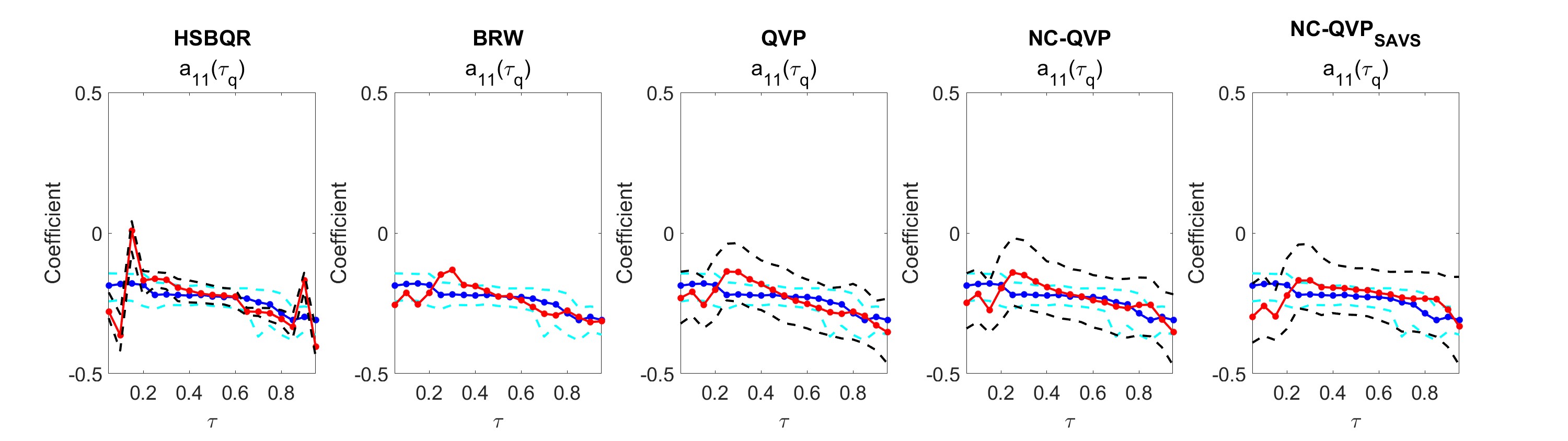}
     \end{subfigure}
     \vfill
     \begin{subfigure}[b]{\textwidth}
         \centering
         \includegraphics[width=\textwidth]{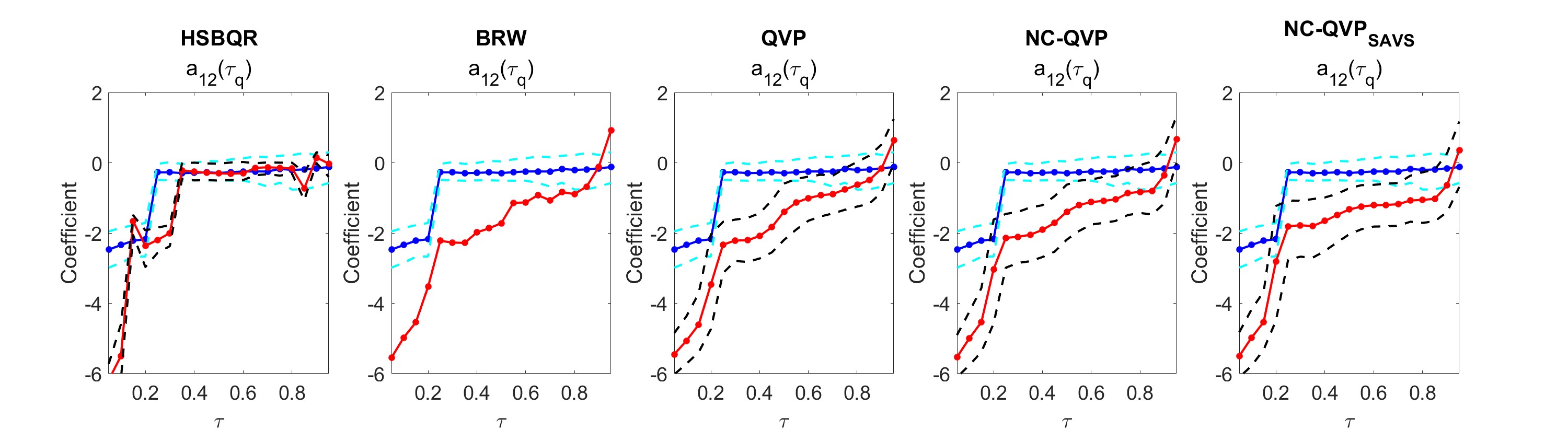}
     \end{subfigure}
     \vfill
     \begin{subfigure}[b]{\textwidth}
         \centering
         \includegraphics[width=\textwidth]{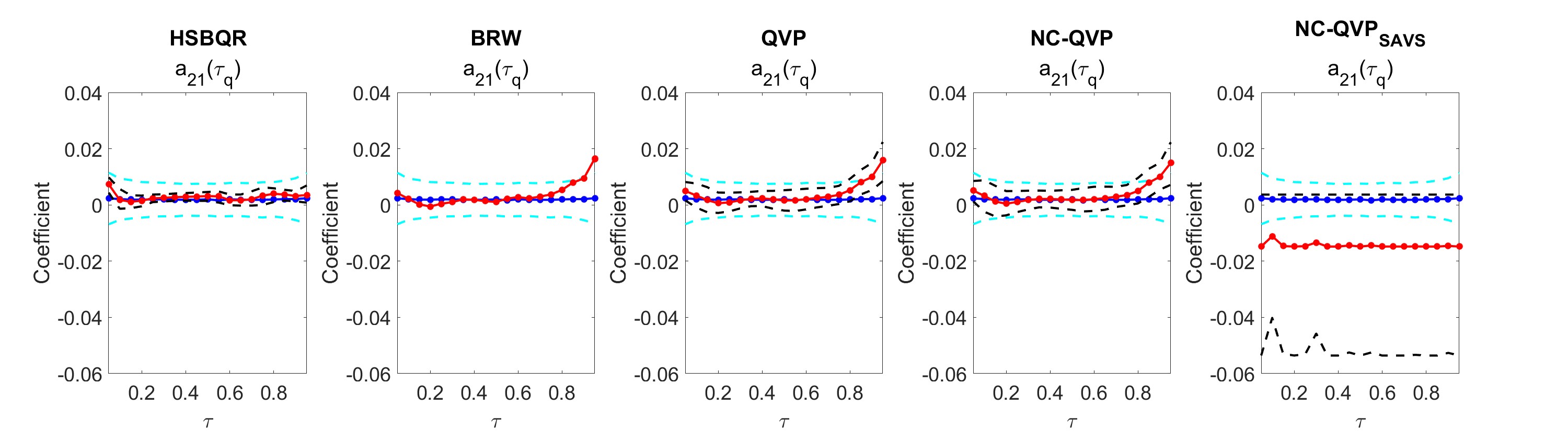}
     \end{subfigure}
     \vfill
     \begin{subfigure}[b]{\textwidth}
         \centering
         \includegraphics[width=\textwidth]{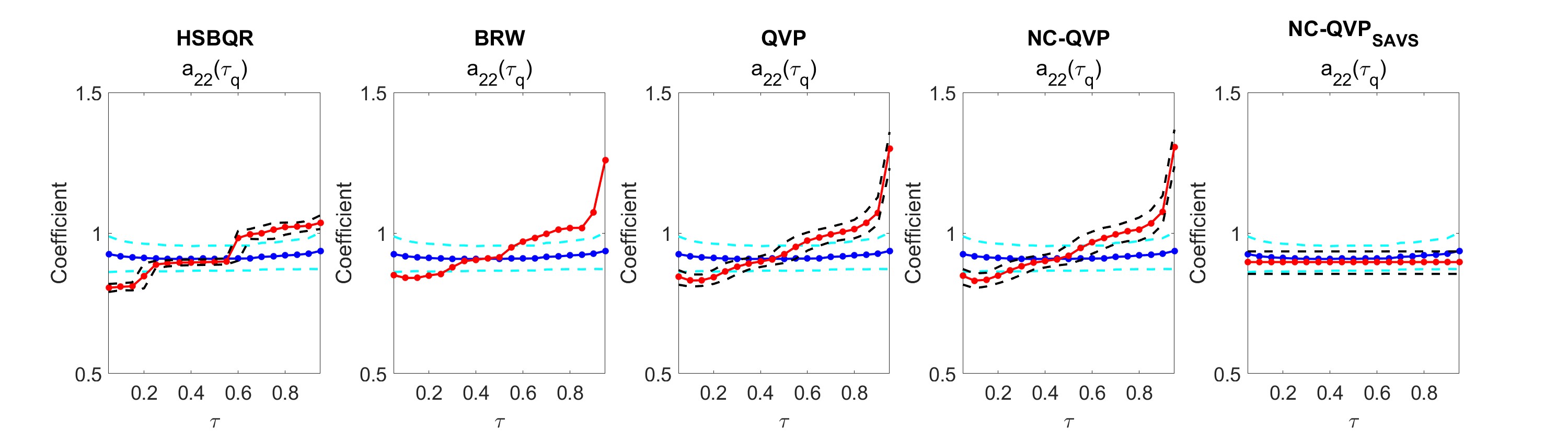}
     \end{subfigure}
        \caption{Lagged effects posteriors}
        \label{fig:CoeffsLag}
\end{figure}
\end{appendices}
\end{document}